\documentclass[twocolumn,twocolappendix,tighten]{aastex701}

\usepackage[T1]{fontenc}
\usepackage{lmodern}

\usepackage{amsmath}
\usepackage{mathtools}
\usepackage{bm}
\usepackage{booktabs}
\usepackage{siunitx}
\usepackage{tikz}
\usepackage{pgfplots}
\usepgfplotslibrary{patchplots}
\pgfplotsset{compat=1.18}

\DeclareSIUnit{\Kcmb}{\mathrm{K_{CMB}}}
\DeclareSIUnit{\Kcmbsq}{\mathrm{K_{CMB}^2}}
\DeclareSIUnit{\pix}{\mathrm{pix}}
\DeclareSIUnit{\jansky}{\mathrm{Jy}}

\newcommand{\splitcell}[1]{\parbox[t]{0.20\textwidth}{\raggedright\strut#1\strut}}
\newcommand{\motivcell}[1]{\parbox[t]{0.74\textwidth}{\raggedright\strut#1\strut}}
\newcommand{\RA}[1]{\ang[
    angle-symbol-degree={^{\mathrm{h}}},
    angle-symbol-minute={^{\mathrm{m}}},
    angle-symbol-second={^{\mathrm{s}}},
    minimum-integer-digits=2,
    quantity-product={}
]{#1}}
\newcommand{\Dec}[1]{\ang[
    minimum-integer-digits=2
]{#1}}

\begin{document}

\defcitealias{polarbear_collaboration_exploration_2024}{PB24}
\defcitealias{p_a_r_ade_measurement_2014}{PB14}

\title{Millimeter-Wavelength Constraints on Circular Polarization of the Cosmic Microwave Background at Arcminute Scales with the Simons Array}

\ExplSyntaxOn
\seq_new:N \g_pb_authors_seq
\NewDocumentCommand{\addauthor}{O{0} m +m}
  { \seq_gput_right:Nn \g_pb_authors_seq { {#1} {#2} {#3} } }
\cs_new:Npn \__pb_order:nnn #1#2#3 {#1}
\cs_new:Npn \__pb_key:nnn #1#2#3 {#2}
\cs_new_protected:Npn \__pb_emit:nnn #1#2#3 { #3 }
\cs_new_protected:Npn \__pb_emit_item:n #1 { \__pb_emit:nnn #1 }
\cs_new_protected:Npn \__pb_compare:nn #1#2
  {
    \int_compare:nNnTF { \__pb_order:nnn #1 } > { \__pb_order:nnn #2 }
      { \sort_return_swapped: }
      {
        \int_compare:nNnTF { \__pb_order:nnn #1 } < { \__pb_order:nnn #2 }
          { \sort_return_same: }
          {
            \str_compare:eNeTF { \__pb_key:nnn #1 } > { \__pb_key:nnn #2 }
              { \sort_return_swapped: } { \sort_return_same: }
          }
      }
  }
\cs_new_protected:Npn \printauthors
  {
    \seq_gsort:Nn \g_pb_authors_seq { \__pb_compare:nn {##1} {##2} }
    \seq_map_function:NN \g_pb_authors_seq \__pb_emit_item:n
  }
\ExplSyntaxOff

\addauthor{arnold}{
\author[orcid=0000-0002-3407-5305,gname=Kam,sname='Arnold']{Kam Arnold}
\affiliation{Department of Astronomy, University of California, San Diego, La Jolla, California 92093, USA}
\email{ksarnold@physics.ucsd.edu}
}

\addauthor{barron}{
\author[orcid=0000-0002-1623-5651,gname=Darcy,sname=Barron]{Darcy R. Barron}
\affiliation{Department of Physics and Astronomy, University of New Mexico, Albuquerque, NM 87131, USA}
\email{dbarron2@unm.edu}
}

\addauthor{chu}{
\author[orcid=0000-0001-5884-911X,gname=Matthew,sname=Chu]{Matthew R. Chu}
\affiliation{Department of Physics, University of California, San Diego, La Jolla, California 92093, USA}
\email[show]{mrchu@ucsd.edu}
\correspondingauthor{Matthew R. Chu}
}

\addauthor{farias}{
\author[orcid=0000-0002-4101-2513,gname=Nicole,sname=Farias]{Nicole Farias}
\affiliation{Department of Physics, University of California, Berkeley, CA 94720, USA}
\email{nicolengfarias@gmail.com}
}

\addauthor{ish}{
\author[orcid=0009-0005-6774-3487,gname=Ethan,sname=Ish]{Ethan H. Ish}
\affiliation{Department of Physics, Westmont University, Santa Barbara, California 93108, USA}
\email{ethanish@gmail.com}
}

\addauthor{ito}{
\author[orcid=0009-0008-3216-4022,gname=Jennifer,sname=Ito]{Jennifer Ito}
\affiliation{Department of Physics, Westmont University, Santa Barbara, California 93108, USA}
\email{jito@westmont.edu}
}

\addauthor{keating}{
\author[orcid=0000-0003-3118-5514,gname=Brian,sname=Keating]{Brian Keating}
\affiliation{Department of Physics, University of California, San Diego, La Jolla, California 92093, USA}
\email{bkeating@ucsd.edu}
}

\addauthor{lee}{
\author[orcid=0000-0003-3106-3218,gname=Adrian,sname=Lee]{Adrian T. Lee}
\affiliation{Department of Physics, University of California, Berkeley, CA 94720, USA}
\email{adrian.lee@berkeley.edu}
}

\addauthor{lonappan}{
\author[orcid=0000-0003-1200-9179,gname=Anto,sname=Lonappan]{Anto I. Lonappan}
\affiliation{Department of Physics, University of California, San Diego, La Jolla, California 92093, USA}
\email{alonappan@ucsd.edu}
}

\addauthor{miyashita}{
\author[orcid=0009-0009-8176-2384,gname=Alec,sname=Miyashita]{Alec Miyashita}
\affiliation{Department of Physics, University of California, Berkeley, CA 94720, USA}
\email{alecmiyashita@berkeley.edu}
}

\addauthor{ng lowry}{
\author[orcid=0009-0006-2653-3034,gname=Lindsay,sname='Ng Lowry']{Lindsay Ng Lowry}
\affiliation{Department of Physics, University of California, Berkeley, CA 94720, USA}
\email{lindsay@nglowry.com}
}

\addauthor{porter}{
\author[orcid=0009-0003-0464-505X,gname=Luke,sname=Porter]{Luke M. Porter}
\affiliation{Department of Physics, Westmont University, Santa Barbara, California 93108, USA}
\email{lporter@westmont.edu}
}

\addauthor{hwang}{
\author[orcid=0009-0007-2462-3762,gname=Kailey,sname=Hwang]{Kailey Hwang}
\affiliation{Department of Physics, Westmont University, Santa Barbara, California 93108, USA}
\email{khwang@westmont.edu}
}

\addauthor{shayan arani}{
\author[orcid=0009-0005-4168-2858,gname=Shahed,sname='Shayan Arani']{Shahed Shayan Arani}
\affiliation{Department of Physics, University of California, San Diego, La Jolla, California 92093, USA}
\email{sshayanarani@ucsd.edu}
}

\addauthor{siritanasak}{
\author[orcid=0000-0001-6830-1537,gname=Praween,sname=Siritanasak]{Praween Siritanasak}
\affiliation{National Astronomical Research Institute of Thailand, Chiangmai, 50180, Thailand}
\email{praween@narit.or.th}
}

\addauthor{hasegawa}{
\author[orcid=0000-0003-1443-1082,gname=Masaya,sname=Hasegawa]{Masaya Hasegawa}
\affiliation{International Center for Quantum-field Measurement Systems for Studies of the Universe and Particles, High Energy Accelerator Research Organization, Tsukuba, Ibaraki 305-0801, Japan}
\affiliation{Institute of Particle and Nuclear Studies, High Energy Accelerator Research Organization, Tsukuba, Ibaraki 305-0801, Japan}
\email{masaya.hasegawa@kek.jp}
}

\addauthor{fujino}{
\author[orcid=0000-0002-1211-7850,gname=Takuro,sname=Fujino]{Takuro Fujino}
\affiliation{International Center for Quantum-field Measurement Systems for Studies of the Universe and Particles, High Energy Accelerator Research Organization, Tsukuba, Ibaraki 305-0801, Japan}
\email{fujinot@post.kek.jp}
}

\addauthor{hirose}{
\author[orcid=0000-0003-3336-134X,gname=Haruaki,sname=Hirose]{Haruaki Hirose}
\affiliation{Institute of Particle and Nuclear Studies, High Energy Accelerator Research Organization, Tsukuba, Ibaraki 305-0801, Japan}
\email{hiroseh224@gmail.com}
}

\addauthor{chinone}{
\author[orcid=0000-0002-3266-857X,gname=Yuji,sname=Chinone]{Yuji Chinone}
\affiliation{International Center for Quantum-field Measurement Systems for Studies of the Universe and Particles, High Energy Accelerator Research Organization, Tsukuba, Ibaraki 305-0801, Japan}
\affiliation{Kavli Institute for the Physics and Mathematics of the Universe (WPI), UTIAS, The University of Tokyo, Kashiwa, Chiba 277-8583, Japan}
\email{chinoney@gmail.com}
}

\addauthor{hazumi}{
\author[orcid=0000-0001-6830-8309,gname=Masashi,sname=Hazumi]{Masashi Hazumi}
\affiliation{Department of Physics and Center for High Energy and High Field Physics (CHiP), National Central University, Taoyuan City 320317, Taiwan}
\email{masashi.hazumi@phy.ncu.edu.tw}
}

\printauthors

\collaboration{all}{The \textsc{Polarbear} Collaboration}

\begin{abstract}
We present constraints on the cosmic microwave background (CMB) circular polarization anisotropy at arcminute angular scales, using dedicated measurements from the Simons Array taken with a quarter-wave plate and rotating half-wave plate. A detection of circular polarization in the CMB would probe a range of proposed generation mechanisms, from Standard Model astrophysics to physics beyond it. At an observing frequency of 90 GHz, we place constraints over the multipole range $300 \leq \ell \leq 2000$, corresponding to angular scales of \qtyrange{5.4}{36}{\arcminute}, in bands of width $\Delta \ell = 100$, roughly tripling the highest multipole previously reached by a $V$-mode power spectrum measurement. Within this domain, we constrain the 95\% confidence upper bound of $\ell (\ell+1) C_\ell^{VV} / (2\pi)$ to range from ${\sim}\qty{22}{\micro\Kcmbsq}$ at the $\ell = 350$ bandpower to ${\sim}\qty{3.5e3}{\micro\Kcmbsq}$ at the $\ell = 1950$ bandpower. These limits are derived from ${\sim}2.5$ months of observations of each of two small regions of the sky located away from the Galactic plane, totaling ${\sim}\qty{12}{\deg^2}$ of effective sky area. Because the methodology of this study marginalizes known foreground contamination, these bounds serve as robust, conservative limits on the true CMB circular polarization signal down to the sensitivity limits of the instrument. Additionally, we provide an experimental and analytical framework for implementing circular polarization observations on other telescopes by utilizing a quarter-wave plate and half-wave plate design.
\end{abstract}


\section{Introduction}
\label{sec:intro}

The cosmic microwave background (CMB) is a mostly unpolarized blackbody radiation signal which is the most distant detectable direct evidence of the Big Bang and its immediate aftermath. Understanding how the post-Big Bang events affect the CMB signal is one of the foremost ongoing scientific goals in the field of cosmology, and experiments have constrained the temperature anisotropies at both small and large angular scales \citep{das_atacama_2014, story_measurement_2013, bennett_nine-year_2013, collaboration_planck_2020}. Most contemporary experiments measuring the CMB focus on characterizing the linear polarization ($E$ and $B$-modes), which can shed light on inflationary physics and other early-Universe phenomena \citep{kamionkowski_statistics_1997, zaldarriaga_all-sky_1997, hu_cmb_1997}. The standard cosmological model, Lambda cold dark matter ($\Lambda$CDM), predicts a measurable linear polarization signal, but no appreciable circular polarization in the CMB. However, there are several potential secondary mechanisms that could produce it.

Before decoupling, a primordial magnetic field acting on the cold pre-decoupling plasma sources $V$ modes in the presence of adiabatic curvature perturbations \citep{giovannini_v-mode_2009}. After decoupling, the most studied channel is Faraday conversion, whereby a magnetic field threading a relativistic plasma converts pre-existing linear polarization, considered both for magnetized galaxy clusters \citep{cooray_is_2003} and for the supernova remnants of Population III (Pop. III) stars \citep{de_circular_2015}. Circular polarization may also be generated after decoupling through scattering of CMB photons off the cosmic neutrino background (C$\nu$B) \citep{mohammadi_evidence_2014} and through the nonlinear photon--photon interactions of the Euler--Heisenberg effective Lagrangian \citep{sawyer_photon-photon_2014, motie_euler-heisenberg_2012}. A smaller set of proposals instead invokes physics beyond the Standard Model, coupling photons to an external vector field through a Chern--Simons term \citep{alexander_generation_2009} or to a cosmological pseudoscalar field acting as dark matter \citep{finelli_rotation_2009}. A comprehensive summary of these origins is provided by \citet{king_circular_2016}. Distinct from all of these are foregrounds, which emit circular polarization of their own rather than acting on CMB photons. Galactic components, such as synchrotron and thermal dust emission, are not expected to contribute appreciably at \qty{90}{\giga\Hz} \citep{king_circular_2016}. The far brighter atmospheric signal, produced by Zeeman splitting of molecular oxygen, is dominant at larger angular scales than those within this study's scope \citep{padilla_two-year_2020, Fujino_2025}, and is in any case removed by the filtering described in Section~\ref{sec:tod}. A diffuse signal observed here would therefore be dominated by converted CMB photons, and we refer to it as CMB circular polarization throughout.

Observational upper limits on the circular polarization anisotropy of the CMB have historically spanned disparate angular scales and employed varied instrumental techniques. The earliest came from Very Large Array (VLA) interferometry at \qty{5}{\giga\Hz}, whose highly localized measurements constrained the root-mean-square fractional circular polarization to $\Delta V/T \le 6.3 \times 10^{-5}$, or ${\sim}\qty{173}{\micro\Kcmb}$ at the widest synthesized beam, on scales $\qty{60}{\arcsecond} \le \theta \le \qty{160}{\arcsecond}$ \citep{partridge_linear_1988}. On much larger angular scales, the Milano Polarimeter (MIPOL) utilized a correlation receiver at \qty{33}{\giga\Hz} to set constraints at scales of \qty{8}{\degree} to \qty{24}{\degree}, bounding the $V$-mode power spectrum at the level of $\qty{1e5}{\micro\Kcmbsq}$ \citep{mainini_improved_2013}. The SPIDER balloon-borne experiment significantly improved these large-scale constraints at \qtylist{95;150}{\giga\Hz}; by leveraging nonidealities in their half-wave plate polarization modulators, SPIDER placed upper limits on scales ${<}\qty{10}{\degree}$ at the level of $\qty{1e2}{\micro\Kcmbsq}$ \citep{nagy_new_2017}. Most recently, the Cosmology Large Angular Scale Surveyor (CLASS) achieved a direct measurement of circular polarization at \qty{40}{\giga\Hz} using a Variable-delay Polarization Modulator. After characterizing and filtering a bright atmospheric oxygen dipole at the level of $\qty{124 \pm 4}{\micro\kelvin}$, CLASS established the tightest limits at the time on the large-scale CMB circular polarization power spectrum \citep{padilla_two-year_2020}. That analysis has since been extended to the CLASS \qtylist{90;150;220}{\giga\Hz} bands, improving the large-scale limits by a factor of ${\sim}5$ and reaching $\ell = 700$ \citep{essinger-hileman_improved_2026}.

In this paper, we present new millimeter-wavelength constraints on the circular polarization of the CMB using data from the \textsc{Polarbear}-2b (PB-2b) receiver of the Simons Array, the successor telescope to \textsc{Polarbear}. This study presents a novel approach to efficiently measuring circular polarization, using a stationary quarter-wave plate (QWP) positioned in front of a constantly spinning half-wave plate (HWP). The technique is available to any experiment already operating linear polarization sensitive detectors and a rotating HWP, along with established CMB mapmaking and power spectrum estimation analysis tools. The remainder of this paper is organized as follows: Section~\ref{sec:instrument} describes the instrument, optical configuration, and observations; Section~\ref{sec:calibration} describes the calibration techniques; Section~\ref{sec:analysis} details the data analysis pipeline and our systematic error characterization; Section~\ref{sec:results} presents our unblinded limits; and we conclude in Section~\ref{sec:discussion}.

\section{Instrument and Observations}
\label{sec:instrument}

The measurements presented in this work utilize data collected by the PB-2b receiver, the second deployed receiver of the Simons Array \citep{arnold_simons_2014, Suzuki2016, stebor_simons_2016}, located in the Atacama Desert in Chile. It features a continuously rotating cryogenic HWP \citep{hill_cryogenic_2020} to modulate incoming polarization and systematically suppress low-frequency atmospheric noise. In this section, we detail the optical and cryogenic architecture of the receiver, the properties of the transition-edge sensor (TES) bolometer array, and the specific observational campaigns that form the basis of our circular polarization constraints.

\subsection{The POLARBEAR-2b Receiver}

\begin{figure*}[htbp!]
    \centering
    \includegraphics[width=\textwidth]{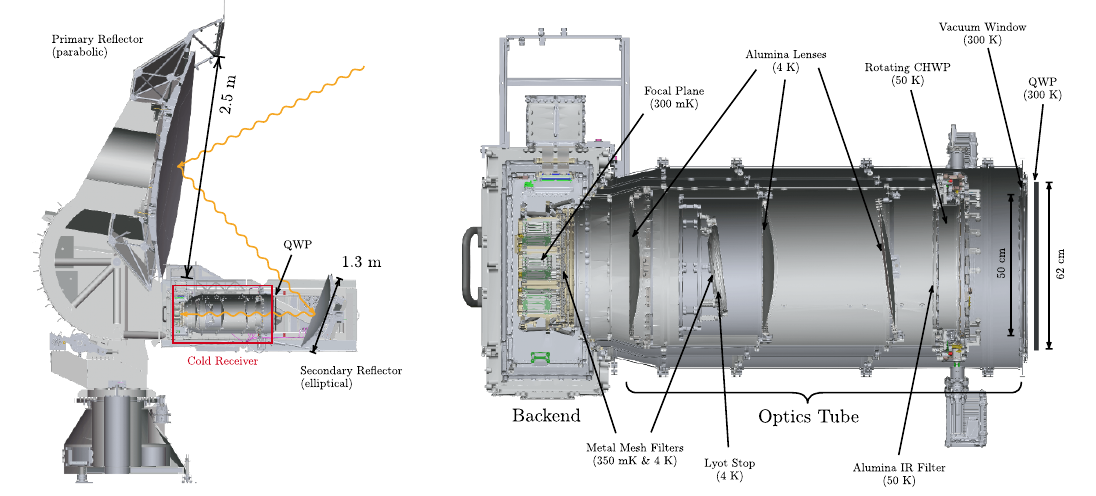}
    \caption{Schematics describing the setup of PB-2b. \emph{Left}: Telescope receiver mounted, with the waved orange arrow indicating path of light rays as they travel across the two mirrors, through the QWP, and into the receiver. \emph{Right}: Interior of the receiver, with the path traced through the QWP at ambient air temperature, through the cold HWP (CHWP), lenses and filters, and into the detectors on the focal plane. The QWP is placed \qty{15}{\centi\meter} from the vacuum window.}
    \label{fig:pb2b_schematic}
\end{figure*}

The Simons Array is located at an altitude of \qty{5200}{\meter} in the Atacama Desert in Chile. The PB-2b receiver was mounted on an off-axis Gregorian telescope featuring a 2.5-meter primary mirror aperture \citep{Suzuki2016, matsuda_cosmic_2017}, shown in Figure~\ref{fig:pb2b_schematic}. The PB-2b focal plane consists of 7,588 TESs coupled to dichroic, dual-polarization pixels utilizing lenslet-coupled sinuous antennas, providing sensitivity to radiation at \qtylist{90;150}{\giga\Hz} \citep{Westbrook_2018} (however, only \qty{90}{\giga\Hz} is utilized for this study as explained in Section~\ref{sec:polmod}), so chosen to avoid foregrounds, specifically atmospheric absorption bands and Galactic synchrotron radiation at lower frequencies. The detectors are read out using a $40{\times}$ digital frequency-domain multiplexing (DfMUX) system \citep{barron_integrated_2021}. The focal plane detectors are split into $Q$-pixels and $U$-pixels, while the latter are rotated by \qty{45}{\degree} from the former.

Following the initial deployment of the PB-2b receiver to the Simons Array in 2022, the instrument's baseline optical performance and readout electronics were successfully validated \citep{lowry_deployment_2022, russell_deployment_2024}. However, in-field evaluations determined that the on-sky sensitivity did not reach the initial design targets. The median detector noise-equivalent temperature (NET) for PB-2b at \qty{90}{\giga\Hz} is ${\sim}\qty{3000}{\micro\Kcmb\sqrt{\second}}$, against a design value of \qty{360}{\micro\Kcmb\sqrt{\second}}, a loss driven by low optical throughput and elevated readout noise \citep{farias_performance_2026}. Furthermore, overall detector performance was hampered by an excess noise component that is strongly suspected to originate from environmental radio frequency (RF) interference at the observing site \citep{farias_performance_2026}. This characterization predates the installation of the QWP of Section~\ref{sec:polmod}, whose ${\sim}\qty{13}{\percent}$ transmission loss \citep{sakaguri_development_2026} further degrades the effective sensitivity of the observations presented here. Regardless, the elevated instrument noise primarily propagates as statistical uncertainty rather than as a systematic bias.

\subsection{Polarization Modulators}
\label{sec:polmod}

To modulate the incoming celestial polarization above unpolarized $1/f$ atmospheric fluctuations, the receiver utilizes a cryogenic HWP. The HWP is situated inside the optics tube at 50 K and rotates continuously at approximately \qty{2.025}{\Hz}, placing the polarization modulation at $4f_\text{HWP} = \qty{8.1}{\Hz}$. While a standard continuously rotating HWP provides modulation of linear polarization (Stokes $Q$ and $U$) and was the primary focus of the initial PB-2b commissioning \citep{russell_deployment_2024}, explicitly isolating the circular polarization (Stokes $V$) signal at high efficiency requires additional optical configuration. For the observations presented in this work, our optical system utilizes both a static QWP and the rotating HWP. The addition of the QWP introduces a known phase retardance that, when coupled with the rotating HWP, allows the incoming Stokes $V$ signal to be modulated and effectively separated from the linear polarization terms in the timestream, as shown in \citet{sakaguri_development_2026}.

The physical characteristics and optical performance of this QWP-HWP system have been extensively evaluated and optimized for the \qty{90}{\giga\Hz} band \citep{sakaguri_development_2026}. The QWP is installed on the sky-side of the receiver and is kept at ambient air temperature. It is constructed from Rexolite 1422 (cross-linked polystyrene)\footnote{C-Lec Plastics, Inc., Philadelphia, PA, USA.}, which is not itself a birefringent material; instead, grooves machined into both faces produce form birefringence, giving different effective indices parallel and perpendicular to the grooves. These grooves are machined with depth and separations to optimize performance at \qty{90}{\giga\Hz}. The \qty{150}{\giga\Hz} band is disfavored in turn because the atmosphere is both brighter and less stable there, sitting between the \qty{118}{\giga\Hz} oxygen line and the wing of the \qty{183}{\giga\Hz} water line \citep{pardo_atmospheric_2001}. Laboratory measurements demonstrate that the QWP achieves a band-averaged circular-to-linear conversion efficiency ($|M^\mathrm{QWP}_{VU}|$) of 99.0\%, with a band-averaged unconverted fraction ($M^\mathrm{QWP}_{UU} = M^\mathrm{QWP}_{VV}$) of $-7.2\%$; because only $I$, $Q$, and $U$ are physically reconstructed downstream of the QWP, with sky $V$ recovered from the $U$ channel, this leakage represents sky $U$ that fails to convert and remains as spurious $U$, directly contaminating the recovered $V$ measurement. Spatial variations across the plate differ by less than 1\% in amplitude and \qty{3}{\degree} in phase \citep{sakaguri_development_2026}. The average transmission is approximately 87\%. When coupled with the cryogenic HWP, a three-stacked sapphire design structure with anti-reflective coatings achieving ${\gtrsim}99\%$ modulation efficiency, the combined system provides highly efficient conversion and subsequent temporal modulation of the incoming Stokes $V$ signal. The QWP has a Mueller matrix,
\begin{equation}
    \renewcommand{\arraystretch}{1.5}
    \bm{M}_\mathrm{QWP} = \begin{pmatrix}
    M^\mathrm{QWP}_{II} & M^\mathrm{QWP}_{QI} & M^\mathrm{QWP}_{UI} & M^\mathrm{QWP}_{VI} \\
    M^\mathrm{QWP}_{IQ} & M^\mathrm{QWP}_{QQ} & M^\mathrm{QWP}_{UQ} & M^\mathrm{QWP}_{VQ} \\
    M^\mathrm{QWP}_{IU} & M^\mathrm{QWP}_{QU} & M^\mathrm{QWP}_{UU} & M^\mathrm{QWP}_{VU} \\
    M^\mathrm{QWP}_{IV} & M^\mathrm{QWP}_{QV} & M^\mathrm{QWP}_{UV} & M^\mathrm{QWP}_{VV} \end{pmatrix},
    \label{eq:qwp_mueller}
\end{equation}
where for an ideal QWP, $M^\mathrm{QWP}_{II} = M^\mathrm{QWP}_{QQ} = M^\mathrm{QWP}_{VU} = -M^\mathrm{QWP}_{UV} = 1$ and $M^\mathrm{QWP}_{IQ} = M^\mathrm{QWP}_{IU} = M^\mathrm{QWP}_{IV} = M^\mathrm{QWP}_{UU} = M^\mathrm{QWP}_{VV} = 0$; the antisymmetric $U\leftrightarrow V$ block encodes the \qty{90}{\degree} phase retardance, so $M^\mathrm{QWP}_{VU} = -M^\mathrm{QWP}_{UV}$ and both have the same magnitude. Here $\bm{M}_\mathrm{QWP}$ describes the QWP alone. The quantities measured on the sky instead belong to the effective Mueller matrix $\bm{M}$ of the full optical chain in the QWP basis, schematically $\bm{M}_\mathrm{rec}\,\bm{M}_\mathrm{HWP}\,\bm{M}_\mathrm{QWP}\,\bm{M}_\mathrm{tel}$, where $\bm{M}_\mathrm{tel}$ represents the telescope mirrors and $\bm{M}_\mathrm{rec}$ the remaining receiver optics (window, filters, and lenses). Its elements $M_{XY}$, written without superscript, follow the same input-to-output index convention as Equation~\eqref{eq:qwp_mueller}. The HWP modulation is treated separately in Section~\ref{sec:tod}, and, following \citet{sakaguri_development_2026}, the polarization efficiency of everything other than the QWP is factored out of $\bm{M}$ as a single scalar $\eta_\text{P}$ that multiplies the polarized signal in that modulation model. The remaining Stokes mixing is dominated by the QWP, so that $\bm{M} \approx \bm{M}_\mathrm{QWP}$, since the cross-polarization of the telescope optics is small and the receiver optics are approximately rotationally symmetric \citep{sakaguri_development_2026}.

Based on astrophysical point-source calibrations in \textcite{sakaguri_development_2026}, we express the measured efficiencies and leakages in terms of the elements of $\bm{M}$, with both $\bm{M}$ and $\bm{M}_\mathrm{QWP}$ normalized to unit transmission so that $M_{II} = M^\mathrm{QWP}_{II} = M^\mathrm{QWP}_{QQ} = 1$. The linear polarization efficiency of the system independent of the QWP, $\eta_\text{P}$, was measured on-sky from Tau A by fitting the observed $Q/I$ fraction as a function of parallactic angle. Both $Q$ and $I$ are measured from the same detector data and so carry the overall transmission as a common factor; their ratio is therefore insensitive to it, and returns $\eta_\text{P}$ in the same normalization as Equation~\eqref{eq:qwp_mueller}. \citet{sakaguri_development_2026} obtain $(46 \pm 2)\%$ from timestreams that were not deconvolved for the detector time constants, so that value also contains the single-pole attenuation of the signal at $4f_\text{HWP}$, $[1 + (2\pi \cdot 4f_\text{HWP}\,\tau)^2]^{-1/2} \approx 0.86$ for $\tau \approx \qty{12}{\milli\second}$. Because the science timestreams here are deconvolved (Section~\ref{sec:stim_cal}), we repeat the Tau A fit on the same observations after the identical deconvolution with stimulator time constants, obtaining $\eta_\text{P} = 0.533 \pm 0.032$. The net circular polarization efficiency and the leakages that enter the converted-$U$ channel are defined from the elements of $\bm{M}$, together with their approximate forms in terms of the QWP alone, as
\begin{subequations}
\begin{align}
    \eta_V &= \eta_\text{P} \left| M_{VU} \right| \approx \eta_\text{P} \left| M^\mathrm{QWP}_{VU} \right|, \label{eq:eta_V} \\
    \epsilon^{UV} &= M_{UU}/M_{VU} \approx M^\mathrm{QWP}_{UU}/M^\mathrm{QWP}_{VU}, \\
    \epsilon^{QV} &= M_{QU}/M_{VU} \approx M^\mathrm{QWP}_{QU}/M^\mathrm{QWP}_{VU}, \\
    \epsilon^{IV} &= M_{IU}/M_{VU} \approx M^\mathrm{QWP}_{IU}/M^\mathrm{QWP}_{VU}. \label{eq:eps_IV}
\end{align}
\end{subequations}
The QWP conversion efficiency is near unity, $M^\mathrm{QWP}_{VU} = -M^\mathrm{QWP}_{UV} = 99.0\%$, so $\eta_V = 0.528 \pm 0.032$ is limited primarily by $\eta_\text{P}$. The retained sky $U$ is constrained on-sky from Tau A to $\epsilon^{UV} = (-11.7 \pm 5.1)\%$, consistent with the laboratory value of $-7.2\%$ quoted above. Because the Tau A fit sees the full optical chain, this value is the effective ratio $M_{UU}/M_{VU}$, whereas the laboratory value is its QWP-only approximation. Leakage from sky $Q$ is not a property of the plate itself, so its QWP-only form above does not apply. It enters through residual miscalibration of the detector polarization angles relative to the QWP axis, which rotates $Q$ into the $U$ channel. Calibrating the angles from the HWPSS (Section~\ref{sec:detector_angles}) intrinsically suppresses this effect. The residual is set by the alignment of the QWP to the telescope, measured at the site to within \qty{0.3}{\degree}, which bounds $\epsilon^{QV} \lesssim 1.1\%$ \citep{sakaguri_development_2026}. Unlike $\epsilon^{UV}$, this bound comes from the alignment measurement rather than from the sky. The remaining term, $I$-to-$V$ leakage $\epsilon^{IV}$, is retained rather than set to zero. An ideal birefringent plate couples intensity only to $Q$, but we do not assume that this structure holds exactly. An additional intensity leakage is generated downstream of the optics, in the detector response itself, and therefore lies outside this Mueller-matrix description; it is discussed in Section~\ref{sec:demod}. Both are absorbed into the effective coefficient measured in Section~\ref{sec:IU_leakage}.

In summary, the static QWP maps the Stokes $V$ on-sky into the instrument-frame Stokes $U$, which the HWP then modulates at four times its rotation frequency. The circular-polarization measurement channel of this analysis is therefore the demodulated $U$ timestream; all subsequent steps construct per-run, per-detector $U_\text{obs}$ maps, estimate their cross spectra $\hat{C}_\ell^{UU,\text{obs}}$ and convert these to $C_\ell^{VV}$, disentangling instrument systematics. The companion $Q_\text{obs}$ channel carries the linear polarization of the measured sky, and is not used in this analysis.

\subsection{Scan Strategy and Observations}
\label{sec:obs}

Two CMB patches, RA23 and RA4.5, were chosen for this analysis, both legacy patches from \textsc{Polarbear} \citep{p_a_r_ade_measurement_2014}. These patches are so named due to the right ascension (RA) of their approximate centers in units of hours, which are for RA23 and RA4.5, $\alpha_\text{RA23} =\RA{23;01;48}$, $\delta_\text{RA23} = \Dec{-32;12;00}$ (J2000) and $\alpha_\text{RA4.5} = \RA{04;40;12}$, $\delta_\text{RA4.5} = \Dec{-45;00;00}$. These locations were chosen to lie away from the Galactic plane, avoiding significant contamination in the sky signal from synchrotron radiation. The dimensions of these patches are ${\sim}\qty{9}{\degree} \times \qty{9}{\degree}$, small enough that we work in the flat-sky approximation for later analysis. The small patch sizes are utilized in order to target sub-degree size angular scales ($300 \leq \ell \leq 2000$). The geometric distortion introduced by the flat-sky approximation is ${\ll}1\%$ and strictly negligible compared to the statistical noise.

An observation of a given patch is taken with a constant elevation scan (CES), which tracks back and forth a minimum and maximum azimuth at a constant elevation. The telescope scans at ${\sim}\qty{0.4}{\degree/\s}$ in azimuth with a sampling rate of ${\sim}\qty{152.6}{\s^{-1}}$; the duration of an observation is typically around \qty{860}{\second}. The CMB patches are observed both rising and setting, at elevations ${>}\qty{30}{\degree}$, high enough to avoid ground pickup, and ${<}\qty{70}{\degree}$, to maintain elevations roughly in the same range as where detector tuning occurs (${\sim}\qty{40}{\degree}$) \citep{sakaguri_development_2026}. Throughout, a detector is called ``tuned'' when it has been successfully biased onto its superconducting transition, and ``optical'' when it is coupled to the sky through its antenna rather than being a dark channel; only detectors that are both can contribute usable data. The scan strategy of observing both when rising and setting provides multiple parallactic angles, which decreases the non-radial systematics of the beam maps (discussed in more detail in Section~\ref{sec:beam}) and the amount of $P \rightarrow V$ leakage (see Section~\ref{sec:leakage}).

We observed RA23 and RA4.5 for about 2.5 months each, in staggered periods. During that time, corruption with the telescope timestamp data resulted in a large amount of data becoming unusable. For RA23 (RA4.5), 114 (99) of the 798 (576) runs with nonzero tuned and optical detectors have technical issues resulting in a total loss of usable detector-observations (the data from a single detector within a single observation), arising from corrupted timestream files, improperly stored identifiers in the database tracking each observation, failed tuning, or faults within the timestream data despite it being saved. In addition to these observation-wide technical failures, a subset of detectors in each observation are not usable because they do not have time constants or absolute gain calibration factors; this is explained in Section~\ref{sec:stim_cal}. For these reasons, 356,385/566,694 (232,866/414,612), about 62\% (56\%) of detector-observations across all observations for RA23 (RA4.5), have usable timestreams for the analysis detailed in Section~\ref{sec:analysis}. A typically behaved observing run yields between 600--700 tuned and optical detectors in the \qty{90}{\giga\Hz} band. These instrument-level losses, together with the data quality cuts imposed after mapmaking, are tabulated in Section~\ref{sec:data_quality}.

\section{Calibrations}
\label{sec:calibration}

This section describes the calibrations that fix the instrumental quantities entering the forward model of Section~\ref{sec:cl_vv}. The pointing model (Section~\ref{sec:pointing}) sets the suppression $P_\ell$ of small-scale power by residual pointing error, and Jupiter observations set the beam window function $B_\ell$ (Section~\ref{sec:beam}). A thermal stimulator provides the detector time constants and the absolute gain $g$ (Section~\ref{sec:stim_cal}), while the detector polarization angles are calibrated from the half-wave plate synchronous signal (HWPSS) fourth harmonic so that every detector can be rotated into the common QWP basis (Section~\ref{sec:detector_angles}). The circular polarization efficiency $\eta_V$ and the leakages $\epsilon^{UV}$ and $\epsilon^{QV}$ follow from the optical chain of Section~\ref{sec:polmod}, while the pipeline transfer function $T_\ell$ and the $I\!\to\!U$ and $P\!\to\!V$ leakage spectra are determined from the processed data and are deferred to Sections~\ref{sec:transfer} and \ref{sec:leakage}. Table~\ref{table:sample} collects all of these, with the role each plays in the likelihood and the section in which it is determined.

\subsection{Pointing}
\label{sec:pointing}

\begin{figure}[htbp]
    \centering
    \includegraphics[width=\linewidth]{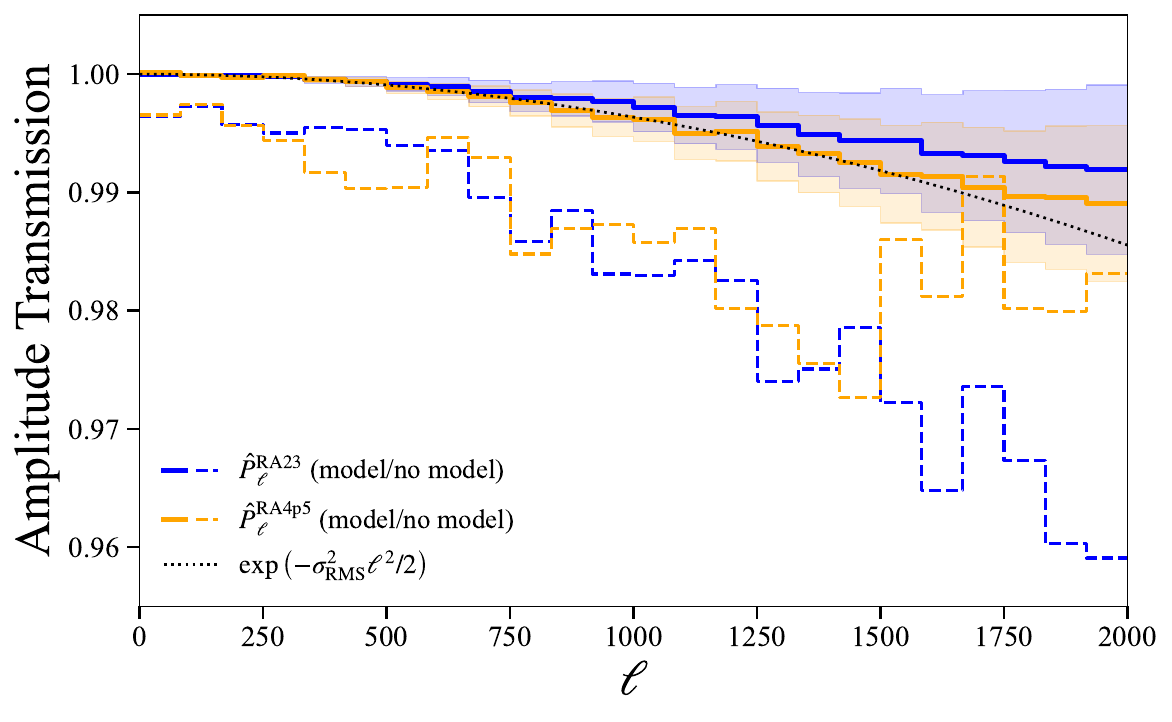}
    \caption{Suppression factor in $\ell$-space with (solid line) and without (dashed line) pointing model correction. The dotted black line shows a Gaussian suppression from the determined RMS for reference; both patches are Gaussian at low-$\ell$ but have less suppression at higher $\ell$.}
    \label{fig:pointing_transfer}
\end{figure}

The PB-2b pointing model is a 14-parameter TPoint-style correction \citep{wallace_rigorous_2002} that maps the encoder azimuth/elevation to the true boresight pointing before the az/el $\to$ RA/Dec conversion, which is a more expansive model than the five-parameter model used in \citetalias{p_a_r_ade_measurement_2014}. The corrections (in arcminutes) decompose into structural, thermal-flexure, solar-flexure, and second-harmonic ``nod'' terms. The 14 coefficients follow \citet{matsuda_cosmic_2017} and are fit by minimizing a $\chi^2$ against a catalog of point-source centroid offsets. The fit leaves a per-axis residual scatter of $\sigma_{\rm az}=14.7''$ and $\sigma_{\rm el}=20.5''$ ($\approx 25''$ total). A few corrections and physical improvements to the model were found to be necessary, detailed in J. Ito et al. (2026, in preparation). These adopt parameters from the pointing models of the Atacama Large Millimeter/submillimeter Array (ALMA) \citep{Mangum2001} and the Very Long Baseline Interferometry (VLBI) Field System \citep{Himwich1993} as adopted by Square Kilometre Array Observatory (SKAO) \citep{katpoint}, reducing the RMS further. However, as the existing model already has reduced pointing suppression significantly from the no-model case, as shown in Figure~\ref{fig:pointing_transfer}, and its residual uncertainty is carried through the likelihood as a sampled parameter, as discussed in Section~\ref{sec:mcmc}, the adjustments were not implemented due to computational restrictions. A comparison of the two models is described in that work, showing comparable performance. The updated model lowers the residual pointing RMS from $25.2''$ to $23.3''$, which changes the pointing transfer function by at most 0.13\% for $\ell \leq 2000$, no more than $0.2\sigma$ of its sampled uncertainty, a negligible difference.

Because the relevant question for this analysis is how the pointing affects the recovered spectrum, we validate the model with geometry-only simulations, isolating the az/el $\to$ RA/Dec geometry to quantify the impact as an effective amplitude transfer function. Concretely, we synthesize a temperature-only sky with the Boltzmann code \texttt{CAMB} \citep{lewis_efficient_2000} at the Planck best-fit cosmology \citep{Planck2018VI}, sample it along the azimuth and elevation actually recorded during the real observations, both with and without the pointing model applied, and coadd the resulting maps following the same procedure used for the data. A pointing error is a purely geometric remapping of where each sample lands on the sky, so its effect on the recovered spectrum does not depend on which Stokes parameter carries the signal; a $T$-only input sky is therefore sufficient, and this is what ``geometry-only'' denotes here. The model transfer function is ${\gtrsim}0.99$ in amplitude out to $\ell\sim2000$. The uncorrected maps yield pointing scatter up to 5\% at $\ell=2000$, compared to ${\lesssim}1\%$ for those with the pointing correction, as shown in Figure~\ref{fig:pointing_transfer}. In addition, by propagating the parameter covariance and cosmic variance through Monte Carlo simulations into $\ell$-space, we find a pointing suppression uncertainty $\delta \hat{P}_\ell \lesssim 1\%$ at the highest bin. Because this term arises from smearing it grows as $\ell^{2}$, so it is negligible below $\ell\sim1000$ but becomes comparable to other spectral sensitivity systematics such as the beam suppression and mapmaking pipeline transfer function (visible in Figure~\ref{fig:sensitivity}) in the highest bandpowers. It is therefore sampled in the likelihood rather than propagated afterwards (Section~\ref{sec:mcmc}).

\subsection{Beam Characterization}
\label{sec:beam}

To estimate the beam window function $B_\ell$ at \qty{90}{\giga\Hz}, we analyze 68 Jupiter observations taken over our observing season, extending the beam analysis in \citet{farias_performance_2026} (see Figure~8 of that work for the inverse-variance average beam map). Jupiter is chosen because it is a bright point-like source that is negligibly polarized at \qty{90}{\giga\Hz}. Each map is apodized with a \qty{2}{\arcminute} two-dimensional cosine taper so it smoothly tapers to zero at the edge, then zero-padded for finer $\ell$-sampling. A 2D fast Fourier transform (FFT) is performed on the resulting padded map, and the result is normalized by the DC mode (effectively the monopole term in a flat-sky approximation). This is then divided by the Fourier transform of the Airy disk to deconvolve the planet's angular size, which is determined from ephemeris data for each observation date. The resulting function in $\ell$-space is squared, radially binned in $\Delta\ell = 100$ rings to match the science binning scheme (with $\ell$ denoting the bin center hereafter), and the square root is taken. Finally, we divide by the pixel-window function $w_\ell^\text{beam} = \mathrm{sinc}\!\left(\frac{\ell\,\Delta x}{2}\right)$, where $\Delta x = \qty{0.001}{\degree}$ is the map resolution.

The resulting beam window function is shown in Figure~\ref{fig:sensitivity}. As noted in Figure~8 from \citet{farias_performance_2026}, the beams are not perfectly radially symmetric, despite the radial-symmetry assumption used in the binning above. This, in conjunction with the statistical uncertainty, is reflected in the error bars in the plot, computed as standard errors of the points in each $\ell$-bin. Observing over a range of azimuths and elevations provides multiple parallactic angles that mitigate the azimuthal artifacts in the beam, as discussed in Section~\ref{sec:obs}. The ellipticity of the \qty{90}{\giga\Hz} beam is quoted at 3\% \citep{farias_performance_2026}, sufficiently low to contribute negligibly to the overall systematic uncertainty, following the \textsc{Polarbear} analyses \citep{collaboration_measurement_2017, collaboration_measurement_2020}.

\subsection{Time Constants and Absolute Gain Calibration}
\label{sec:stim_cal}

PB-2b detectors are about four times slower than \textsc{Polarbear}'s, with a median effective time constant of \qty{12.1}{\milli\second} at \qty{90}{\giga\Hz} \citep{farias_performance_2026} compared to ${\lesssim}\qty{3}{\milli\second}$ \citep{p_a_r_ade_measurement_2014, collaboration_measurement_2017}. This imposes non-negligible attenuation and phase retardance on the polarization signal at the modulation frequency $4f_\text{HWP}$, and the timestreams therefore require deconvolution before demodulation. As discussed in \citet{farias_performance_2026}, effective time constants are determined by fitting the first-harmonic amplitude of a chopped thermal signal (known hereafter as a stimulator) to a single-pole model at a series of seven frequencies; \qty{5}{\Hz}, \qty{9}{\Hz}, \qty{13}{\Hz}, \qty{19}{\Hz}, \qty{27}{\Hz}, \qty{37}{\Hz}, and \qty{44}{\Hz}; so chosen to avoid the significant HWP harmonics. Fit parameters from this methodology are saved for later use as long as they produce plausibly physical values (single-pole model amplitude ${\leq}200$ analog-to-digital converter (ADC) counts and $\tau \le \qty{500}{\milli\second}$) and have at least three surviving frequency points within the sweep. These calibration runs are taken immediately before and after each science observation, and the effective time constant used is their mean, $\tau = (\tau_\text{before} + \tau_\text{after})/2$. We deconvolve per timestream before demodulation by Fourier-transforming and dividing by the single-pole model evaluated at this time constant.

Absolute gain calibration factors convert the instrument's measured current units (ADC counts) into \si{\Kcmb}. The stimulator, operating simultaneously with a Jupiter observation, is used to determine these factors. Jupiter's known thermodynamic temperature, measured by \textcite{Planck2017LII}, is first converted into a Rayleigh--Jeans brightness temperature and then into an equivalent \si{\Kcmb} temperature fluctuation. This conversion incorporates the Simons Array instrument characterization \citep{Westbrook_2018, Suzuki2016}, including the effective detector bandpasses and Lyot efficiencies ($\eta_\text{Lyot} = 0.537$ and $0.849$ for the \qtylist{90;150}{\giga\Hz} bands, respectively). Furthermore, to account for sources with different spectral indices, color-correction scaling factors are calculated internally by integrating the source spectra over the instrument's simulated diplexer bandpass. The \qty{90}{\giga\Hz} bandpass is flat, spanning \qtyrange{76.1}{94.3}{\giga\Hz}, utilizing the Fourier-transform spectrometer measurement adopted by \textcite{sakaguri_development_2026}, which gives a band center of \qty{85.18}{\giga\Hz} and a bandwidth of \qty{18.25}{\giga\Hz} (a fractional bandwidth of 0.214). The color correction is normalized to a Rayleigh--Jeans source, $\alpha = 2$ for a flux density $S \propto \nu^{\alpha}$, which is appropriate for Jupiter. The spectrum relevant to CMB anisotropies is instead $\mathrm{d}B_\nu/\mathrm{d}T$, whose effective index is $\alpha_\text{eff} = 4 + x - 2x e^{x}/(e^{x}-1) \approx 1.6$ at \qty{90}{\giga\Hz}, where $x = h\nu/k_\text{B}T_\text{CMB}$; the resulting correction relative to the Rayleigh--Jeans normalization is ${\approx}0.8\%$. The QWP conversion efficiency $|M^\mathrm{QWP}_{VU}(\nu)|$ does not enter this absolute gain calibration, which is an intensity measurement. Its frequency dependence across the band instead enters through the band-averaged value used in $\eta_V$ (Equation~\eqref{eq:eta_V}). As in the typical planet scans, we fit a 2D Gaussian to the source in ADC counts. By comparing the amplitude of this Gaussian fit and the measured beam size to the expected temperature and solid angle of Jupiter, a conversion factor in \si{\Kcmb}/ADC is determined. Ten such Jupiter scans with the stimulator were used to build the absolute gain calibration template database.

As a secondary calibration source, we use the HWPSS, a spurious non-sky signal generated due to imperfections in the HWP, following a procedure similar to the relative gain calibration in \citet{sakaguri_development_2026}. We build a template of calibration factors from observations with reliable stimulator calibration, determining, per detector, a conversion between the amplitude of the HWPSS fourth harmonic (at $4f_\text{HWP}$) and the equivalent stimulator amplitude. Likewise, we estimate the effective time constant from the $4f_\text{HWP}$ phase. The $4f_\text{HWP}$ phase is primarily set by the physical angle between each detector and the QWP axis, but the time constant introduces an additional phase offset. We build a template for the $\tau=0$ phase corresponding to the physical angle; the difference between each observation's $4f_\text{HWP}$ phase and this template gives the effective time constant. This method agrees well with the stimulator-based approach, as validated on observations for which both stimulator calibration and the HWPSS $4f_\text{HWP}$ signal are available. This alternative method is utilized when there are technical issues with the stimulator that prevent it from being used as a reliable calibration source.

The time constant during an observation may differ from the mean $\tau$ used for deconvolution to the extent that it drifts between the two calibrations, which are ${\sim}\qty{2}{\hour}$ apart. Taking the true value to lie uniformly between them gives $\sigma_\tau = |\tau_\text{after} - \tau_\text{before}|/\sqrt{12}$ for each detector-observation, which the single-pole model converts into a fractional error on the deconvolution gain of $x^2/(1+x^2)\,\sigma_\tau/\tau$, with $x = 2\pi \cdot 4f_\text{HWP}\,\tau$. The drift is largely common to the array within one calibration interval, so we treat each interval's error as coherent across its detectors and independent between intervals, and average over each patch with the map weights. Time constants from the HWPSS phase are measured within the observation and carry no such term. The result is a unit-mean gain factor $g_\tau$, with uncertainty \qty{0.37}{\percent} for RA23 and \qty{0.62}{\percent} for RA4.5, which multiplies the data in the same way as $g$. An overall bias in the stimulator time constants does not enter $g_\tau$, as it cancels against the calibration of $\eta_\text{P}$ (Section~\ref{sec:polmod}).

Because the gain enters the measured spectrum only as an overall factor of $g^2$, a fractional calibration error rescales $C_\ell$ by twice that fraction at every multipole---its effect is therefore analytic rather than something that must be propagated by simulation, and we account for it explicitly through the residual gain coefficient $g$ defined in Equation~\eqref{eq:gain}. All science observations analyzed here were taken in the QWP configuration, and the applied calibration is therefore the mean over only the four planet observations available in that configuration; their \qty{8.9}{\percent} common-mode scatter averages down to \qty{4.45}{\percent}, which dominates the budget. The remaining terms are the Jupiter brightness-temperature model of \textcite{Planck2017LII} (${\sim}\qty{1.0}{\percent}$), the per-detector calibration scatter (\qty{7.6}{\percent} across 706 detectors, or \qty{0.29}{\percent} after averaging), and the HWPSS template fit error (${<}\qty{0.1}{\percent}$). Summing in quadrature gives a total absolute gain uncertainty of $\sigma_g = \qty{4.6}{\percent}$, corresponding to a \qty{9.1}{\percent} multiplicative scale uncertainty on the quoted limits. Rather than propagate this through the pipeline, we carry it into the model of Section~\ref{sec:cl_vv} as a prior on $g$.

\subsection{Detector Polarization Angles}
\label{sec:detector_angles}

The QWP converts sky $V$ into the Stokes $U$ of its own frame, for which the efficiency and leakage terms of Equations~\eqref{eq:eta_V}--\eqref{eq:eps_IV} are defined. Circular polarization is invariant under rotation about the line of sight, so were $V$ the only signal present no orientation reference would be needed. It is the linear polarization, spin-2 and hence basis-dependent, that forces one---a detector whose polarization axis lies at an angle $\theta_\text{det}^{(j)}$ to the QWP axis measures $Q$ and $U$ mixed by $2\theta_\text{det}^{(j)}$, so without correcting for it each detector would fold an uncontrolled fraction of sky linear polarization into its circular polarization channel. We therefore rotate every detector into the common QWP basis, both to keep sky linear polarization out of the measured $U$ and because $\epsilon^{UV}$, $\epsilon^{QV}$, and $\epsilon^{IV}$ are themselves ratios of Mueller elements about the QWP axis, quantifying the leakage only once the data are expressed in that frame.

Because the HWPSS fourth-harmonic signal is physically related to the QWP transmission \citep{sakaguri_development_2026, farias_performance_2026} and is dominant over the other harmonics by several orders of magnitude, we can calibrate these angles directly from the timestream data. This permits us to rotate from the individual detector orientations into the common QWP basis. For each detector, the mean-subtracted timestream prior to HWPSS removal is binned into 128 uniform bins in the encoder-measured HWP angle $\phi$ over $[0, 2\pi)$. Because the samples are sorted by measured angle rather than by time, no constant rotation rate is assumed and departures from the nominal rotation rate are absorbed by the binning. The resulting binned HWPSS profile is then fit by least squares to a harmonic series in $\phi$,
\begin{equation}
    d^{(j)}(\phi) = \sum_{n=1}^{6} \left[ A_n^{(j)} \sin (n\phi) + B_n^{(j)} \cos (n\phi) \right].
\end{equation}
The physical alignment angle per detector is determined from the fourth-harmonic amplitudes as
\begin{equation}
    \theta_\text{det}^{(j)} = \frac{1}{2} \operatorname{arctan2} \left( A_4^{(j)}, B_4^{(j)} \right),
\end{equation}
which is defined on $(-\pi/2, \pi/2]$.

\section{Data Analysis}
\label{sec:analysis}

\begin{figure*}[t!]
    \centering
    \includegraphics[width=\textwidth]{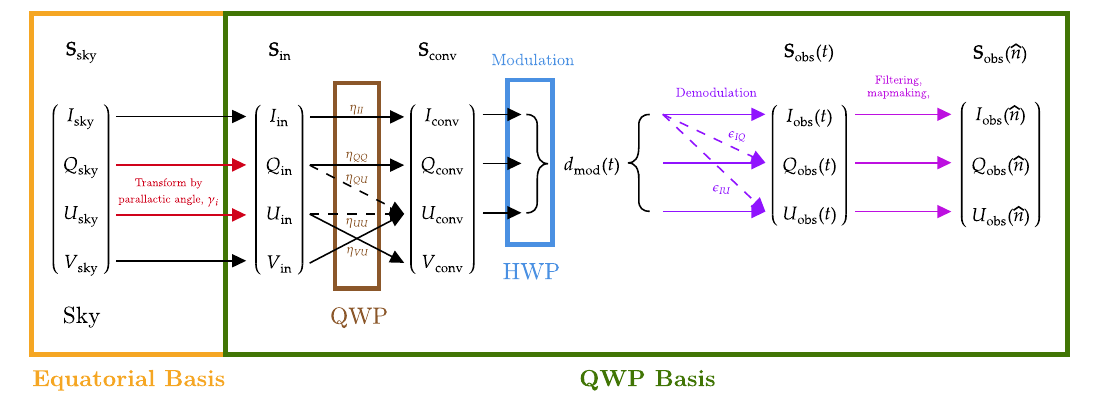}
    \caption{Schematic describing the Stokes parameters at each part of the observing and analysis pipeline, with indicated leakages and efficiencies and associated bases. This is the path traced by Figure~\ref{fig:pb2b_schematic}. The subscripts label successive stages along this path: ``sky'' denotes the true celestial signal; ``in'' the same signal transformed into the telescope basis; ``conv'' the Stokes parameters after the QWP has applied its phase delay, converting $V$ into $U$ and vice versa; ``mod'' the $I$, $Q$, and $U$ of the converted signal modulated by the rotating HWP into a detector timestream; and ``obs'' the signal actually recorded by the detectors, either in the time domain as a function of $t$ or in the map domain as a function of $\hat{n}$. This figure is based on Figure~7 from \textcite{sakaguri_development_2026}.}
    \label{fig:stokes}
\end{figure*}

This section describes the analysis pipeline in the order in which the data flow through it. Timestreams are calibrated, cleaned, and demodulated into per-detector Stokes components (Section~\ref{sec:tod}); these are binned into per-observation maps (Section~\ref{sec:mapmaking}) and screened by the data quality cuts of Section~\ref{sec:data_quality}. The surviving maps are combined into cross-spectra, yielding the measured $UU$ bandpowers and their covariance (Section~\ref{sec:power_spectra}), whose internal consistency is then established with the null test suite of Section~\ref{sec:null_tests}. The remaining sections quantify the terms needed to invert those bandpowers into a circular polarization spectrum: the pipeline transfer function (Section~\ref{sec:transfer}) and the instrumental leakages (Section~\ref{sec:leakage}). These are assembled into the forward model of Section~\ref{sec:cl_vv}, which is inverted in the likelihood of Section~\ref{sec:mcmc} to give $C_\ell^{VV}$. Figure~\ref{fig:stokes} traces the Stokes parameters through this chain and defines the subscripts used throughout.

\subsection{Blinding Scheme}
\label{sec:blinding}

To prevent analysis choices from being biased, consciously or otherwise, toward the final circular polarization result, we developed the analysis pipeline on a benchmark dataset with full access to its power spectra, then applied it to the science data under a blinding scheme that hides the $VV$ bandpowers. The benchmark set comprises ${\sim}10\%$ of the usable RA23 data, on which we had unblinded access to the recovered spectra. These observations were selected on observing conditions and detector yield alone, requiring more than 500 tuned and optical detectors and a precipitable water vapor (PWV; a measure of water vapor in the line of sight to the zenith) below \qty{3}{\milli\meter}, so that pipeline development was carried out on data of reliably good quality rather than on observations chosen for any property of their recovered spectra. All analysis choices were initially set using this benchmark set alone, including the time-domain filtering parameters, glitch-flagging and HWPSS-removal procedures, data-quality cut thresholds (Section~\ref{sec:data_quality}), the power-spectrum estimator configuration (Section~\ref{sec:power_spectra}), and the leakage and transfer-function treatments (Sections~\ref{sec:transfer} and \ref{sec:leakage}). Because the RA4.5 observations were acquired chronologically later, and to maintain a uniform analysis pipeline across both sky patches, we did not define a separate benchmark dataset for RA4.5. Instead, the analysis pipeline developed on the RA23 benchmark was applied directly to the RA4.5 dataset, validated by the success of the RA4.5 null tests (Section~\ref{sec:null_tests}). A separate set of RA4.5 observations was set aside for the transfer-function simulations of Section~\ref{sec:transfer}. It was not used to develop or validate the pipeline, so it was not excluded from the RA4.5 science data.

We then applied this pipeline to the remaining ${\sim}90\%$ of the data. Throughout this stage the $VV$ bandpowers themselves were blinded; while the intermediate pipeline products (filtered timestreams, detector and coadded maps, and the various calibration and leakage estimates) remained accessible, the recovered science spectra were withheld from inspection. The null tests of Section~\ref{sec:null_tests} were run in this blinded configuration for both patches, with only the summary probabilities-to-exceed (PTEs) exposed and the underlying bandpowers hidden, so that internal-consistency failures could be diagnosed without sight of the science spectra. Failures were addressed with additional data-quality cuts and adjustments to the pipeline, made in parallel and based only on these PTEs and on secondary data products such as the timestreams and maps; the $VV$ bandpowers, and hence the upper limits derived from them, remained blinded throughout. The unblinding sequence was fixed in advance rather than settled as results became available. The pipeline was frozen, with no analysis choice or data cut changed thereafter, once the null-test passage criteria of Section~\ref{sec:null_tests} were met, and only then were the $VV$ bandpowers revealed. The benchmark observations are themselves excluded from the final science dataset, at a ${\sim}10\%$ sensitivity cost, so that the limits reported in Section~\ref{sec:results} derive entirely from data whose spectra were unseen during pipeline construction.

\subsection{Timestream Processing}
\label{sec:tod}

In this section we will detail the data processing that occurs from the intake of raw modulated detector data to isolated Stokes timestreams, including the time-domain filters meant to improve the quality of the data. A schematic of how the data passes through each component, both physically in the telescope and later in the analysis pipeline, is shown in Figure~\ref{fig:stokes}.

\subsubsection{Calibration and Modulation}

The raw detector data are first converted from ADC counts into \si{\Kcmb} using the procedures in Section~\ref{sec:stim_cal}. That conversion is a single scalar responsivity value per detector, which propagates to being a scalar multiplier in $\ell$-space. We parameterize any residual error in it with a dimensionless coefficient $g$,
\begin{equation}
    d_\text{mod}^{(j)} (t) = g\, d_\text{mod}^{(j),\text{true}} (t),
    \label{eq:gain}
\end{equation}
defined as the ratio of the applied calibration factor to the true one, where $d_\text{mod}^{(j),\text{true}}$ is what a perfectly calibrated detector would record. The calibration of Section~\ref{sec:stim_cal} has already been applied to the data, so $g$ is unity by construction and only quantifies how far that calibration may be from the truth, with a width given by the absolute gain uncertainty derived there. As mentioned in Section~\ref{sec:stim_cal}, we account for its residual uncertainty once, in spectrum space (Section~\ref{sec:cl_vv}), to avoid propagating the parameter through hundreds of thousands of individual detector timestreams. 

The time constants determined from the same procedures are also deconvolved at first order. We also mask the first and last second of each subscan, a single left- or right-going sweep in azimuth at constant elevation, to exclude transients from the turnaround between sweeps. At this point, we also determine the corresponding RA and declination (Dec) based on each detector's position in the focal plane and the focal plane's boresight pointing at a given timestamp, and these are corrected according to the pointing model from Section~\ref{sec:pointing}. The HWP encoder data is also reconstructed per the procedure discussed in \citet{bixler_mechanical_2026}, which is used in the following subsection to demodulate into individual Stokes components.

As discussed in \citet{sakaguri_development_2026}, the HWP modulates polarized signal according to the following equation for each detector $j$,
\begin{multline}
    d_\text{mod}^{(j)} (t) = I_\text{conv}^{(j)} (t) + \ \\ \mathrm{Re} \left[ \left( \eta_\text{P} \left( Q_\text{conv}^{(j)} (t) + i U_\text{conv}^{(j)} (t) \right) + \epsilon^{IP}_\text{det} I_\text{conv}^{(j)} (t) \right) \right. \\ \left. e^{-i\left[4\theta_\text{HWP}(t) - 2\theta_\text{det}^{(j)} \right]} \right] + A_\text{HWPSS}^{(j)} (t) + N_\text{w}^{(j)} (t) + N_\text{atm}^{(j)} (t),
\end{multline}
where ``$\text{conv}$'' represents the Stokes components after passing through the QWP, $\theta_\text{HWP} (t)$ is the angle of the HWP as it rotates, $\theta_\text{det}$ is the detector's polarization angle relative to the QWP axis (Section~\ref{sec:detector_angles}), $A_\text{HWPSS} (t)$ is the HWPSS signal, $N_\text{w} (t)$ is white noise, and $N_\text{atm} (t)$ is low-frequency noise caused by changes in atmospheric loading. In addition, as noted in Section~\ref{sec:polmod}, the detector response is not strictly linear, so a second-order term mixes the HWP-synchronous signal with the unmodulated intensity and produces sidebands about $4f_\text{HWP}$ that survive demodulation. This leaks the intensity timestream into the demodulated Stokes components, which we denote with the complex leakage parameter $\epsilon^{IP}_\text{det}$, where the subscript denotes that it originates in the detector response rather than the optics. The same term was identified in \textsc{Polarbear} as an excess $I\!\to\!P$ leakage and derived from the bolometer response model \citep{takakura_performance_2017, takakura_characterization_2017}.

\subsubsection{HWPSS Removal}

The first step before demodulating is to remove the HWPSS, which we model as a sum of harmonics of the HWP angle with slowly varying complex amplitudes, $A_\text{HWPSS}(t) = \sum_m \text{Re}\left[ d_m(t)\,e^{im\theta_\text{HWP}(t)} \right]$. Following \citet{polarbear_collaboration_constraints_2026}, for each $m = 1, \dots, 12$ we band-pass the timestream within $\pm\qty{0.1}{\Hz}$ of $mf_\text{HWP}$ (1023-tap finite impulse response filter), demodulate by $2e^{-im\theta_\text{HWP}(t)}$, and low-pass at \qty{0.1}{\Hz} to obtain $d_m(t)$, then subtract the reconstructed harmonic before proceeding to the next. Notably, $m=4$ is included, so any polarization sky signal that varies slowly enough to lie within $\pm \qty{0.1}{\Hz}$ of $4f_\text{HWP}$ is subtracted along with the HWPSS, while faster HWP-modulated polarization signal content (at $4f_\text{HWP}$) is preserved for downstream demodulation. The effect this has on real data is encapsulated when we calculate the pipeline transfer function, which will be discussed in Section~\ref{sec:transfer}.

\subsubsection{Filtering and Glitch Removal}
\label{sec:filter1}

The modulated data is put through two polynomial filters, one on the entire timestream and one per subscan. Both utilize Chebyshev polynomials, the former with $n=5$ and the latter with $n=1$. These are designed to remove the low-frequency noise primarily associated with fluctuations in atmospheric loading and the gain drifts that accompany them. These dominate the low-frequency content of the modulated timestream because the bulk atmospheric emission is unpolarized and therefore not modulated by the HWP.

In addition, during timestream processing, we identified populations of transient glitches and scan-synchronous spikes. Transient glitches are well documented in bolometric detector arrays generally, where energetic particle interactions with the silicon detector wafers produce events spanning a wide range of amplitudes and time constants \citep{catalano_impact_2014}. The scan-synchronous component is instead tied to the observing environment, arising from far-sidelobe ground pickup and from radio frequency interference (RFI) coupling into the multiplexed readout \citep{barron_integrated_2021}, the latter also being the suspected origin of the excess noise reported for PB-2b \citep{farias_performance_2026}. Because the precise coupling mechanisms of out-of-band RFI and ground emission cannot be easily modeled from first principles, we employ a masking procedure and replacement with Gaussian noise for the purposes of demodulation. First, the autocorrelation function (ACF) is used as a convergence criterion; glitches are repeatedly flagged and the ACF is repeatedly recomputed, cleaning until the maximum ACF in the 5--\qty{400}{\second} lag window drops below 0.08 or after four iterations. Each pass flags samples whose median-subtracted absolute deviation exceeds $3\sigma$, then extends each flagged region outward from its peak to the nearest local minimum on either side to capture the glitch's full rise and fall, saving the mask. The masked regions are replaced with a linear baseline interpolated between the five-sample means on either side, plus a realization of white Gaussian noise scaled to that detector's \qtyrange{6.5}{7.5}{\Hz} noise floor. The noise is multiplied by a half-sine window that goes to zero at the first and last masked sample, so that no noise discontinuity is spliced against the surrounding real data. This Gaussian-filled timestream is used for demodulation, but those timestamps remain masked for later mapmaking.

\subsubsection{Demodulation}
\label{sec:demod}

This section primarily follows the framework described in \textcite{takakura_performance_2017}. To demodulate, we apply a bandpass filter spanning $4f_\text{HWP} \pm \qty{3.8}{\Hz}$ to the modulated signal, isolating the polarization modulation at $4f_\text{HWP}$. This effectively removes the intensity signal, as with the scan speed of \qty{0.4}{\degree/\s} the lower bound of our bandpass is above the beam-imposed diffraction limit of the intensity timestream (${\sim}\qty{1.8}{\Hz}$ with beam size \qty{4.9}{\arcminute} \citep{farias_performance_2026}). This allows us to retain resolving power while not contaminating the modulated polarized signal with low-frequency noise. The remaining signal after HWPSS removal and bandpass filtering is, with $\Phi^{(j)}(t) = 4\theta_\text{HWP} (t) - 2 \theta_\text{det}^{(j)}$ for brevity (where the detector polarization angle $\theta_\text{det}^{(j)}$ is calibrated per Section~\ref{sec:detector_angles}),
\begin{equation}
\label{eq:demod_bp}
\begin{split}
    &d_{\text{bp}}^{(j)} (t) = \text{Re} \Big[ \big( \eta_\text{P} \big( Q_\text{conv}^{(j)} (t) + i U_\text{conv}^{(j)} (t) \big) \\ \
    &\; + \epsilon^{IP}_\text{det} I_\text{conv}^{(j)} (t) \big) e^{-i\Phi^{(j)}(t)} \Big] + N_{\text{w},\text{bp}}^{(j)} (t) + N_\text{atm,bp}^{(j)} (t) \\
    &= \frac{1}{2} \Bigg\{ \Big[ \eta_\text{P} \big( Q_\text{conv}^{(j)} (t) + i U_\text{conv}^{(j)} (t)\big) \\
    &\; + \epsilon^{IP}_\text{det} I_\text{conv}^{(j)} (t) \Big] e^{-i\Phi^{(j)}(t)} \\
    &\; + \Big[ \eta_\text{P} \big( Q_\text{conv}^{(j)} (t) - i U_\text{conv}^{(j)} (t) \big) + {\epsilon^{IP}_\text{det}}^* I_\text{conv}^{(j)} (t) \Big] e^{i\Phi^{(j)}(t)} \Bigg\} \\
    &\; + N_{\text{w},\text{bp}}^{(j)} (t) + N_\text{atm,bp}^{(j)} (t)
\end{split}
\end{equation}
where ``$\text{bp}$'' denotes a bandpass-filtered timestream. This is then shifted to baseband, i.e., the polarization signal is moved from the $4f_\text{HWP}$ carrier to zero frequency, by multiplying Equation~\eqref{eq:demod_bp} by $2e^{+4i\theta_\text{HWP} (t)}$, resulting in
\begin{equation}
\begin{split}
    d_{\text{bb}}^{(j)} (t) &= \Big[ \eta_\text{P} \big( Q_\text{conv}^{(j)} (t) + i U_\text{conv}^{(j)} (t) \big) \\
    &\quad \ + \epsilon^{IP}_\text{det} I_\text{conv}^{(j)} (t) \Big] e^{2i\theta_\text{det}^{(j)}}  + \Big[ \eta_\text{P} \big( Q_\text{conv}^{(j)} (t) \\
    &\quad \ - i U_\text{conv}^{(j)} (t) \big) + {\epsilon^{IP}_\text{det}}^* I_\text{conv}^{(j)} (t) \Big] e^{i\left[8\theta_\text{HWP}(t) - 2\theta_\text{det}^{(j)}\right]} \\
    &\quad \ + 2 N_{\text{w},\text{bp}}^{(j)} (t) e^{4i\theta_\text{HWP}(t)} \\
    &\quad \ + 2 N_\text{atm,bp}^{(j)} (t) e^{4i\theta_\text{HWP}(t)},
\end{split}
\end{equation}
where ``$\text{bb}$'' stands for ``baseband.'' A low-pass filter is then applied at 3.8 Hz, which removes the term shifted up to $8\theta_\text{HWP} (t)$. The remaining modulated timestream is,
\begin{equation}
\begin{split}
    d_{\text{LPF}}^{(j)} (t) &= \Big[ \eta_\text{P} \big( Q_\text{conv}^{(j)} (t) + i U_\text{conv}^{(j)} (t) \big) \\
    &\quad \ + \epsilon^{IP}_\text{det} I_\text{conv}^{(j)} (t) \Big] e^{2i\theta_\text{det}^{(j)}} + 2 N_{\text{w},\text{bp}}^{(j)} (t) e^{4i\theta_\text{HWP}(t)} \\
    &\quad \ + 2 N_\text{atm,bp}^{(j)} (t) e^{4i\theta_\text{HWP}(t)}.
\end{split}
\end{equation}
The timestream is then rotated into the QWP axis basis by multiplying by $e^{-2i\theta_\text{det}^{(j)}}$,
\begin{equation}
\begin{split}
    d_\text{QWP}^{(j)} (t) &= e^{-2i\theta_\text{det}^{(j)}}\, d_{\text{LPF}}^{(j)} (t) \\
    &= \eta_\text{P} \big( Q_\text{conv}^{(j)} (t) + i U_\text{conv}^{(j)} (t) \big) + \epsilon^{IP}_\text{det} I_\text{conv}^{(j)} (t) \\
    &\quad + 2 \big[ N_{\text{w},\text{bp}}^{(j)} (t) + N_\text{atm,bp}^{(j)} (t) \big] e^{i \Phi^{(j)} (t)},
\end{split}
\end{equation}
with $\Phi^{(j)} (t)$ as defined above. This removes the detector-angle dependence of the sky terms exactly, so the real and imaginary components may be saved directly as $Q_\text{obs}^{(j)} (t)$ and $U_\text{obs}^{(j)} (t)$ without any residual dependence on $\theta_\text{det}^{(j)}$. The intensity $I_\text{obs}^{(j)} (t)$ is recovered separately, from the unmodulated low-frequency component of the timestream (denoted ``lp'') rather than from the band around $4f_\text{HWP}$, so it carries no $\eta_\text{P}$ factor and retains the full low-frequency atmospheric noise. Together,
\begin{subequations}
\begin{align}
    I_\text{obs}^{(j)} (t) &= I_\text{conv}^{(j)} (t) + N_{\text{w},I} (t) + N_{\text{atm},I} (t), \\
    Q_\text{obs}^{(j)} (t) &= \eta_\text{P} Q_\text{conv}^{(j)} (t) + \epsilon^{IQ}_\text{det} I_\text{conv}^{(j)} (t) \nonumber \\
    &\quad + N_{\text{w}, Q} (t) + N_{\text{atm},Q} (t), \\
    U_\text{obs}^{(j)} (t) &= \eta_\text{P} U_\text{conv}^{(j)} (t) + \epsilon^{IU}_\text{det} I_\text{conv}^{(j)} (t) \nonumber \\
    &\quad + N_{\text{w}, U} (t) + N_{\text{atm},U} (t),
    \label{eq:obs_U}
\end{align}
\end{subequations}
where the intensity leakage coefficients are the real and imaginary parts of $\epsilon^{IP}_\text{det}$,
\begin{equation}
    \epsilon^{IQ}_\text{det} = \text{Re} [ \epsilon^{IP}_\text{det} ], \qquad
    \epsilon^{IU}_\text{det} = \text{Im} [ \epsilon^{IP}_\text{det} ],
    \label{eq:IU_leakage}
\end{equation}
and the noise terms follow the same demodulation as the sky signal. The intensity inherits the low-pass-filtered noise directly, $N_{\text{w/atm},I} = N_{\text{w/atm},\text{lp}}^{(j)}(t)$, while the polarization channels carry the bandpassed noise projected onto the two demodulation quadratures, $N_{\text{w/atm},Q} = 2N_{\text{w/atm},\text{bp}}^{(j)}(t)\cos\Phi^{(j)}(t)$ and $N_{\text{w/atm},U} = 2N_{\text{w/atm},\text{bp}}^{(j)}(t)\sin\Phi^{(j)}(t)$.

To express this in terms of the Stokes components on the sky in the QWP basis, we use the effective matrix of Section~\ref{sec:polmod}, $\mathbf{S}_\text{conv} = \bm{M}\,\mathbf{S}_\text{in}$. For the observed Stokes $U$,
\begin{multline}
    U_\text{obs}^{(j)} (t) = \eta_\text{P} \Big[ M_{IU} I_\text{in}^{(j)} (t) + M_{QU} Q_\text{in}^{(j)} (t) \\
    \ + M_{UU} U_\text{in}^{(j)} (t) + M_{VU} V_\text{in}^{(j)} (t) \Big] \\
    \ + \epsilon^{IU}_\text{det} I_\text{in}^{(j)} (t) + N^{(j)} (t),
\end{multline}
with $I_\text{conv}^{(j)} (t) \approx I_\text{in}^{(j)} (t)$, assuming negligible polarization leakage to intensity \citep{sakaguri_development_2026}. Using Equations~\eqref{eq:eta_V}--\eqref{eq:eps_IV}, the observed Stokes $U$ becomes
\begin{multline}
U_\text{obs}^{(j)} (t) = \eta_V \left[ V_\text{in}^{(j)} (t) + \epsilon^{QV} Q_\text{in}^{(j)} (t) + \epsilon^{UV} U_\text{in}^{(j)} (t) \right] \\
\ + \left( \epsilon^{IU}_\text{det} + \eta_V \epsilon^{IV} \right) I_\text{in}^{(j)} (t) + N^{(j)} (t).
\label{eq:obs_U_sky}
\end{multline}

The intensity term of Equation~\eqref{eq:obs_U_sky} collects two contributions of distinct origin: $\epsilon^{IU}_\text{det}$, generated downstream of the optics in the nonlinear detector response, and $\eta_V \epsilon^{IV}$, any static intensity leakage in the optics. We assume a value for neither, and therefore define the total intensity-to-$U$-channel leakage, including the transferred $I$-to-$V$ as well as the downstream $I$-to-$U$, to be
\begin{equation}
    \epsilon_\text{tot}^{IU} \equiv \epsilon^{IU}_\text{det} + \eta_V \epsilon^{IV},
    \label{eq:eps_ItoU}
\end{equation}
as the single quantity carried through the remainder of this work. The observed Stokes $U$ timestream is then, in final form,
\begin{multline}
    U_\text{obs}^{(j)} (t) = \eta_V \Big[ V_\text{in}^{(j)} (t) + \epsilon^{QV} Q_\text{in}^{(j)} (t) \\
    \ + \epsilon^{UV} U_\text{in}^{(j)} (t) \Big] + \epsilon_\text{tot}^{IU} I_\text{in}^{(j)} (t) + N^{(j)} (t).
    \label{eq:obs_U_final}
\end{multline}
Because $\epsilon_\text{tot}^{IU}$ is measured as a ratio of beam window functions (Section~\ref{sec:IU_leakage}), it is in general multipole dependent, and is written $\epsilon_{\ell,\text{tot}}^{IU}$ wherever it appears in $\ell$-space.

As mentioned earlier, Equation~\eqref{eq:IU_leakage} will be discussed in more detail in Section~\ref{sec:leakage}. From the bandpass, the two noise terms are between $\pm \qty{3.8}{\Hz}$ in baseband, which corresponds to between \qtyrange{4.3}{11.9}{\Hz} in the modulated regime. Generally the atmospheric low-frequency noise shows up below \qty{1}{\Hz}, which is primarily decimated by the bandpass, and while the white noise is flat in frequency space, it becomes independent in $Q$ and $U$ and is independent detector-to-detector \citep{kusaka_modulation_2014}, so it averages down with more observing time.

\subsubsection{Additional Filtering}
\label{sec:filter2}

In the event that low-frequency noise is not fully mitigated by the initial bandpass filtering, we reapply the polynomial filters described in Section~\ref{sec:filter1}. This residual noise is mostly unpolarized and driven by inhomogeneously distributed, turbulent water vapor and line-of-sight temperature gradients. In addition, because this secondary filtering stage acts on the demodulated timestreams, it also suppresses circularly polarized emission from atmospheric oxygen via Zeeman splitting \citep[][see also Appendix~\ref{app:circpol}]{Hanany_2003, spinelli_2011, Petroff_2020, Fujino_2025}. Unlike water vapor, oxygen is well-mixed, producing a signal that varies smoothly with the telescope's pointing relative to Earth's magnetic field. While this signal is carried at $4f_\text{HWP}$ through the initial filtering stages, demodulation downconverts it to low frequencies. Because the telescope sweeps monotonically in azimuth during a subscan, a low-order polynomial filter in the time domain acts effectively as a low-order expansion in azimuth, absorbing the bulk of this smoothly varying atmospheric signal. Any surviving residuals remain tied to telescope azimuth rather than celestial coordinates, and are subsequently removed by the scan-synchronous filter (see Appendix~\ref{app:circpol}). 

To address this remaining azimuth-locked pickup, such as ground spillover and far-sidelobe emission that recur at fixed pointings, we apply a scan-synchronous filter. Because left- and right-going sweeps generally couple differently to the local topography, the timestreams for each detector are divided into individual subscans and processed independently by scan direction. Within each direction, the subscans are pooled to construct an empirical template of the detector's azimuthal response. The data are binned in fixed \qty{0.2}{\degree} azimuth intervals, and the median of each bin is taken to suppress transient noise and outliers; bins with insufficient samples are discarded. We then fit a Chebyshev polynomial of degree $n=5$ to this profile, weighted by the intra-bin variance so that well-determined bins govern the fit. Finally, this smooth, direction-dependent azimuth template is evaluated at each sample's true pointing and subtracted directly from the timestream.

\subsection{Mapmaking}
\label{sec:mapmaking}

The result at the end of this processing is a set of timestream data, consisting of $\left\{ \text{RA}^{(j)} (t) \right\}$, $\left\{ \text{Dec}^{(j)} (t) \right\}$, and $\left\{ U_\text{obs}^{(j)}(t) \right\}$, for which we have associated turnaround buffer and glitch masks. For each detector in each observation we construct flat-sky maps of its associated timestream. A tangent-plane (gnomonic) projection is defined about the field center, spanning \qty{9}{\degree} on a side at a fixed pixel scale with width \qty{0.04}{\degree}, and detectors are mapped independently before being combined. The best-fit planar gradient in sky coordinates is then subtracted from each detector's retained samples, suppressing residual large-scale scan-synchronous and atmospheric gradients that survive the time-domain filtering. The cleaned samples are accumulated into the map by inverse-variance weighting, scaled according to the number of timestamp hits within a given pixel. Following collection of all detector maps within a single observation, the coadded observation map is computed as the per-pixel inverse-variance-weighted mean of the detector maps, utilizing the weight map saved alongside each map. This produces one map per CES, yielding 448 maps for RA23 and 341 for RA4.5 that survive the data quality cuts of Section~\ref{sec:data_quality}.

At \qty{0.04}{\degree/\pix} (\qty{2.4}{\arcminute}) over a $\qty{9}{\degree} \times \qty{9}{\degree}$ field, the map oversamples the science band. The pixel Nyquist multipole, $\ell_\text{Nyq} = \qty{180}{\degree}/(\Delta \theta) \approx 4500$, sits well above the highest analyzed multipole $\ell_\text{max} = 2000$ (a factor ${\sim}2.25$ in margin), so aliasing from pixelization is negligible over the analyzed multipoles. Pixelization does still smooth the map, suppressing power by the azimuthally averaged pixel window function,
\begin{multline}
    W_\ell^2 = \frac{1}{2\pi} \int_0^{2\pi} \text{sinc}^2 \left( \frac{\ell \Delta \theta}{2} \cos \varphi \right) \\ \text{sinc}^2 \left( \frac{\ell \Delta \theta}{2} \sin \varphi \right) \ d\varphi,
    \label{eq:pix_window}
\end{multline}
where $\varphi$ is the relative azimuthal angle of the 2D Fourier mode. The value of this function is ${\sim}0.996$ at $\ell = 300$ and ${\sim}0.850$ at $\ell = 2000$. This suppression is folded into the measurement of the transfer function, discussed in Section~\ref{sec:transfer}, as the simulated maps are generated at the same resolution as the observed maps. The pixel window therefore enters the forward model of Section~\ref{sec:cl_vv} exactly once, through $T_\ell$. The explicit $W_\ell$ of Equation~\eqref{eq:pix_window} is used only in Equation~\eqref{eq:pv_leakage}, to remove a second copy from a quantity that would otherwise carry one.

\subsection{Data Quality Cuts}
\label{sec:data_quality}

We impose additional data quality cuts to ensure that known systematics do not propagate into the final analysis, motivated by the null test suite of Section~\ref{sec:null_tests}. We split these cuts into both observation-level and detector-level cuts, as was done in the null tests. The yields in the following table are relative to the previous cut. Instrument loss refers to detector-observations lost to the technical failures described in Section~\ref{sec:obs}, such as corrupted timestream files, improperly stored database identifiers, or failed detector tuning. It also includes detectors that lack time constants or absolute gain calibration factors. The two observation cuts are whether a given run has associated temperature and PWV from the Atacama Pathfinder EXperiment (APEX, ${\sim}\qty{5}{\km}$ from the PB-2b site) and focal plane temperatures within \qtyrange{0.30}{0.38}{\kelvin}. APEX weather readings were required to run null tests with associated criteria and the focal plane temperature ranges exclude a focal plane with elevated temperatures. With the detector splits, we eliminate detectors with extreme non-physical limits of calibration parameters. Gain calibration factors must be negative (due to TES polarity) and ${\geq}-\qty{20}{\milli\Kcmb}/\mathrm{ADC}$, time constant must be ${<}\qty{50}{\milli\second}$, and measured white noise must be between \qty{10}{\micro\Kcmb/\sqrt{Hz}} and \qty{100}{\milli\Kcmb/\sqrt{Hz}}. These ensure the calibration factors are physical and the saved maps are within expected bounds. Finally, we make selections based on quantities derived from the maps themselves. Detector maps with known issues, such as spatial artifacts resulting from filtering, HWPSS removal, or glitch masking, are intentionally not used. These only appear in the RA23 patch. Runs with pixels that point to the edge of the patch footprint are not used; this is typically reflective of the boresight getting stuck and not tracking properly as the sky rotates. Finally, detector maps with abnormally high RMS, ${>}\qty{500}{\milli\Kcmb}$, are also dropped, as a final removal of any detector maps that are unphysical yet survive the earlier cuts.

The results of these cuts, as well as the earlier losses due to instrument technical issues, are represented in Table~\ref{tab:data_selection}.

\begin{table}[htbp]
\caption{Summary of data retention yields from saved detector-observations. The first row reflects the instrument losses described in Section~\ref{sec:obs}, both observation-wide and from detectors lacking calibration factors; all subsequent yields are relative to the preceding cut.}
\label{tab:data_selection}
\begin{tabular*}{\columnwidth}{@{\extracolsep{\fill}}lcc}
\hline\hline
Selection Type                 & RA23    & RA4.5   \\ \hline
\textit{Instrument Loss Yield} & \textit{62.9\%} & \textit{56.2\%} \\
\textit{Observation selection} & \textit{82.0\%} & \textit{83.2\%} \\
\quad Weather data exists      & 98.1\%  & 95.6\%  \\
\quad Focal plane temperature  & 83.6\%  & 87.0\%  \\
\textit{Detector selection}    & \textit{89.3\%} & \textit{83.2\%} \\
\quad Gain calibration factor  & 92.3\%  & 84.6\%  \\
\quad Time constant            & 99.6\%  & 99.4\%  \\
\quad White noise              & 97.1\%  & 99.0\%  \\
\textit{Map selection}         & \textit{94.1\%} & \textit{91.3\%} \\
\quad Map artifacts            & 96.1\%  & 100\%   \\
\quad Boresight pointing drift & 99.9\%  & 92.5\%  \\
\quad High variance            & 98.1\%  & 98.8\%  \\
\textit{Overall Yield}         & 43.3\%  & 35.5\%  \\
\hline\hline
\end{tabular*}
\end{table}

\subsection{Power Spectrum Estimation}
\label{sec:power_spectra}

\subsubsection{MASTER Estimator}
\label{sec:master}

The angular auto-power spectrum is estimated from a set of cross-power spectra, computed from pairs of independently coadded maps using the flat-sky implementation of the MASTER (Monte-Carlo Apodised Spherical Transform EstimatoR) pseudo-power-spectrum formalism via \textsc{NaMaster} and its Python wrapper \texttt{pymaster} \citep{hivon_master_2002, alonso_unified_2019}. Because each pair of maps is built from two disjoint sets of observations, their instrumental noise is uncorrelated and approximately averages to zero in the cross-correlation, so the resulting bandpowers are an unbiased estimate of the sky signal free of the noise-bias term that contaminates a single-map auto-spectrum. Each map is treated as a small patch of the celestial sphere projected onto a flat grid in right ascension and declination; the physical angular extent of the grid is set by the patch size, with the right-ascension axis rescaled by the cosine of the central declination to account for the convergence of meridians, so that the two-dimensional Fourier modes correspond to well-defined angular multipoles.

Before the transform, each map is lightly conditioned. Unobserved pixels are set to zero, the pixel-value distribution is iteratively sigma-clipped to obtain a robust per-map noise amplitude used to weight that pair relative to others, and a weighted monopole is subtracted so that the mean within the observed region vanishes. The coverage mask for each field is the region of valid coverage, defined as pixels with hit count ${>}\qty{-5}{\decibel}$ of the maximum. The threshold is set logarithmically rather than as a linear fraction because the hit count of a constant-elevation scan falls off across more than an order of magnitude between the deeply covered center of the field and its turnaround-dominated perimeter, so a decibel cut places the boundary on the natural scale of the coverage profile. The \qty{-5}{\decibel} value, retaining pixels above ${\sim}\qty{32}{\percent}$ of the peak hit count, balances sky area against edge quality. Pixels below it carry the highest per-pixel noise and the largest residuals from the ground-template and polynomial filtering of Section~\ref{sec:tod}, so they contribute little statistical weight relative to the systematic risk they introduce; they also lie where the coverage boundary becomes most irregular, which broadens the mode-coupling kernel and degrades the conditioning of the binned coupling matrix. For RA4.5 a \qty{0.2}{\degree}-radius circular hole is additionally cut at the position of the bright point source J0440$-$4333 (see Appendix~\ref{app:point_source_validation}) \citep{bonato_almacal_2018}, the only significant potentially contaminating point source within either footprint \citep{sakaguri_development_2026}, so that its compact emission does not contribute to the bandpowers. The mask is apodized with a \qty{0.2}{\degree}-wide cosine taper. At five map pixels, this width is well resolved by the grid, so its smooth roll-off replaces the slowly decaying mode-coupling sidelobes of a sharp edge with a compact kernel that keeps the binned coupling matrix well conditioned. It nonetheless remains small compared to both the ${\sim}\qty{2}{\degree}\times\qty{3}{\degree}$ effective footprint retained by the hit-count threshold and the ${\sim}\qty{1.2}{\degree}$ angular wavelength of the lowest analyzed multipole ($\ell = 300$). The apodized footprint is not applied as a binary weight; it is multiplied by the per-pixel inverse-variance weight map from mapmaking (Section~\ref{sec:mapmaking}), and it is the power spectrum of this combined weight map that sets the mode-coupling matrix.

The masked, apodized maps are then carried through the MASTER estimator, which deconvolves the mode coupling induced by the finite apodized footprint using the coupling matrix computed from the power spectrum of the mask itself \citep{hivon_master_2002}. Because that matrix is singular at the level of individual multipoles, the spectrum is first averaged into uniform bandpowers over the range where the binned matrix can be inverted. No beam is deconvolved at this stage, so these bandpowers remain beam-smoothed. The beam is removed only in the final inversion of Section~\ref{sec:cl_vv}.

In the context of this study, this MASTER estimation procedure is primarily used to determine the measured circular polarization spectra as cross-spectra between pairs of independently coadded observation maps. However, we also use this estimator within the context of simulated maps, in order to estimate transfer functions (see Section~\ref{sec:transfer}) and leakages (see Section~\ref{sec:leakage}).

\subsubsection{Measured $UU$ Spectra}
\label{sec:meas_UU}

The quantity accessible directly from the maps is the $UU$ power spectrum of $U_\text{obs}$, not the circular polarization spectrum itself. As shown in Equation~\eqref{eq:obs_U_final}, the observed $U$ carries the sky $V$ signal converted by the QWP, but scaled by the circular polarization efficiency $\eta_V$ and accompanied by leakage from sky $I$, $Q$, and $U$, as well as suppression by the instrument beam, residual pointing error, and the filtering applied to the timestreams. Recovering $C_\ell^{VV}$ therefore requires measuring the observed $UU$ bandpowers first and then disentangling each of these contributions from them. This section presents that measurement; the transfer function and leakage terms are quantified in Sections~\ref{sec:transfer} and \ref{sec:leakage}, and the inversion that yields $C_\ell^{VV}$ is carried out in Section~\ref{sec:cl_vv}.

We estimate the average of the cross-spectra from the set of coadded observation maps, which are immune to the noise bias of an auto-spectrum because the noise in two different maps is uncorrelated. Forming cross-spectra between the individual CES maps of Section~\ref{sec:mapmaking} would require of order $10^5$ pairs per patch. These are therefore first combined into a smaller number of deeper maps. Each patch is observed in blocks of several consecutive CESes, and the maps within a block are coadded together. Each CES contributes to exactly one such map, so no observation is shared between any two of them and their noise remains uncorrelated. This reduces the 448 CES maps of RA23 and the 341 of RA4.5 to $M_\text{patch}^\text{RA23} = 75$ and $M_\text{patch}^\text{RA4.5} = 61$ independent maps, respectively, which we index by $a \in \{1, \dots, M_\text{patch}\}$. We then form the cross-power spectra of all $N_\text{patch}=M_\text{patch}(M_\text{patch}-1)/2$ distinct map pairs, indexed by $k \in \{1, \dots, N_\text{patch}\}$, where the pair $k$ is formed from the two distinct maps $a$ and $a'$. Each pair spectrum is computed with the flat-sky MASTER estimator on monopole-subtracted, apodized maps and binned into $\Delta\ell=100$ bandpowers over $300\le\ell\le2000$. The pairs are not equally informative, because the maps entering them differ in depth and different pairs overlap over different fractions of the patch. A uniform average would therefore be suboptimal, and the bandpower estimate is instead formed as an inverse-variance weighted mean over pairs,
\begin{equation}
  \bar{\hat{C}}_\ell^{UU,\text{obs}} = \frac{\sum_{k} w_k\,\hat{C}_\ell^{UU,\text{obs},(k)}}{\sum_{k} w_k},
  \label{eq:cl_mean}
\end{equation}
which is the quantity plotted as the data points. Throughout, a hat denotes a quantity estimated from the data and a bar an average, so that $\hat{C}_\ell$ is a single measured cross-spectrum and $\bar{\hat{C}}_\ell$ its mean over pairs; spectra carrying a bar alone, such as the leakage terms of Section~\ref{sec:leakage}, are averages over simulations rather than measurements. The one exception is $\hat{n}$, which denotes a direction on the sky.

The weight $w_k$ is a single $\ell$-independent inverse-noise-variance factor per pair,
\begin{equation}
  w_k = \frac{f_{\rm sky}^{\rm eff}(k)}{\sigma_a^2\,\sigma_{a'}^2},
  \label{eq:weight}
\end{equation}
where $\sigma_a$ and $\sigma_{a'}$ are per-map noise proxies for the two maps forming pair $k$, each obtained by taking the standard deviation of the unmasked pixels, discarding those deviating by more than $4\sigma$, and adopting the standard deviation of the pixels that remain. The effective sky fraction $f_{\rm sky}^{\rm eff}(k)$ is the mean over pixels of the product of those two maps' apodized coverage masks. Because the cross-spectrum variance scales as $\sigma_a^2\sigma_{a'}^2/f_{\rm sky}^{\rm eff}$, Equation~\eqref{eq:weight} is the corresponding inverse-variance weight; the common sky-signal contribution to the map variance is shared by every pair and cancels in Equation~\eqref{eq:cl_mean}, so a scalar weight encoding only per-pair noise and mask overlap well approximates the optimal weight.

Estimating the uncertainty on $\bar{\hat{C}}_\ell^{UU,\text{obs}}$ requires accounting for correlations between the pair spectra, because the $N_\text{patch}$ pairs are formed from a common set of $M_\text{patch}$ maps, with each map entering $M_\text{patch}-1$ pairs. In principle this couples pairs that share a map, which would make the maps, rather than the pairs, the independent units. Two such pairs are correlated, however, only through a component common to their unshared maps. These maps are dominated by instrument noise, which is independent from map to map, and each is itself a coadd of separate observations, so their atmospheric fluctuations are independent as well. The only component common to every map is the sky signal itself. Theoretical predictions place CMB circular polarization well below our sensitivity \citep{king_circular_2016, farias_performance_2026}, so this shared component is expected to be negligible. We confirm this directly by estimating the shared-map covariance $\zeta_1 \equiv \mathrm{Cov}\big(\hat{C}_\ell^{UU,\text{obs},(k)},\, \hat{C}_\ell^{UU,\text{obs},(k')}\big)$ between distinct pairs $k$ and $k'$ that have one of their two maps in common, which is consistent with zero in every $\ell$-bin; the pairs are therefore effectively independent. We accordingly estimate the covariance of $\bar{\hat{C}}_\ell^{UU,\text{obs}}$ with a pair-level bootstrap. In each of $N_\text{boot}=10^{3}$ bootstrap realizations, indexed by $r \in \{1, \dots, N_\text{boot}\}$ and labeled throughout by the superscript $(r)$, we draw $N_\text{patch}$ pairs with replacement and recompute the weighted mean of Equation~\eqref{eq:cl_mean} over the resampled set,
\begin{equation}
  \bar{\hat{C}}_\ell^{UU,\text{obs},(r)}
    = \frac{\sum_{k \in \mathcal{B}^{(r)}} w_k\, \hat{C}_\ell^{UU,\text{obs},(k)}}
           {\sum_{k \in \mathcal{B}^{(r)}} w_k},
\end{equation}
where $\mathcal{B}^{(r)}$ is the multiset of pair indices drawn in realization $r$, so that $\bar{\hat{C}}_\ell^{UU,\text{obs},(r)}$ is the weighted mean recovered from that realization. The bandpower covariance is the sample covariance of the $N_\text{boot}$ resampled means,
\begin{subequations}
\begin{align}
  \Sigma^{UU,\text{obs}}_{\ell\ell'}
    &= \frac{1}{N_\text{boot}-1}\sum_{r=1}^{N_\text{boot}}
      \Delta^{(r)}_\ell \, \Delta^{(r)}_{\ell'}, \\
  \Delta^{(r)}_\ell &\equiv \bar{\hat{C}}_\ell^{UU,\text{obs},(r)}
      - \big\langle \bar{\hat{C}}_\ell^{UU,\text{obs}} \big\rangle_r ,
\end{align}
\end{subequations}
where $\langle\cdot\rangle_r$ averages over realizations; the uncertainty per $\ell$-bin is $\sqrt{\Sigma^{UU,\text{obs}}_{\ell\ell}}$. The covariance is therefore measured directly from the resampled scatter, rather than assumed from an analytic form evaluated at a fiducial spectrum; the Gaussian likelihood of Section~\ref{sec:mcmc} uses only this measured second moment. We verified that the alternative map-level bootstrap, in which whole maps rather than pairs are resampled and each pair is weighted by the product of the number of times its two maps were drawn, inflates the bandpower uncertainties $\sqrt{\Sigma^{UU,\text{obs}}_{\ell\ell}}$ by a factor of ${\sim}1.8$; since $\zeta_1\simeq0$, this excess reflects the scatter in those draw counts from one realization to the next rather than genuine map-to-map correlation, and would bias the covariance high.

\subsection{Null Tests}
\label{sec:null_tests}

We assess the internal consistency of the power spectrum, measured in Section~\ref{sec:power_spectra}, with a suite of sign-flip jackknife null tests, following the framework developed for earlier \textsc{Polarbear} analyses \citep{polarbear_collaboration_exploration_2024, polarbear_collaboration_constraints_2026}. Each test is defined by a criterion $c$ that partitions the data into two halves according to a particular systematic. The null test criteria are listed in Appendix~\ref{app:null_tests} and are adapted from \citetalias{polarbear_collaboration_exploration_2024}, with a few additional criteria specific to PB-2b. Differencing the two half-coadds cancels the common sky signal, so the resulting null bandpower spectrum $\hat{C}^{\rm null}_b(c)$ (estimated via MASTER as discussed in Section~\ref{sec:master}) is consistent with zero in the absence of systematic contamination or miscalibration. Throughout this section $b$ indexes the same $\Delta\ell = 100$ bandpowers over $300 \le \ell \le 2000$ used elsewhere in this work; we label bins by the discrete index $b$ rather than their center $\ell$ only to keep the sums and maxima over bins compact. The null tests and the data quality cuts of Section~\ref{sec:data_quality} were developed together rather than in sequence---a criterion that failed motivated a candidate cut or pipeline adjustment, which was applied before the full suite was rerun, iterating until every criterion passed. The cuts quoted in Section~\ref{sec:data_quality} are the endpoint of that loop, and the PTEs reported below are those of the final iteration.

The null distribution for each criterion is estimated directly from the data by Monte Carlo simulation. We generate $N = 200$ random sign-flip realizations $\hat{C}^{{\rm rand},i}_b(c)$, $i = 1, \dots, N$, each constructed identically to the data null spectrum $\hat{C}^{\rm null}_b(c)$ except that the observations are assigned to the two halves at random rather than by the criterion $c$. Each realization therefore inherits the noise, filtering, and survey geometry of the data but contains no coherent split-dependent signal. From this ensemble we form the per-criterion mean $\mu_b(c)$ and standard deviation $\sigma_b(c)$, the standardized per-bin residuals of the data, $\chi_{c,b} = \big(\hat{C}^{\rm null}_b(c) - \mu_b(c)\big)/\sigma_b(c)$, and the bandpower covariance $\mathbf{C}_c$, which defines a covariance-weighted statistic for each criterion,
\begin{equation}
  \chi^2_{\rm null}(c) = \big(\hat{\mathbf{C}}^{\rm null}(c) - \boldsymbol{\mu}(c)\big)^{\!\mathsf{T}}
  \,\mathbf{C}_c^{-1}\,\big(\hat{\mathbf{C}}^{\rm null}(c) - \boldsymbol{\mu}(c)\big),
  \label{eq:chi2crit}
\end{equation}
where boldface denotes vectors over the bandpower index $b$. Evaluated on the data split, Equation~\eqref{eq:chi2crit} yields the observed statistic $\chi^2_{\rm null}(c)$, while the identical expression applied to each random shuffle $i$ (replacing $\hat{\mathbf{C}}^{\rm null}(c)$ with $\hat{\mathbf{C}}^{{\rm rand},i}(c)$) yields its null distribution $\{\chi^{2,i}_{\rm rand}(c)\}_{i=1}^{N}$. Because the same statistics are evaluated on every shuffle, each quantity carries its own empirical null distribution and we make no assumption about its analytic form. A PTE is read off as
\begin{equation}
  {\rm PTE} = \frac{1 + N_{\ge}}{N + 1},
  \label{eq:pte}
\end{equation}
where $N_{\ge}$ is the number of shuffle realizations whose value of the statistic is at least as extreme as the observed (data) value, so that the fraction of trials (data included) at least as extreme as the observed value is well defined and unbiased at finite $N$.

To avoid reporting many correlated PTEs, we compress the full set of residuals into four summary statistics, each sensitive to a distinct failure mode. These are the grand mean $\langle\chi_{\rm null}\rangle_{c,b}$, which detects a coherent additive bias and is evaluated two-sided; the largest single excursion $\max_{c,b}\chi^2_{\rm null}$, sensitive to one anomalous bandpower; the worst criterion $\max_{c}\chi^2_{\rm null}(c)$ from Equation~\eqref{eq:chi2crit}, sensitive to one bad split; and the total $\sum_{c}\chi^2_{\rm null}(c)$, which captures a diffuse, low-level excess spread across many splits. Each statistic is referred to its own shuffle distribution via Equation~\eqref{eq:pte} to yield a PTE (Table~\ref{tab:null_summary}).

\begin{figure*}[ht!]
    \centering
    \includegraphics[width=\textwidth]{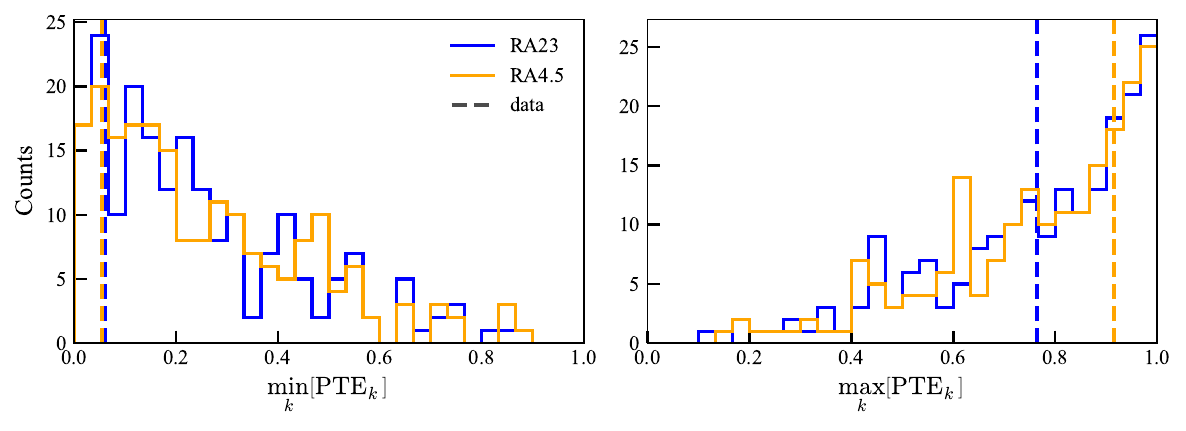}
    \caption{Look-elsewhere calibration of the null tests for RA23 (blue) and RA4.5 (orange). \emph{Left:} distribution over the $N=200$ random sign-flip shuffles of the most significant of the four summary-statistic PTEs, $\min_k {\rm PTE}_k$; \emph{right:} the least significant, $\max_k {\rm PTE}_k$. Dashed vertical lines mark the values measured in the data, and the corrected meta-PTEs of Equation~\eqref{eq:metapte} are simply the shuffle fractions at or beyond them (to the left in the left panel, to the right in the right), and are tabulated alongside the individual statistics in Table~\ref{tab:null_summary}. In both patches the data fall well within the null distributions, showing neither an anomalously low minimum PTE (left) nor the clustering near unity that would signal over-consistency (right).}
    \label{fig:null_summary}
\end{figure*}

Because four statistics are examined simultaneously, the most extreme of them is biased toward small values even under a clean null, and the minimum PTE cannot be interpreted at face value. We therefore treat the ``worst of four,'' ${\rm PTE}^{\min} = \min_k {\rm PTE}_k$, and the ``best of four,'' ${\rm PTE}^{\max} = \max_k {\rm PTE}_k$, as test statistics in their own right and calibrate them against the null. The identical four-statistic reduction is applied to each shuffle, producing distributions of ${\rm PTE}^{\min}$ and ${\rm PTE}^{\max}$ from which we compute the look-elsewhere--corrected meta-PTEs,
\begin{subequations}
\begin{align}
{\rm PTE}^{\min}_{\rm meta}
  &= \frac{1 + \#\{\,i : {\rm PTE}^{\min}_i \le {\rm PTE}^{\min}_{\rm data}\,\}}{N + 1}, \\
{\rm PTE}^{\max}_{\rm meta}
  &= \frac{1 + \#\{\,i : {\rm PTE}^{\max}_i \ge {\rm PTE}^{\max}_{\rm data}\,\}}{N + 1}
\end{align}
  \label{eq:metapte}
\end{subequations}
By construction the shuffles ``look elsewhere'' exactly as the analysis does, so the multiple-testing penalty is incorporated automatically, without assuming Gaussianity or independence among the four statistics. The condition for passage is that both meta-PTEs fall within $(0.05, 0.95)$. Each is uniformly distributed under a clean null, so either tail is informative, and the two bounds catch complementary failures. A low ${\rm PTE}^{\min}_{\rm meta}$ or a high ${\rm PTE}^{\max}_{\rm meta}$ indicates residual contamination; in the first case the least consistent of the four statistics is more extreme than the shuffle ensemble permits, and in the second even the most consistent one is not consistent enough. A high ${\rm PTE}^{\min}_{\rm meta}$ or a low ${\rm PTE}^{\max}_{\rm meta}$ instead signals the opposite pathology of over-consistency, the hallmark of error bars that are too conservative. Both patches pass, and we find ${\rm PTE}^{\min}_{\rm meta} = 0.19\,(0.16)$ and ${\rm PTE}^{\max}_{\rm meta} = 0.58\,(0.30)$ for RA23 (RA4.5), with the smallest individual PTE in either patch being $0.055$ (Table~\ref{tab:null_summary}, Figure~\ref{fig:null_summary}).

\begin{table}[htbp]
\caption{Summary of the PB-2b $VV$ null tests for the RA23 and RA4.5 patches. }
\label{tab:null_summary}
\begin{tabular*}{\columnwidth}{@{\extracolsep{\fill}}lccccc}
\hline\hline
 & \multicolumn{2}{c}{RA23} & \multicolumn{2}{c}{RA4.5} \\
\cline{2-3} \cline{4-5}
Statistic & Value & PTE & Value & PTE \\ \hline
$\langle \chi_\mathrm{null} \rangle_{c,b}$ & 0.024 & 0.765 & 0.020 & 0.915 \\
$\max_{c,b}\, \chi^2_\mathrm{null}$ & 11.22 & 0.195 & 12.00 & 0.145 \\
$\max_{c}\, \chi^2_\mathrm{null}(c)$ & 34.86 & 0.060 & 33.81 & 0.055 \\
$\sum_{c}\, \chi^2_\mathrm{null}(c)$ & 543.3 & 0.265 & 533.1 & 0.315 \\ \hline
$\mathrm{PTE}^{\min}_\mathrm{meta}$ & \multicolumn{2}{c}{0.190} & \multicolumn{2}{c}{0.160} \\
$\mathrm{PTE}^{\max}_\mathrm{meta}$ & \multicolumn{2}{c}{0.575} & \multicolumn{2}{c}{0.300} \\
\hline\hline
\end{tabular*}
\end{table}

\subsection{Transfer Functions}
\label{sec:transfer}

The combination of ground-template removal, polynomial and subscan filtering, and demodulation suppresses and mixes power in the recovered maps; this is an effect we characterize with an end-to-end simulation transfer function $T_\ell$. For each patch, given that true sky $V$ is an unknown signal, we generate $N_{\rm sim}=100$ realizations of a known input sky signal, in this case white noise smoothed with a Gaussian kernel of \qty{5}{\arcminute} standard deviation so that it carries power across the full range of analyzed multipoles. The simulated sky is then scanned into timestream using the real pointing and detector set, with noise added following the measured PB-2b noise spectrum but scaled down in amplitude. The resulting timestreams are propagated through the identical map-making pipeline applied to the data (time-constant deconvolution, turnaround masking, degree-five polynomial and subscan-polynomial filtering, ground-template and notch HWPSS removal, demodulation, and per-subscan binning and coaddition). Retaining noise at all keeps the stochastic character of the timestreams that the noise-dependent stages of the pipeline respond to, such as glitch flagging, sigma-clipping, and inverse-variance weighting, while scaling it down reduces the scatter on $T_\ell$ and so the number of realizations needed for it to converge. Aside from this reduced noise level, the only difference from the data run is the substitution of the known input sky for the measured timestreams, so the resulting output map $m^{\rm out}$ carries the full multiplicative and mode-mixing response of the analysis to a known input $m^{\rm in}$.

We estimate the transfer function from the cross-spectrum of the processed output with the input realization. For each input sky realization $s \in \{1,\dots,N_{\rm sim}\}$ we first average over a set of observations $i \in \{1,\dots,N_{\rm obs}\}$ reserved for these simulations, which is the benchmark set for RA23 and a separate set for RA4.5 (Section~\ref{sec:blinding}),
\begin{equation}
  \hat{T}^{(s)}_\ell = \frac{1}{N_{\rm obs}} \sum_{i=1}^{N_{\rm obs}} \frac{ \hat{C}_\ell^{\,m^{\rm out}_{(i,s)}\times m^{\rm in}_{(s)}} }{ \hat{C}_\ell^{\,m^{\rm in}_{(s)}\times m^{\rm in}_{(s)}} },
  \label{eq:transfer_per_sim}
\end{equation}
and then average over realizations,
\begin{equation}
  T_\ell = \frac{1}{N_{\rm sim}} \sum_{s=1}^{N_{\rm sim}} \hat{T}^{(s)}_\ell ,
  \label{eq:transfer}
\end{equation}
binned in the same bandpowers as our science spectra. Here $m^{\rm in}_{(s)}$ is the input sky of realization $s$ and $m^{\rm out}_{(i,s)}$ is the map recovered by propagating that realization through observation $i$; the index $i$ runs over that set rather than over the full data set. Using the output$\times$input cross-spectrum rather than the output auto-spectrum makes the estimator insensitive to any signal uncorrelated with the input (e.g.\ map noise). Normalizing by the per-realization input auto-spectrum, rather than the ensemble-mean or theory spectrum, cancels the sample variance of each realization in the ratio. The effective number of modes per bandpower, measured directly from the scatter of the input auto-spectra, is $N_\ell \approx 90$--700 over the range considered, so this removes a $\sqrt{2/N_\ell} \approx 5$--15\% per-realization fluctuation and reduces the variance of $T_\ell$ by 20--25\%. Because $T_\ell$ is formed bin by bin as this ratio, it characterizes the response of the filtering pipeline rather than any property of the sky, and is insensitive to the choice of input spectrum provided each bandpower carries sufficient power. 

Circular polarization is a spin-0 field with a single auto-spectrum and no $E$/$B$ decomposition, and the analysis correspondingly operates on the single demodulated $U$ map, so one transfer function suffices; the lensed-$\Lambda$CDM $TT/TE/EE/BB$ realizations used to evaluate the $P\!\to\!V$ leakage are described separately in Section~\ref{sec:PV_leakage}.

The uncertainty on $T_\ell$ is estimated from the scatter among the $\hat{T}^{(s)}_\ell$,
\begin{multline}
  \mathrm{Cov}\left(T_\ell, T_{\ell'}\right) \\ \ = \frac{1}{N_{\rm sim}\left(N_{\rm sim}-1\right)} \sum_{s=1}^{N_{\rm sim}} \left( \hat{T}^{(s)}_\ell - T_\ell \right) \left( \hat{T}^{(s)}_{\ell'} - T_{\ell'} \right).
\end{multline}
Because every observation in the set processes the same set of input skies, any realization-dependent bias in $\hat{T}^{(s)}_\ell$ is common to all $i$ and averages down only with $N_{\rm sim}$; the independent samples are therefore the input skies rather than the observations, and the covariance must be taken over $s$ after the average over $i$ has been performed. %

\begin{figure}[htbp]
    \centering
    \includegraphics[width=\linewidth]{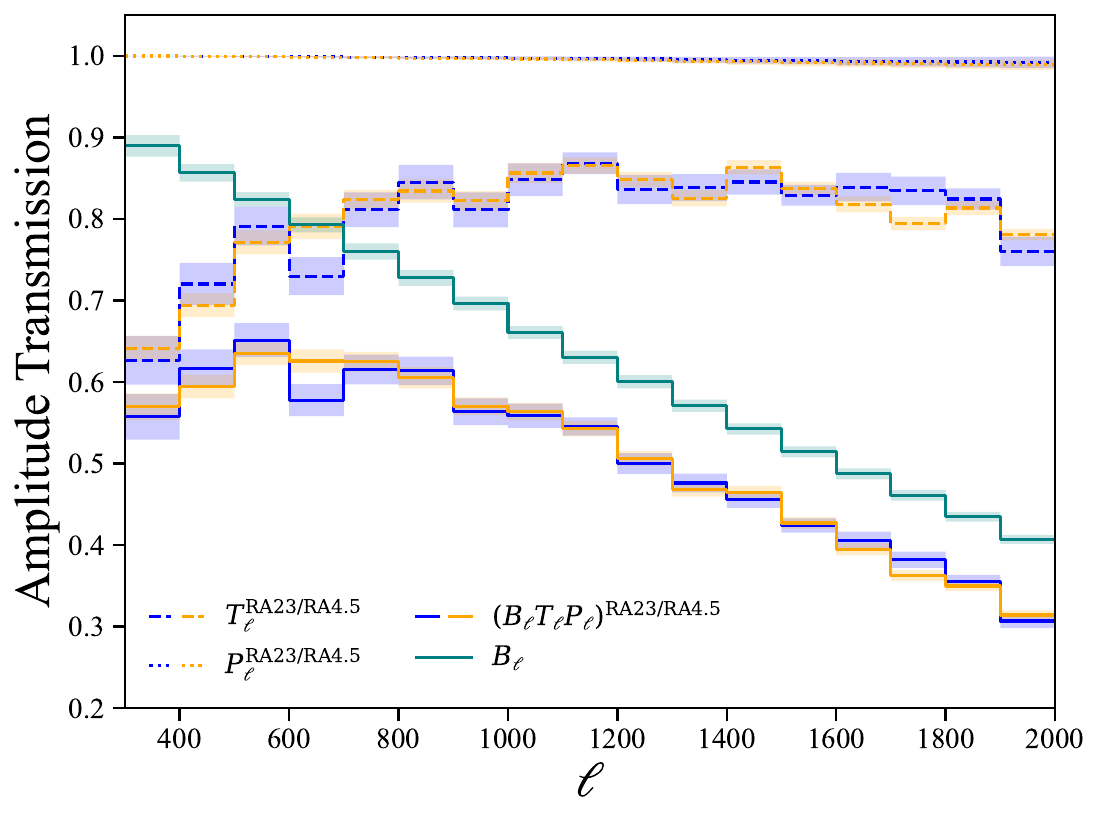}
    \caption{Per-mode amplitude transmission of the $VV$ analysis versus multipole. Dashed steps show the simulation transfer function $T_\ell$ (Equation~\eqref{eq:transfer}) for RA23 (blue) and RA4.5 (orange), the dotted lines represent the pointing transfer function for each patch (blue and orange), and the solid teal line represents the independently measured beam window function $B_\ell$. The solid blue and orange curves represent the product $T_\ell P_\ell B_\ell$ per patch, with $1\sigma$ bands from the combined $T_\ell$, $P_\ell$, and $B_\ell$ uncertainties. The two patches are nearly identical.}
    \label{fig:sensitivity}
\end{figure}

The transfer function is determined independently for each patch. $\bar{\hat{C}}_\ell^{UU,\text{obs}}$ is deconvolved by $T_\ell$, together with the beam and the circular polarization efficiency $\eta_V$, before the upper limits are derived, and its covariance is propagated into the final error budget; the results are shown together with $B_\ell$ in Figure~\ref{fig:sensitivity}. Additionally, Figure~\ref{fig:sensitivity} demonstrates low-$\ell$ suppression, which validates that the polynomial and scan-synchronous filtering stages (described in Sections~\ref{sec:filter1} and \ref{sec:filter2}, respectively) effectively marginalize the low-$\ell$ contamination associated with atmosphere, most notably the circularly polarized signal induced by $\mathrm{O}_2$ Zeeman splitting and the loading fluctuations induced by variable water vapor.

\subsection{Leakage Estimation}
\label{sec:leakage}

\begin{figure*}[ht!]
    \centering
    \includegraphics[width=\linewidth]{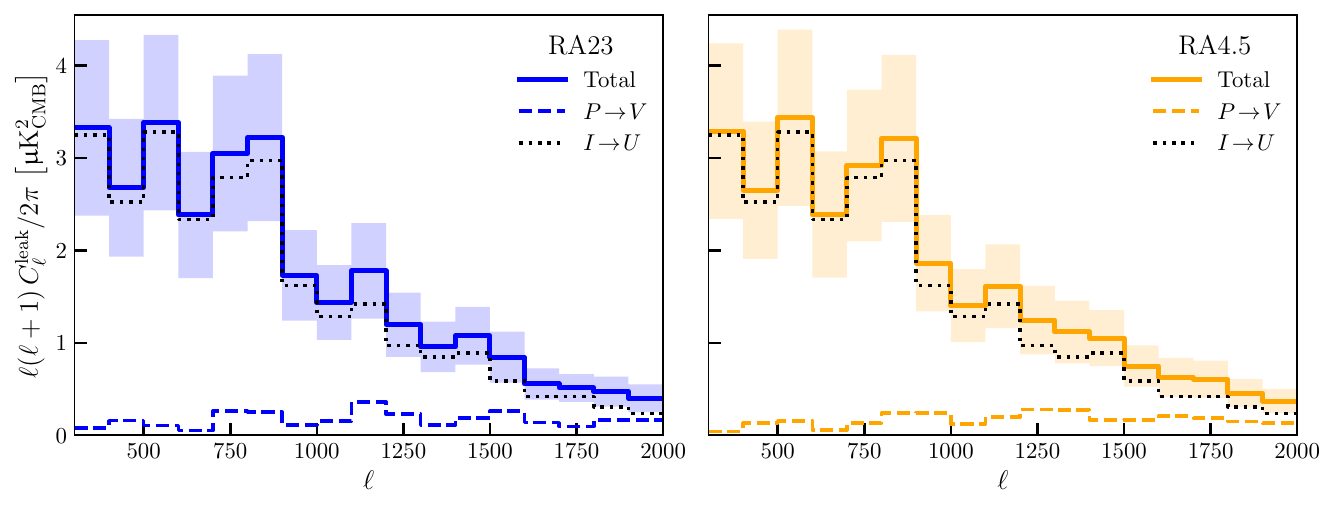}
    \caption{Leakages determined from Sections~\ref{sec:IU_leakage} and \ref{sec:PV_leakage}. The left is for RA23, and the right RA4.5. The joint contribution of both leakage sources is shown as the solid step plot with associated uncertainty ranges, accounting for $\eta_V = 0.528 \pm 0.032$. The uncertainties in the plots are propagated from the circular polarization uncertainty and variance of the individual leakage contributors.}
    \label{fig:leakages}
\end{figure*}

In \citet{sakaguri_development_2026}, the leakages between Stokes parameters due to instrument systematics were discussed. In this section we will derive bandpower intensity leakage and forecast the polarization leakage based on the scan strategy of PB-2b. The contributions of both leakage sources are shown in Figure~\ref{fig:leakages}, where
\begin{equation}
C_\ell^\text{leak} = |\eta_V|^2 \bar{C}_\ell^{P\to V} + C_\ell^{I\to U},
\end{equation}
based on the leakage terms present in Equation~\eqref{eq:backward_model}.

\subsubsection{Total $I\!\to\!U$ Leakage in $\ell$-Space}
\label{sec:IU_leakage}

As mentioned earlier, \citet{sakaguri_development_2026} report a small but non-negligible intensity leakage into the $V$ channel. However, there is additional leakage from intensity generated in the detector response \citep{takakura_characterization_2017}, downstream of the optics and therefore outside the Mueller-matrix description; this is the $\epsilon^{IU}_\text{det}$ of Equation~\eqref{eq:IU_leakage}. The quantity measured here is therefore the total $\epsilon_{\ell,\text{tot}}^{IU}$ of Equation~\eqref{eq:eps_ItoU}, which additionally absorbs the static optical leakage $\eta_V \epsilon^{IV}$. We measure it directly from observations of Jupiter, which at \qty{90}{\giga\Hz} is a bright and, to good approximation, unpolarized source (both in linear and circular). Any $U$ response in its co-added map is therefore dominated by instrumental leakage of $I$. We co-add all \qty{90}{\giga\Hz} Jupiter maps in $I_\text{obs}$ and $U_\text{obs}$, with Stokes data transformed into the QWP basis, apodize the finite map with a \qty{2}{\arcminute} taper, zero-pad, and Fourier transform; both transforms share a common normalization (the $I$ zero-mode), which cancels in the leakage ratio below. In essence, this is the same procedure as the estimation of the instrumental beam, detailed in Section~\ref{sec:beam}, and uncertainty in each $\ell$-bin is propagated similarly.

As a result, the total $I\!\to\!U$ leakage is the ratio of the two amplitudes,
\begin{equation}
  \epsilon_{\ell,\text{tot}}^{IU} = \frac{B_\ell^{U}}{B_\ell^{I}},
\end{equation}
where $B_\ell^{I/U}$ are the beam functions of the demodulated Stokes component maps $I_\text{obs}$ and $U_\text{obs}$, respectively, following the same procedure in Section~\ref{sec:beam}. The uncertainty on the leakage in each multipole bin is estimated with a leave-one-out jackknife over the individual planet observations, re-coadding Stokes $I$ and $U$ and re-running the full pipeline with each observation removed in turn. Across $300 \le \ell \le 2000$, we find $\bar{\epsilon}_{\mathrm{tot}}^{IU} \sim (1.8 \pm 0.3)\%$. This is the total leakage into the $V$-carrying channel, containing both the optical term of Section~\ref{sec:polmod} and the detector-response term of Section~\ref{sec:demod}; we do not attempt to separate the two. We carry $\epsilon_{\ell,\text{tot}}^{IU}$ and its uncertainty in $\ell$-space as systematic terms in the $VV$ error budget. 

We estimate the total leakage in $\ell$-space,
\begin{equation}
    C_\ell^{I\to U} = \left| \epsilon_{\ell,\text{tot}}^{IU} \right|^2 C_\ell^{TT,\text{sky}},
\end{equation}
where $C_\ell^{TT,\text{sky}}$ is computed from CAMB \citep{lewis_efficient_2000}, using best-fit cosmological parameters from Planck \citep{Planck2018VI}. Measured $I$ maps from PB-2b were not utilized, due to high low-frequency noise from weather contaminating the demodulated intensity maps. Because $I$ and $V$ are spin-0 fields, they are invariant under the rotation represented by the detector transform.

\subsubsection{Overall $P$-to-$V$ Leakage}
\label{sec:PV_leakage}

As discussed in Section~\ref{sec:polmod}, residual instrumental coupling converts a fraction of the linear polarization into the circular-polarization channel, so the measured $V$ map contains a leakage term $\epsilon^{QV}\,Q_\text{in} + \epsilon^{UV}\,U_\text{in}$, where $\epsilon^{UV}$ was measured and $\epsilon^{QV}$ bounded in \citet{sakaguri_development_2026}, as described in Section~\ref{sec:polmod}. Both are defined from the effective Mueller matrix in Equations~\eqref{eq:eta_V}--\eqref{eq:eps_IV}. At the power-spectrum level, with pixel window function $W_\ell^2$ from Equation~\eqref{eq:pix_window}, this contributes
\begin{multline}
  \bar{C}_\ell^{\,P\to V} = \frac{1}{W_\ell^2} \Big[ \left|\epsilon^{QV}\right|^2\,\bar{C}_\ell^{QQ,\text{in}} + \left|\epsilon^{UV}\right|^2\,\bar{C}_\ell^{UU,\text{in}} \\
  \ + 2\,\epsilon^{QV}\epsilon^{UV}\,\bar{C}_\ell^{QU,\text{in}} \Big],
  \label{eq:pv_leakage}
\end{multline}
for each patch, where the bar indicates this is a measured quantity from simulations, rather than a deterministic, analytical systematic. The explicit factor of $1/W_\ell^2$ is what prevents the pixel window from being counted twice. These leakage spectra are estimated from simulated theory skies synthesized with the same pixel grid and resolution as the data, so they carry the pixel window of that grid. The forward model of Equation~\eqref{eq:backward_model} then multiplies them by $\left| T_\ell \right|^2$, which already includes the pixel window by construction (Section~\ref{sec:mapmaking}). Dividing it out here leaves the pixel window applied exactly once, to the sky signal and the leakage alike. Propagating it into the $VV$ budget requires the auto- and cross-spectra of the linear polarization as it actually projects onto our maps due to the multitude of QWP orientations relative to the sky per observation. Because the detectors measure polarization in the QWP frame, the relevant $Q$ and $U$ are the sky signal rotated by the parallactic angle and then averaged over the full set of observations, which partially cancels the leakage wherever the patch is observed over a range of parallactic angles.

We estimate these spectra from simulations. For each realization we draw a Gaussian lensed-$\Lambda$CDM sky ($TT/TE/EE/BB$; flat-sky $Q$ and $U$ synthesized on the survey patch geometry with the Boltzmann code wrapper \texttt{CAMB} \citep{lewis_efficient_2000}), using best-fit cosmological parameters from Planck \citep{Planck2018VI}. From the real observations we build effective rotation maps, accumulating the weighted parallactic-angle factors over all observations $i$,
\begin{equation}
  C(\hat{n}) = \frac{\sum_i w_i\,\cos 2\gamma_i}{\sum_i w_i}, \
  S(\hat{n}) = \frac{\sum_i w_i\,\sin 2\gamma_i}{\sum_i w_i},
\end{equation}
where $\gamma_i$ is the parallactic angle at the patch center for observation $i$. For the sake of this leakage analysis, $Q_\text{in}$ and $U_\text{in}$ are treated as spin-0 fields, assuming independent leakage per Stokes component. The effective instrument-frame polarization is then $\tilde{Q} = Q\,C + U\,S$ and $\tilde{U} = -Q\,S + U\,C$, restricted to the well-observed region (pixels within \qty{-5}{\decibel} of the peak hit count). Their mask-deconvolved (MASTER) auto- and cross-spectra give one realization of $C_\ell^{QQ,\text{in}}$, $C_\ell^{UU,\text{in}}$, and $C_\ell^{QU,\text{in}}$.

Repeating over $10^4$ sky realizations yields the mean leakage spectra ($\bar{C}_\ell^{QQ,\text{in}}$, $\bar{C}_\ell^{UU,\text{in}}$, $\bar{C}_\ell^{QU,\text{in}}$) and their bandpower covariance, which are the products carried forward (per patch). Combined with the calibrated $\epsilon^{QV}$ and $\epsilon^{UV}$ through Equation~\eqref{eq:pv_leakage}, they give the $P\!\to\!V$ leakage contribution and its uncertainty in the final $VV$ error budget.

\subsection{Estimation of $C_\ell^{VV}$}
\label{sec:cl_vv}

With the understanding of the instrument systematics and leakages, we can now extract $V_\text{sky} (\hat{n})$. Starting with Equation~\eqref{eq:obs_U_final}, the mapmaking procedure in Section~\ref{sec:mapmaking} is utilized to produce one map per observation $(i)$. This brings the domain from time $(t)$ to map pixel space, $(\hat{n})$. Given that the instrument systematics happen in map space, the effects of the beam, pointing model, and transfer function are represented as operators $\mathcal{B}$, $\mathcal{P}$, and $\mathcal{T}$, respectively.
\begin{align}
    U_\text{obs}^{(i)} (\hat{n}) &= \eta_V \left[ V_\text{in}^{(i)} (\hat{n}) + \epsilon^{QV} Q_\text{in}^{(i)} (\hat{n}) + \epsilon^{UV} U_\text{in}^{(i)} (\hat{n}) \right] \nonumber \\
    &\quad \ + \epsilon_\text{tot}^{IU} I_\text{in}^{(i)} (\hat{n}) \nonumber \\
    &= \mathcal{BPT} \Big\{ \eta_V \Big[ V_\text{sky}^{(i)} (\hat{n}) + \epsilon^{QV} Q_\text{sky}^{(i)} (\hat{n}) \nonumber \\
    &\quad \ + \epsilon^{UV} U_\text{sky}^{(i)} (\hat{n}) \Big] + \epsilon_\text{tot}^{IU} I_\text{sky}^{(i)} (\hat{n}) \Big\}.
    \label{eq:obs_U_map}
\end{align}
Averaging Equation~\eqref{eq:obs_U_map} across all cross-spectra following the methodology of Section~\ref{sec:meas_UU}, the resulting $\ell$-space equation relates the noise-free observed-frame spectrum $C_\ell^{UU,\text{obs}}$, of which $\bar{\hat{C}}_\ell^{UU,\text{obs}}$ of Equation~\eqref{eq:cl_mean} is the measurement, to the sky spectrum,
\begin{multline}
    C_\ell^{UU,\text{obs}} = g^2 g_\tau^2 \left| P_\ell \right|^2 \left| B_\ell \right|^2 \left| T_\ell \right|^2 \\
    \times \Big[ \left| \eta_V \right|^2 \left(C_\ell^{VV,\text{sky}} + \bar{C}_\ell^{P\to V} \right) + C_\ell^{I\to U} \Big],
\label{eq:backward_model}
\end{multline}
with the leakage spectra defined in Section~\ref{sec:leakage}. Solving for the sky $VV$ spectrum,
\begin{multline}
    C_\ell^{VV,\text{sky}} = \frac{1}{\left|\eta_V\right|^2} \Bigg [ \frac{C_\ell^{UU,\text{obs}}}{g^2 g_\tau^2 \left| B_\ell \right|^2 \left| P_\ell \right|^2 \left| T_\ell \right|^2} \\
    - C_\ell^{I\to U} \Bigg] - \bar{C}_\ell^{P\to V} .
    \label{eq:forward_model}
\end{multline}
The absolute gain, $g$, of Equation~\eqref{eq:gain} enters on the measured spectrum alone. The beam, pointing, transfer function, and leakage coefficients are each ratios of maps built from the same calibrated timestreams, so $g$ cancels identically in all of them, while the simulated input spectra $\bar{C}_\ell^{X,\text{in}}$ are generated in true \si{\Kcmb} and carry no instrumental gain. In the joint fit of Section~\ref{sec:mcmc}, $g$ is common to both patches and so cancels exactly in the transfer-function ratio of Equation~\eqref{eq:joint_pred}. It enters the RA4.5 prediction only through the small patch-difference leakage term. Because the limits scale as $g^{2}$, we sample it as a free parameter with the Gaussian prior $g = 1.000 \pm \sigma_g$ of Section~\ref{sec:stim_cal}, giving it its own chain dimension rather than fixing it and marginalizing afterwards. The time-constant factor $g_\tau$ of Section~\ref{sec:stim_cal} enters in the same way, but it is specific to each patch and small, so it is not sampled; it is drawn with the fixed nuisances after sampling (Section~\ref{sec:mcmc}). In the joint fit, $g_\tau$ is drawn independently for each patch, and the two draws are averaged in each bandpower with weights given by each patch's statistical inverse variance on $C_\ell^{VV}$, computed from the diagonal of its $\hat C_\ell^{UU,\mathrm{obs}}$ covariance and the nominal $T_\ell$ and $P_\ell$.

\subsection{Likelihood and Upper Limits}
\label{sec:mcmc}

The bandpowers are compared to the model of Section~\ref{sec:cl_vv} through a Gaussian likelihood. Writing $\mathbf{d}_p = \bar{\hat{C}}_\ell^{UU,\text{obs},p}$ for the measured bandpowers of patch $p$ (Equation~\eqref{eq:cl_mean}), $\boldsymbol{\Sigma}_p$ for their pair-level bootstrap covariance (Section~\ref{sec:meas_UU}), and $\mathbf{m}_p(\boldsymbol{\theta})$ for the predicted noise-free spectrum,
\begin{multline}
    -2\ln\mathcal{L}(\boldsymbol{\theta}) = \sum_{p} \big[\mathbf{d}_p - \mathbf{m}_p(\boldsymbol{\theta})\big]^{\mathsf{T}} \boldsymbol{\Sigma}_p^{-1} \\
    \times \big[\mathbf{d}_p - \mathbf{m}_p(\boldsymbol{\theta})\big] - 2\ln\pi(\boldsymbol{\theta}),
    \label{eq:likelihood}
\end{multline}
where $\pi(\boldsymbol{\theta})$ is the prior. It is Gaussian on the calibrated parameters $\eta_V$, $g$, $a_P$, $B_\ell$, and $T_\ell$ (Table~\ref{table:sample}), truncated by the positivity bounds given below, and uniform on $C_\ell^{UU,\text{obs}}$. The sampled parameters $\boldsymbol{\theta}$ are the three global scalars, the circular polarization efficiency $\eta_V$, the absolute gain $g$, and the pointing amplitude $a_P$ of Equation~\eqref{eq:pl_amp}, together with the per-bandpower latent spectrum $C_\ell^{UU,\text{obs}}$, beam $B_\ell$, and pipeline transfer function $T_\ell$; the model is $\mathbf{m}_{\mathrm{RA23}} = C_\ell^{UU,\text{obs}}$ for the patch in which the signal is parameterized, and Equation~\eqref{eq:joint_pred} for RA4.5. It is essential to distinguish the two Gaussians here: the covariance $\boldsymbol{\Sigma}_p$ is the likelihood, constraining the latent spectrum by the data, whereas $\eta_V$, $g$, $a_P$, $B_\ell$, and $T_\ell$ enter through $\pi(\boldsymbol{\theta})$ as Gaussian priors set by their external calibration measurements (Table~\ref{table:sample}).

The coadded map is the true sky convolved with the distribution of pointing displacements across the coadded observations, so in harmonic space the pointing error multiplies the spectrum by the characteristic function of that distribution. A displacement common to every observation is then a rigid translation, which contributes pure phase and drops out exactly, while binning in $|\ell|$ retains only the trace of the displacement covariance, whose anisotropy is suppressed to ${\sim}10^{-4}$ of the effect at these scales. The pointing uncertainty therefore carries a single degree of freedom, the dispersion $\sigma_\mathrm{coh}$ induced across observations by the uncertainty in the fitted pointing parameters ($12.6''$ for RA23 and $12.1''$ for RA4.5), on the fixed shape $\exp(-\sigma_\mathrm{coh}^2\ell^2/2)$, and we sample its amplitude,
\begin{equation}
    P_\ell(a_P) = \hat{P}_\ell + a_P\,\delta P_\ell, \qquad a_P \sim \mathcal{N}(0,1),
    \label{eq:pl_amp}
\end{equation}
where $\hat{P}_\ell$ is the measured transfer function of Section~\ref{sec:pointing} and $\delta P_\ell = \hat{P}_\ell\big[\exp(-\sigma_\mathrm{coh}^2\ell^2/2) - 1\big]$ is its $1\sigma$ uncertainty template. This single degree of freedom is what distinguishes $P_\ell$ from $T_\ell$, whose filtering is mode-selective rather than a convolution, so that no single scalar controls its curve and each bandpower must be sampled separately. Diagonalizing the measured covariances confirms the distinction---one eigenmode carries $99.8\%$ of the variance of $\mathrm{Cov}(P_\ell)$, against the 15 of 17 modes needed for $\mathrm{Cov}(T_\ell)$. A single $a_P$ is shared by both patches, with per-patch templates, because one pointing-model fit underlies both.  We sample Equation~\eqref{eq:likelihood} with the affine-invariant ensemble sampler \textsc{emcee} \citep{foreman-mackey_emcee_2013}. The remaining, weakly coupled systematics ($g_\tau$, $\epsilon^{UV}$, $\epsilon^{QV}$, $\epsilon_{\ell,\text{tot}}^{IU}$, and $\bar{C}_\ell^{QQ/QU/UU,\text{in}}$) are held at their central values during sampling and marginalized afterwards by drawing nuisance realizations from their priors for every posterior sample and recomputing Equation~\eqref{eq:forward_model}. Hard physical limits $\eta_V, g, B_\ell, T_\ell, P_\ell > 0$ are imposed throughout. We discard the first 30\% of each chain as burn-in and assess convergence with the integrated autocorrelation time $\tau_\mathrm{int}$, estimated separately for every sampled parameter. We require $N_\mathrm{step}/\tau_\mathrm{int,max} \gtrsim 50$, where $\tau_\mathrm{int,max}$ is the largest $\tau_\mathrm{int}$ over all sampled parameters, including the global $\eta_V$ as well as the per-bandpower ones, and both quantities are expressed in stored steps.

Because individual bandpowers may fluctuate negative, we set upper limits with the unified Feldman--Cousins construction \citep{feldman_unified_1998} rather than a positivity-truncated posterior, which collapses asymmetrically against the boundary. From the marginalized posterior we take the per-bin mean $\hat{\mu}_\ell$ and standard deviation $\sigma_\ell$, treat $\hat{\mu}_\ell \sim \mathcal{N}(\mu_\ell,\sigma_\ell^2)$ with the physical bound $\mu_\ell \ge 0$, and obtain the 95\% interval $[\mu_\ell^{\rm lo}, \mu_\ell^{\rm hi}]$ from the likelihood-ratio ordering. We quote $\mu_\ell^{\rm hi}$ as the 95\% upper limit; $\mu_\ell^{\rm lo} > 0$ would flag a bandpower inconsistent with $C_\ell^{VV}=0$ (a two-sided detection band).

We analyze both cases:
\paragraph{Per-patch fits.} RA23 and RA4.5 are analyzed independently, each with a $3+3N_b$-dimensional chain ($\eta_V$, $g$, and $a_P$ plus the per-bandpower $C_\ell^{UU,\text{obs}}$, $B_\ell$, $T_\ell$; $N_b=17$, so $54$ dimensions). We run $512$ walkers for $3\times10^{4}$ stored steps, storing every third sample; both chains are well converged ($N_\mathrm{step}/\tau_\mathrm{int,max} > 100$).
\paragraph{Joint fit.} Assuming the circular-polarization signal is statistically isotropic, a single $C_\ell^{VV}$ is common to both patches, so a joint fit combines their constraining power. We parameterize the signal in the RA23 frame and predict the RA4.5 spectrum from it,
\begin{multline}
  C_\ell^{UU,\text{obs},\mathrm{RA4.5}}
    = \left(\frac{T_\ell^{\mathrm{RA4.5}} P_\ell^\text{RA4.5}(a_P)}{T_\ell^{\mathrm{RA23}} P_\ell^\text{RA23}(a_P)}\right)^{\!2}
      C_\ell^{UU,\text{obs},\mathrm{RA23}} \\
    \ + g^2 B_\ell^2 \left( T_\ell^{\mathrm{RA4.5}} P_\ell^\text{RA4.5}(a_P) \right)^2 \eta_V^2
      \big[(\epsilon^{QV})^2\Delta_\ell^{QQ} \\
    \ + (\epsilon^{UV})^2\Delta_\ell^{UU} + 2\epsilon^{QV}\epsilon^{UV}\Delta_\ell^{QU}\big],
  \label{eq:joint_pred}
\end{multline}
where $\Delta_\ell^{X} = \bar{C}_\ell^{X,\mathrm{RA4.5}}/W_{\ell,\mathrm{RA4.5}}^2 - \bar{C}_\ell^{X,\mathrm{RA23}}/W_{\ell,\mathrm{RA23}}^2$ is the patch-difference of each leakage spectrum, each deconvolved by its own pixel window for the reason given below Equation~\eqref{eq:pv_leakage}; the two differ because the right-ascension rescaling of Section~\ref{sec:master} gives each patch a slightly different effective pixel solid angle. The two patches contribute the two terms of the sum in Equation~\eqref{eq:likelihood}. The pointing transfer functions are evaluated at the shared amplitude of Equation~\eqref{eq:pl_amp}, so the coherent pointing error largely cancels in the ratio. Only the difference in fractional susceptibility, $2\big(\delta P_\ell^{\mathrm{RA4.5}}/\hat{P}_\ell^{\mathrm{RA4.5}} - \delta P_\ell^{\mathrm{RA23}}/\hat{P}_\ell^{\mathrm{RA23}}\big)$, survives. At $\ell=1950$ this moves the ratio by $0.11\%$, against the $0.72\%$ by which $P_\ell$ itself moves, a sevenfold cancellation. This is the expected behavior, as a single pointing model underlies both patches and its common mode should not affect their relative calibration; the per-patch fits retain the full $0.72\%$. Because the RA4.5 prediction depends on the transfer-function ratio, $T_\ell$ is sampled separately for each patch (raising the chain to $3+4N_b = 71$ dimensions) so that the $T_\ell$-ratio uncertainty propagates rather than being held fixed, which would make the RA4.5 prediction artificially sharp. The larger parameter space lengthens the autocorrelation time, so the chain must be run correspondingly longer. We use $256$ walkers for $7\times10^{4}$ stored steps, retaining every sixth sample, which is roughly 4.7 times as many raw steps per walker as the per-patch fits. The heavier thinning leaves $N_\mathrm{step}/\tau_\mathrm{int,max}$ unchanged, since it rescales both quantities equally, but keeps the stored chain within memory; like the individual patch likelihood estimation, this ratio is also ${>}100$. The joint fit yields the tightest limits, as both patches inform the common $C_\ell^{VV}$.

To gauge the relative importance of each component in the $C_\ell^{VV}$ error budget, we propagate the measured uncertainty of each parameter through the inversion of Equation~\eqref{eq:forward_model} one at a time. For a given parameter we draw $10^4$ Monte Carlo realizations from its measurement distribution (Gaussian, or uniform for $\epsilon^{QV}$) while holding all others at their central values, and take the per-bin standard deviation of the resulting $C_\ell^{VV}$ as that parameter's contribution to the uncertainty; varying all parameters simultaneously gives the total $\delta C_\ell^{VV}$, to which the individual terms add in quadrature. Because the standard deviation is sign-symmetric, this budget is unaffected by bandpowers that fluctuate below zero, and as no positivity is imposed it reproduces the unconstrained uncertainty that the Feldman--Cousins construction (Section~\ref{sec:mcmc}) consumes.

\begin{figure*}[ht!]
    \centering
    \includegraphics[width=\linewidth]{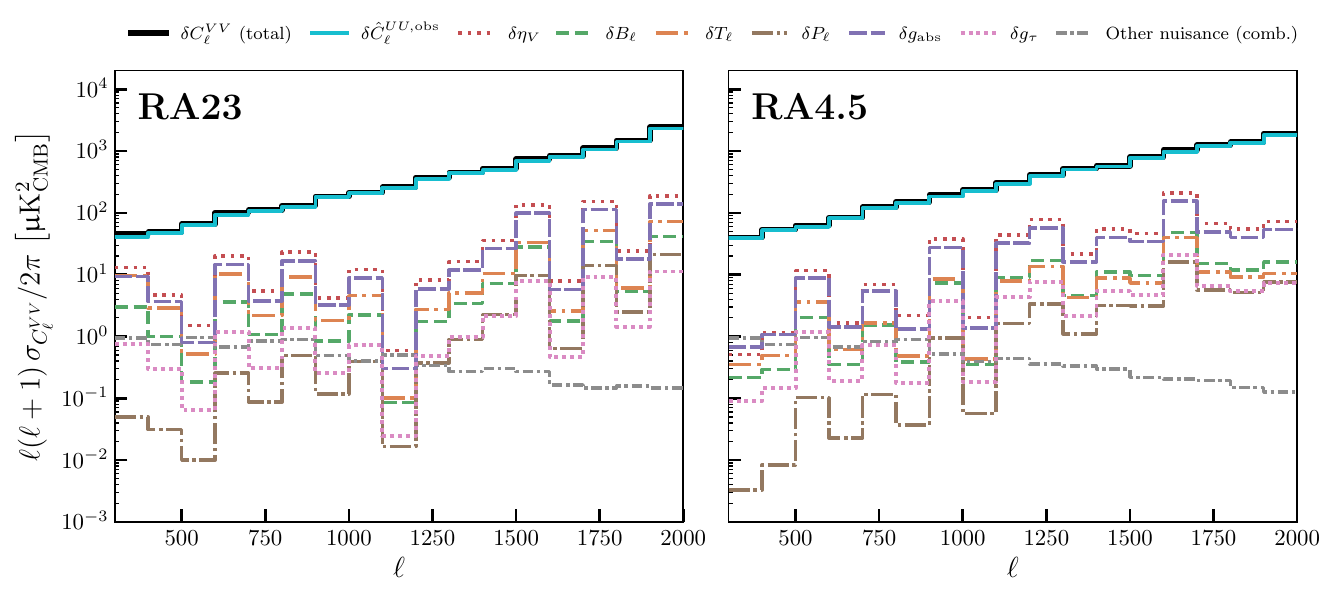}
    \caption{Relative contribution of overall uncertainty in a Monte Carlo realization distribution of $10^4$ draws. This result shows the uncertainty is significantly dominated by the statistical uncertainty of the measured spectra; the next most significant contributions, from the absolute gain $g$ and the circular polarization efficiency $\eta_V$, are comparable, with $\eta_V$ slightly larger, and typically lie one to two orders of magnitude below it. The sampled parameters and their error ranges are detailed in Table~\ref{table:sample}.}
    \label{fig:uncertainty_breakdown}
\end{figure*}

Figure~\ref{fig:uncertainty_breakdown} shows the relative contributions of each parameter to the overall uncertainty, separating the six parameters that carry their own chain dimension, the latent spectrum $C_\ell^{UU,\text{obs}}$, the circular polarization efficiency $\eta_V$, the absolute gain $g$, the beam $B_\ell$, the transfer function $T_\ell$, and the pointing transfer function $P_\ell$, from the quantities propagated after sampling: $g_\tau$, shown separately, and the $I\!\to\!U$ and $P\!\to\!V$ leakages, shown as a single combined curve. The median contribution of $g_\tau$ is 0.3\% (0.7\%) of the statistical term for RA23 (RA4.5), below every sampled calibration, which is why it is propagated rather than sampled. In both patches the statistical term $C_\ell^{UU,\text{obs}}$ traces the total at every multipole, while all calibration, beam, transfer, and leakage contributions lie one to three orders of magnitude below it across $350\le\ell\le1950$. The upper limits are therefore statistics-limited, as no instrumental systematic contributes appreciably to the $C_\ell^{VV}$ error, and the result is robust to their calibration.

\begin{table*}[htbp]
\caption{Inputs of the likelihood estimator, Equation~\eqref{eq:likelihood}, and the model it evaluates, Equation~\eqref{eq:forward_model}. The spectral values are roughly estimated to demonstrate relative significance, and are approximately the same in both patches, except $g_\tau$, quoted as RA23 / RA4.5; no central value is quoted for the measured bandpowers because they scatter about zero, so only their scale is meaningful. Each quantity is classified by the role it plays: the measured bandpowers are the data; sampled parameters carry their own chain dimension, with the stated distribution acting as a prior except for $C_\ell^{UU,\text{obs}}$, which is constrained by the likelihood itself; fixed nuisances are held at their central values during sampling and marginalized afterwards. Uncertainties are $1\sigma$ and carry the units of the adjacent value; that of $B_\ell$ is set per bandpower by the scatter of the Jupiter observations rather than by a single number. The ``Dim.'' column gives the number of chain dimensions each sampled quantity contributes, with $N_b=17$ bandpowers and $T_\ell$ counted per patch, so that the per-patch chains are $3+3N_b=54$ dimensional and the joint chain of Section~\ref{sec:mcmc}, which carries a separate $T_\ell$ for each patch, is $3+4N_b=71$ dimensional. $P_\ell$ enters through the single amplitude of Equation~\eqref{eq:pl_amp} rather than one dimension per bandpower, and the uncertainty quoted for it is that at the highest bandpower. The last column gives the section in which each quantity is determined.}
\label{table:sample}
\begin{tabular*}{\textwidth}{@{\extracolsep{\fill}}lccccc}
\hline\hline
Quantity & Value & Uncertainty & Role & Dim. & Section \\ \hline
$\bar{\hat{C}}_\ell^{UU,\text{obs}}$ & --- & ${\sim}0.001 \ \unit{\micro\kelvin\squared}_\mathrm{CMB}$ & Data & --- & \ref{sec:meas_UU} \\ \hline
$C_\ell^{UU,\text{obs}}$ & --- & --- & Sampled (likelihood) & $N_b$ & --- \\
$\eta_V$ & $0.528$ & $0.032$ & Sampled (prior) & $1$ & \ref{sec:polmod} \\
$g$ & $1.000$ & $0.046$ & Sampled (prior) & $1$ & \ref{sec:stim_cal} \\
$B_\ell$ & ${\sim}0.4$--$0.9$ & per bandpower & Sampled (prior) & $N_b$ & \ref{sec:beam} \\
$T_\ell$ & ${\sim}0.4$--$0.8$ & $0.1$ & Sampled (prior) & $N_b$ & \ref{sec:transfer} \\
$P_\ell$ & ${\sim}0.99$--$1.00$ & $0.007$ & Sampled (prior) & $1$ & \ref{sec:pointing} \\ \hline
$\epsilon_{\ell,\text{tot}}^{IU}$ & ${\sim}1.8\%$ & $0.3\%$ & Fixed nuisance & --- & \ref{sec:IU_leakage} \\
$g_\tau$ & $1.000$ & $0.0037$ / $0.0062$ & Fixed nuisance & --- & \ref{sec:stim_cal} \\
$\epsilon^{QV}$ & ${\lesssim}1.1\%$ & --- & Fixed nuisance & --- & \ref{sec:polmod} \\
$\epsilon^{UV}$ & $-11.7\%$ & $5.1\%$ & Fixed nuisance & --- & \ref{sec:polmod} \\
$\bar{C}_\ell^{QQ,\mathrm{in}}$ & ${\lesssim}0.004 \ \unit{\micro\kelvin\squared}$ & $0.001$ & Fixed nuisance & --- & \ref{sec:PV_leakage} \\
$\bar{C}_\ell^{UU,\mathrm{in}}$ & ${\lesssim}0.004 \ \unit{\micro\kelvin\squared}$ & $0.001$ & Fixed nuisance & --- & \ref{sec:PV_leakage} \\
$\bar{C}_\ell^{QU,\mathrm{in}}$ & ${\lesssim}250 \ \unit{\nano\kelvin\squared}$ & $100$ & Fixed nuisance & --- & \ref{sec:PV_leakage} \\
\hline\hline
\end{tabular*}
\end{table*}

\section{Results}
\label{sec:results}

\begin{figure*}[ht!]
    \centering
    \includegraphics[width=\linewidth]{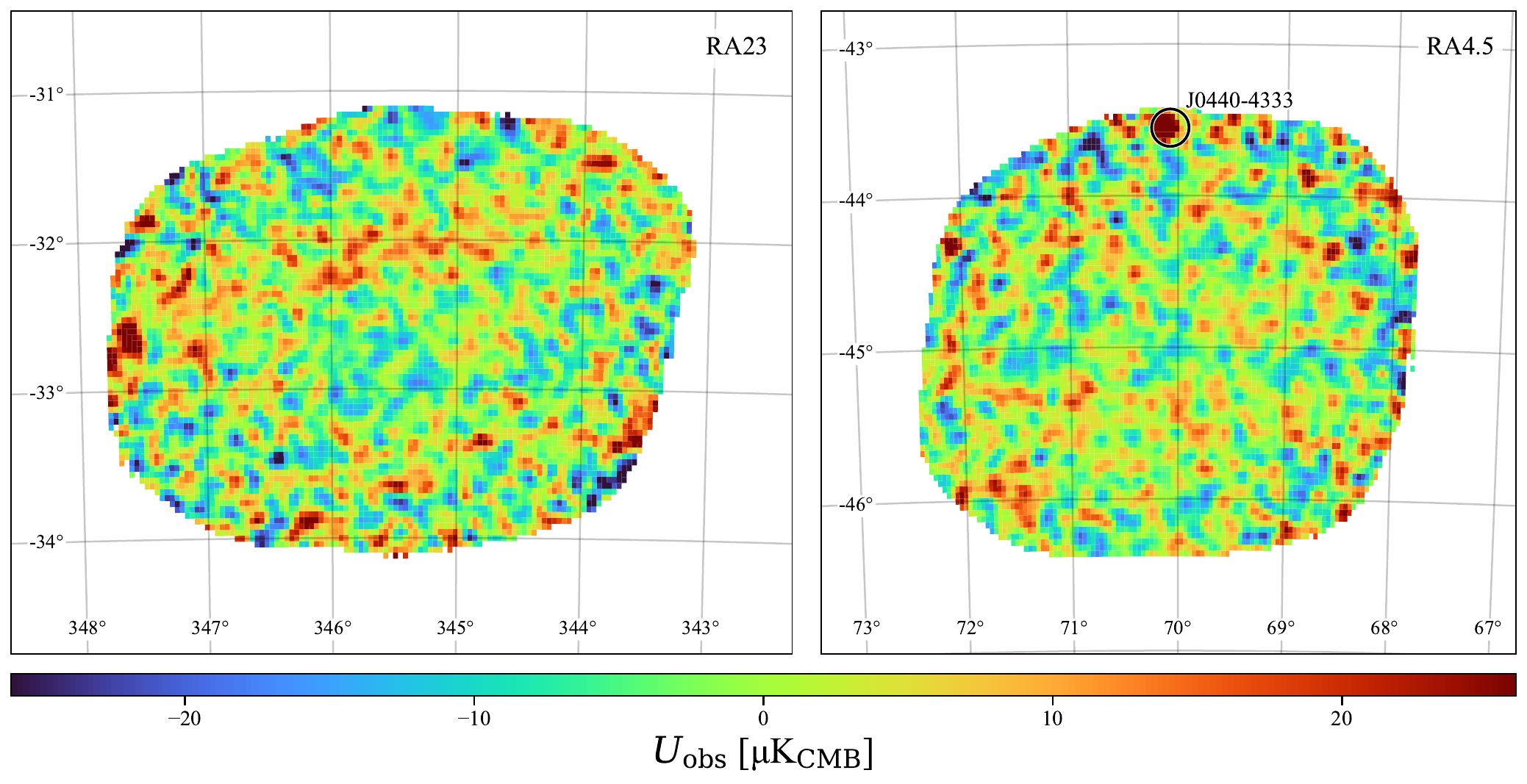}
    \caption{Coadded $U_\text{obs}$ maps of the RA23 (left) and RA4.5 (right) patches, in gnomonic projection with an RA/Dec graticule. $U_\text{obs}$ is the observed Stokes $U$ essentially as it comes out of the detector timestreams. It retains all leakages and has not been deconvolved for the beam, transfer function, or pointing. The maps are coadded over all observations and smoothed with a \qty{5}{\arcminute}-FWHM Gaussian, and both panels share a common color scale. The footprint is set by a \qty{-5}{\decibel} cut on the pixel hit count relative to its maximum, as described in Section~\ref{sec:master}. These maps are noise-dominated. The point source J0440$-$4333, circled in the RA4.5 panel, is discussed in Appendix~\ref{app:point_source_validation}.}

    \label{fig:cmb_maps}
\end{figure*}

\begin{figure*}[ht!]
    \centering
    \includegraphics[width=\linewidth]{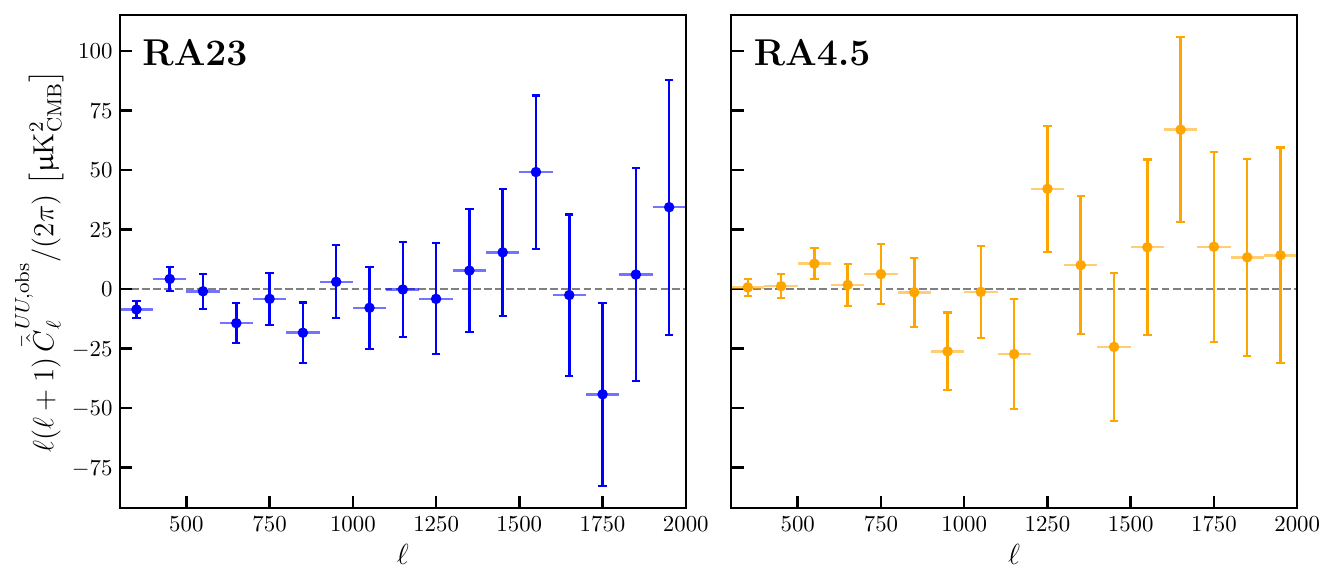}
    \caption{Measured $\ell (\ell+1) \bar{\hat{C}}_\ell^{UU,\text{obs}}/(2\pi)$ for the RA23 (\emph{left}) and RA4.5 (\emph{right}) fields, shown as $\Delta\ell = 100$ bandpowers over $300 \le \ell \le 2000$. Points mark the bandpower values; vertical bars are the $\pm1\sigma$ bandpower uncertainties obtained from the pair-level bootstrap of Section~\ref{sec:meas_UU}, while the horizontal bars denote each bandpower's $\ell$-width and are not uncertainties. The dashed line is the null hypothesis of zero signal.}
    \label{fig:vv_spectra}
\end{figure*}

\begin{figure*}[!t]
    \centering
    \includegraphics[width=\linewidth]{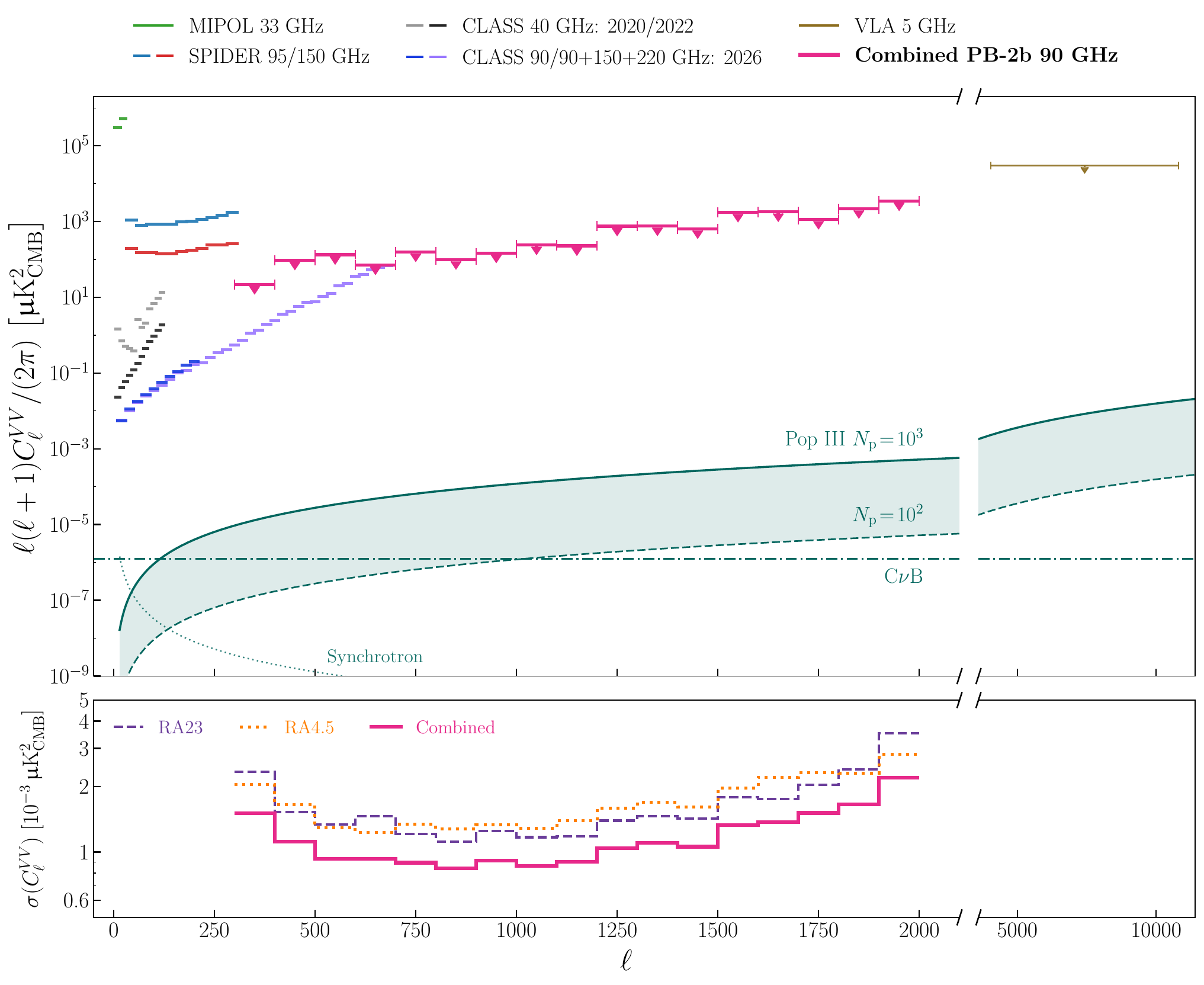}
    \caption{\textbf{Upper panel}: $95\%$ confidence upper limits on the CMB circular-polarization power spectrum, $\ell(\ell+1)C_\ell^{VV}/2\pi$. Downward arrows denote the $95\%$ upper limit in each bin; horizontal bars indicate the multipole width of the bandpower and are not uncertainties. The PB-2b measurements (this work) are shown for combined (bold magenta) upper limit constraints, together with previous limits from MIPOL \citep{mainini_improved_2013}, SPIDER \citep{nagy_new_2017}, CLASS \citep{padilla_two-year_2020, eimer_class_2024, essinger-hileman_improved_2026}, and VLA \citep{partridge_linear_1988} (at much higher $\ell$, note the break in the $\ell$-axis). The paired SPIDER colors denote \qtylist{95;150}{\giga\Hz}, the paired CLASS \qty{40}{\giga\Hz} colors the 2020 and 2022 analyses, and the paired 2026 CLASS colors the \qty{90}{\giga\Hz} band and the \qtylist{90;150;220}{\giga\Hz} co-add, which assumes the circular polarization follows the CMB frequency dependence. Also shown are representative predictions from \citet{king_circular_2016}, evaluated at \qty{90}{\giga\Hz}: Faraday conversion by Pop. III supernova remnants and scattering off the cosmic neutrino background. The simulated intrinsic full-sky circular polarization emission of Galactic synchrotron, a foreground, is shown as well \citep{king_circular_2016}. The two Pop. III curves correspond to $N_p = 100$ and $N_p = 1000$, where $N_p$ is the number of Pop. III stars per halo, a poorly constrained quantity to which $C_\ell^{VV}$ is quadratically sensitive. \textbf{Lower panel}: the corresponding $1\sigma$ bandpower sensitivities of the RA23, RA4.5, and combined PB-2b datasets, i.e. the standard deviation of the deconvolved $C_\ell^{VV}$ posterior, shown as $C_\ell$ rather than $\ell(\ell+1)C_\ell/2\pi$ so that the comparison between datasets is not dominated by the steep $\ell(\ell+1)$ rise.}
    \label{fig:vv_upper}
\end{figure*}

We find no evidence for circular polarization in either field. The coadded $U_\text{obs}$ maps entering the analysis are shown in Figure~\ref{fig:cmb_maps}; both are consistent with noise, with no visible structure apart from the point source J0440$-$4333 in RA4.5. Figure~\ref{fig:vv_spectra} shows the measured detector $UU$ power spectra, $\ell(\ell+1)\bar{\hat{C}}_\ell^{UU, \text{obs}}/2\pi$, for the RA23 and RA4.5 patches across $300 \lesssim \ell \lesssim 2000$, where the error bars are the bandpower uncertainties of Section~\ref{sec:meas_UU} and include none of the aforementioned leakages or systematics. As shown by Figure~\ref{fig:uncertainty_breakdown}, the uncertainty is dominated by statistical uncertainty of the measured $U$, which we know from \citet{sakaguri_development_2026} is primarily sky $V$. From the measured spectra, $\tilde{\chi}^2 = 1.14$ (PTE $=0.304$) for RA23 and $1.00$ (PTE $=0.450$) for RA4.5, consistent with the null hypothesis. We therefore place $95\%$ confidence upper limits on $C_\ell^{VV}$, constructed bin-by-bin with the Feldman--Cousins unified prescription \citep{feldman_unified_1998}, which respects the physical boundary $C_\ell^{VV}\ge 0$. The resulting limits on $\ell(\ell+1)C_\ell^{VV}/2\pi$ are given in Table~\ref{tab:vv_upper} and Figure~\ref{fig:vv_upper}; the combined two-field limit ranges from \qty{21.8}{\micro\Kcmbsq} at the bin centered at $\ell=350$ to \qty{3.49e3}{\micro\Kcmbsq} at the bin centered at $\ell=1950$. Figure~\ref{fig:vv_upper} also shows representative predicted signals from Pop. III supernova remnants, the C$\nu$B, and Galactic synchrotron. These lie roughly six orders of magnitude in $\ell(\ell+1)C_\ell^{VV}/2\pi$ below the limits reported here, equivalently a factor of ${\sim}10^{3}$ in circular polarization amplitude, so detecting any of them remains far out of reach at this sensitivity. Furthermore, the contribution from Galactic synchrotron as represented in Figure~\ref{fig:vv_upper}, is estimated in \citet{king_circular_2016} from full-sky simulations utilizing \textsc{Hammurabi} \citep{waelkens_hammurabi_2009}. Unlike the Pop. III and C$\nu$B sources, Galactic synchrotron is not isotropic and homogeneous, and given that RA23 and RA4.5 are specifically chosen to lie away from the Galactic plane, this Galactic synchrotron signal is a maximum estimate. In practice, the emission signal would likely be much smaller relative to the other two sources within RA23 and RA4.5, therefore the actual on-sky $V$ signal within this $\ell$-range and at an observing frequency of \qty{90}{\giga\Hz} can justifiably be claimed to consist primarily of CMB photons. Therefore, it is reasonable to claim that the upper limit constraints placed by this study are primarily on the CMB circular polarization signal with minimal contamination from foregrounds.

\begin{table}[htbp]
\caption{PB-2b $95\%$ C.L. upper limits on $\ell(\ell+1)C_\ell^{VV}/2\pi$ (\unit{\micro\Kcmbsq}), in $\Delta\ell=100$ bandpowers. RA23 and RA4.5 are the individual fields; ``Combined'' is the joint two-field limit.}
\label{tab:vv_upper}
\begin{tabular*}{\columnwidth}{@{\extracolsep{\fill}}cccc}
\hline\hline
$\ell_b$ & RA23 & RA4.5 & Combined \\ \hline
350  &  24.3 &  82.3 &  21.8 \\
450  &   134 &   113 &  95.7 \\
550  &   115 &   215 &   133 \\
650  &  72.8 &   176 &  71.5 \\
750  &   172 &   290 &   157 \\
850  &   108 &   271 &  98.7 \\
950  &   384 &   150 &   147 \\
1050 &   311 &   426 &   239 \\
1150 &   482 &   292 &   228 \\
1250 &   615 &  1380 &   756 \\
1350 &   952 &  1130 &   763 \\
1450 &  1200 &   688 &   642 \\
1550 &  2340 &  1820 &  1750 \\
1650 &  1430 &  3440 &  1800 \\
1750 &  1020 &  2710 &  1140 \\
1850 &  2730 &  2850 &  2180 \\
1950 &  5520 &  3850 &  3490 \\
\hline\hline
\end{tabular*}
\end{table}

These constraints extend the measured $V$-mode power spectrum to $\ell = 2000$, or angular scales of \qtyrange{5.4}{36}{\arcminute}, roughly three times the highest multipole previously reached. Prior bandpower limits probe larger scales and lower multipoles: MIPOL at $\ell\lesssim30$ \citep{mainini_improved_2013}, SPIDER over $33<\ell<307$ at \qtylist{95;150}{\giga\Hz} \citep{nagy_new_2017}, and CLASS at \qty{40}{\giga\Hz} over $\ell\lesssim125$ \citep{padilla_two-year_2020, eimer_class_2024} and subsequently to $\ell\lesssim700$ at \qtylist{90;150;220}{\giga\Hz} \citep{essinger-hileman_improved_2026}. The Very Large Array observations of \citet{partridge_linear_1988} reach smaller angular scales still, but bound the fractional circular polarization within individual fields rather than the angular power spectrum. Our limits are weaker at the maximum $\ell$ in absolute \unit{\micro\Kcmbsq} than the large-scale SPIDER and CLASS constraints; $\ell (\ell+1) C_\ell^{VV}/(2\pi)$ rises steeply toward high $\ell$, and the instrument noise grows after beam and transfer-function deconvolution, but they are complementary, covering multipoles above $\ell \approx 700$ that have not previously been accessible to a $V$-mode power spectrum measurement.

The bin-to-bin scatter of the limits in Figure~\ref{fig:vv_upper} is an intrinsic feature of noise-dominated measured spectra of Figure~\ref{fig:vv_spectra} rather than a numerical artifact. The posterior chains are well converged (Gelman--Rubin $\hat{R}<1.001$ in every bin, with a Monte Carlo uncertainty on each limit of ${\lesssim}0.4\%$), so each bandpower's limit is fixed by where that bin's measured power fluctuates relative to zero. The limits vary by ${\approx}0.2\,$dex (a factor ${\sim}1.6$) from bin to bin about a smooth trend, comparable to the ${\approx}0.23\,$dex scatter of the CLASS \qty{40}{\giga\Hz} limits \citep{padilla_two-year_2020} and in contrast to the much smoother SPIDER limits (${\approx}0.03\,$dex). This difference reflects noise structure, not sampling methodology. SPIDER utilizes $V \to P$ leakage in its HWP to measure circular polarization, and so uncertainty is dominated by a HWP calibration coupling that is strongly correlated across $\ell$ \citep{nagy_new_2017}. This suppresses bin-to-bin scatter, whereas the PB-2b and CLASS limits are dominated by per-bin statistical noise.

The underlying bandpower sensitivities, shown in the lower panel of Figure~\ref{fig:vv_upper}, are better behaved than the limits themselves. The combined dataset is the most sensitive in every bandpower, with each individual field exceeding it by an average factor of ${\sim}1.4$, consistent with the $\sqrt{2}$ expected from combining two fields of comparable depth. The upper limits do not follow this ordering bin by bin, however, because a Feldman--Cousins limit depends on the central value as well as on its uncertainty.

\section{Conclusions}
\label{sec:discussion}

We present results on the circular polarization of the sky at \qty{90}{\giga\Hz} from a series of targeted observations with PB-2b for two specific patches over $300 \leq \ell \leq 2000$, at \unit{\micro\Kcmb} sensitivity. As established in Section~\ref{sec:intro}, these limits constrain the total diffuse circular polarization of those fields, which includes the circularly polarized CMB signal and trace foregrounds not removed by location choice, masking, and filtering. Both contributions lie far below the instrument noise that dominates the measured bandpowers, so the limits reported here are set by that noise floor rather than by any detected sky signal. This extends the measured $V$-mode power spectrum into the few-arcminute regime, reaching multipoles roughly three times higher than previous measurements, and improves on the prior upper limit placed by SPIDER \citep{nagy_new_2017} at \qty{95}{\giga\Hz} by nearly two orders of magnitude in the single bandpower where those two analyses overlap. The CLASS constraints of \citet{essinger-hileman_improved_2026} overlap this analysis over $300 \lesssim \ell \lesssim 700$; above that range these are the only $V$-mode bandpower limits reported to date.

The current limitations of CMB circular polarization studies are that no current experiment utilizing a QWP has the sensitivity necessary for a statistically significant detection \citep{king_circular_2016}. Figure~\ref{fig:vv_upper} conclusively demonstrates this, as the limits reported here sit roughly six orders of magnitude in $\ell(\ell+1)C_\ell^{VV}/2\pi$ above the predicted signals. In addition, because the measurement is noise-dominated, with the bandpower uncertainty scaling inversely with integration time, closing that gap with PB-2b would demand of order $10^{6}$ times the observing time presented here, or an equivalent growth in detector count. Focusing study on lower observing frequency, however, could reduce the discrepancy between noise sensitivity and signal magnitude without requiring an unprecedented improvement in instrument sensitivity or detector count. The Faraday conversion contribution scales as $\nu^{-3}$, so the factor of ${\sim}7\times10^{2}$ in amplitude gained between \qty{90}{\giga\Hz} and \qty{10}{\giga\Hz} accounts for nearly the entire gap. \citet{king_circular_2016} correspondingly forecast a signal-to-noise ratio above unity for a \qty{10}{\meter} telescope observing for 20 months at \qty{10}{\giga\Hz}, provided $N_p \gtrsim 100$. In addition, the single observing band utilized in this experiment does not permit differentiation between potential sources, meaning more exotic sources of circular polarization cannot be rigorously identified among known contaminants, such as Milky Way synchrotron radiation.

CMB circular polarization provides an interesting window into early-Universe physics not encapsulated by standard linear polarization measurements probing $\Lambda$CDM. While very few other experiments with comparable sensitivity to the Simons Array are measuring microwave-band circular polarization directly, similar procedures to those of \citet{nagy_new_2017} might be able to constrain it via leakage from linear polarization. For example, the Simons Observatory (SO) Small Aperture Telescopes (SATs) utilize rotating cold HWPs with nearly identical design to that of PB-2b, which can be used to characterize $V \to P$ leakage to even higher sensitivity than that of SPIDER's limits \citep{nagy_new_2017}. Because such a measurement would rely on the small $V \to P$ leakage of the HWP rather than on the direct conversion afforded by a QWP, the effective circular polarization efficiency of the SATs is likely to fall well below the $\eta_V = 0.528 \pm 0.032$ achieved here. That penalty may be offset by raw sensitivity, as each mid-frequency SAT carries 12{,}292 detectors and the SAT survey is projected to reach a combined map depth of \qty{2}{\micro\kelvin}-arcmin at \qtylist{93;145}{\giga\Hz} over five years \citep{galitzki_simons_2024}, orders of magnitude higher than what was achievable with PB-2b. In addition, the SATs will observe at six distinct frequency bands (\qtylist{30;40;90;150;220;280}{\giga\Hz}), which may allow discrimination between the contributions of circular polarization sources as a function of frequency, most importantly foregrounds. This allows simultaneous probing of multiple circular polarization sources and foregrounds, which differ in their dependence on observing frequency. These prospects would allow for research in this field to persist beyond the end of PB-2b's active observation.

\begin{acknowledgments}
This work is supported by the Gordon and Betty Moore Foundation Grant No. GBMF7939, the Simons Foundation Grant No. 034079, the National Science Foundation grants AST-1207892 and AST-1440338, the World Premier International Research Center Initiative (WPI) of MEXT, Japan, and the John Templeton Foundation grant 58724. In Japan, this work was supported by JSPS KAKENHI grant Nos. 18H05539 and 19H00674, and the JSPS Core-to-Core Program JPJSCCA20200003. Work at LBNL is supported in part by the U.S. Department of Energy, Office of Science, Office of High Energy Physics, under Contract No. DE-AC02-05CH11231. Calculations were performed on the Central Computing System, owned and operated by the Computing Research Center at KEK, and the National Energy Research Scientific Computing Center, which is supported by the Department of Energy under Contract No. DE-AC02-05CH11231. Simons Array operated at the James Ax Observatory in the Parque Astronomico Atacama in Northern Chile under the stewardship of the Agencia Nacional de Investigaci\'on y Desarrollo (ANID). We acknowledge support from the ASI-COSMOS Network \verb|(cosmosnet.it)| and from the INDARK INFN Initiative \verb|(web.infn.it/CSN4/IS/Linea5/InDark)|. Support from the Ax Center for Experimental Cosmology at UC San Diego is gratefully acknowledged. M. Hazumi is partially supported by the Yushan Fellow Program by the Ministry of Education (MOE), Taiwan. YC acknowledges support from JSPS KAKENHI Grant Number JP24K00667.
\end{acknowledgments}

\appendix

\section{Null Test Criteria}
\label{app:null_tests}

We adopt a similar strategy for null splits as in \textcite{polarbear_collaboration_exploration_2024}, both for detector-based and observation-based splits, however, we adopt several additional criteria specific to the performance of PB-2b. As with \textsc{Polarbear}, some of these splits are not necessarily evenly split into halves. For splits which are repeated from the detector-based splits, we average over all detectors within a given observation, then split on the median of all observation averages in the observing season, which probes changes in instrument performance over time. The full set of criteria is listed in Table~\ref{tab:null_criteria}.

\begin{table*}[htbp]
\caption{Null test splits used to assess the internal consistency of the measured spectra, following the strategy of \textcite{polarbear_collaboration_exploration_2024} with additional criteria specific to PB-2b.}
\label{tab:null_criteria}
\begin{tabular*}{\textwidth}{@{\extracolsep{\fill}}ll}
\hline\hline
Split & Motivation \\ \hline
\multicolumn{2}{l}{\textbf{Detector-based splits}} \\
\splitcell{A/B pixel handedness} & \motivcell{The sinuous antennae come in two symmetric handednesses; probes the frequency dependence of the polarization angle, which is relevant here through linear-to-circular leakage.} \\
\splitcell{$Q$/$U$ type pixel} & \motivcell{Detectors are aligned horizontally/vertically or rotated by \qty{45}{\degree}. Especially crucial given that the QWP modulates $V$ into the $U$ parameter the detectors measure.} \\
\splitcell{Top/bottom detector} & \motivcell{The two detectors within each antenna, oriented along the two polarization vectors of the pixel.} \\
\splitcell{Focal plane position} & \motivcell{Left/right and upper/lower physical location, probing location-dependent optical systematics not captured by the pointing model.} \\
\splitcell{Gain calibration factor} & \motivcell{Per-observation gain factor, without distinguishing stimulator from HWPSS calibration.} \\
\splitcell{Effective time constant} & \motivcell{Per-observation detector speed, again without distinguishing the calibration method.} \\
\splitcell{Calibration source} & \motivcell{Whether a detector draws its calibration mostly from the stimulator or from the HWPSS.} \\
\splitcell{$4f_\text{HWP}$ amplitude} & \motivcell{Dominated by contamination from the QWP; validates optical and polarization systematics as well as the time-domain HWPSS removal.} \\
\splitcell{$2f_\text{HWP}$ amplitude} & \motivcell{Several orders of magnitude weaker than the $4f_\text{HWP}$ amplitude, but associated with $I \rightarrow P$ leakage, which is small but not negligible.} \\
\splitcell{White noise} & \motivcell{Time-domain white noise level, probing noise modeling and data selection.} \\
\splitcell{Input glitch ACF} & \motivcell{Glitch autocorrelation prior to cleaning.} \\
\hline
\multicolumn{2}{l}{\textbf{Observation-based splits}} \\
\splitcell{Median detector quantities} & \motivcell{Observation-level analogues of the detector splits above: gain calibration factor, $4f_\text{HWP}$ amplitude and phase, $2f_\text{HWP}$ amplitude and phase, white noise, and effective time constant.} \\
\splitcell{Calibration source} & \motivcell{Which method calibrates the majority of detectors, since stimulator calibration is wholly or partly missing for some observations.} \\
\splitcell{``Glitchy'' runs} & \motivcell{Median fraction of detectors with glitches exceeding an autocorrelation threshold.} \\
\splitcell{1st/2nd half of season} & \motivcell{A time-based split probing overall drift in instrument performance across the observing season.} \\
\splitcell{Median PWV} & \motivcell{Loading and general weather systematics.} \\
\splitcell{PWV standard deviation} & \motivcell{Variability of PWV within a single observation, probing rapid changes in loading.} \\
\splitcell{Sun/Moon distance} & \motivcell{Angular distance to the Sun or Moon, probing far-sidelobe contamination.} \\
\splitcell{Elevation} & \motivcell{Median scan elevation, i.e. the atmospheric column along the line of sight.} \\
\splitcell{Bolometer tuning strategy} & \motivcell{Tuning to a fixed fractional resistance of $0.75$ versus an algorithm setting a different fractional resistance per detector.} \\
\splitcell{Ambient air temperature} & \motivcell{Components held at air temperature, notably the QWP, may have emission or transmission properties correlated with temperature.} \\
\splitcell{Focal plane temperature} & \motivcell{Sets the thermal noise level, which could bias the gain and leave residual sky signal in the noise maps.} \\
\splitcell{Pointing correction} & \motivcell{Magnitude of the geometric correction required by the pointing model, which depends on the patch azimuth and elevation when observed.} \\
\hline\hline
\end{tabular*}
\end{table*}

\section{Atmospheric Polarization}
\label{app:circpol}
\begin{figure}[htbp]
    \centering
    \includegraphics[width=\linewidth]{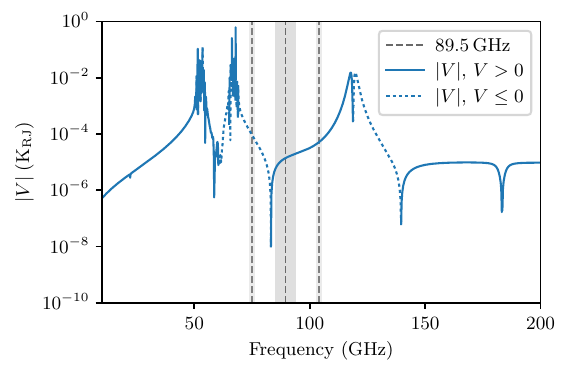}
    \caption{Frequency dependence of the absolute value of the atmospheric circular polarization evaluated at zenith. The dark dashed vertical line marks the measured band center at \qty{89.5}{\giga\Hz}, and the surrounding dark shaded region indicates its uncertainty. The lighter dashed vertical lines mark the nominal lower and upper band edges, while the adjacent light shaded regions indicate the uncertainty associated with the measured bandwidth. The model is made using the software suite of \citet{Petroff_2020}, adapted to the PB-2b observation date and frequency.}
    \label{fig:frequency_dependence}
\end{figure}

Atmospheric circular polarization emission is the result of Zeeman splitting of oxygen due to Earth's magnetic field \citep{Lenoir_atmospheric_V, Keating_atmosphere_V, Hanany_2003, spinelli_2011, Petroff_2020, Fujino_2025} and presents a major foreground for CMB circular polarization measurements. Through the Zeeman effect, Earth's magnetic field splits the rotational energy levels of atmospheric oxygen, allowing transitions that emit polarized radiation. The imbalance between the right- and left-handed circularly polarized components produces a net circular polarization signal near the oxygen resonance lines. As such, CMB experiments observe at frequencies away from these resonance lines, which are well characterized (see Figure~\ref{fig:frequency_dependence}). However, even away from the resonances, atmospheric emission is still a dominant foreground and therefore must be accounted for.

\begin{figure}[htbp]
    \centering
    \includegraphics[width=\linewidth]{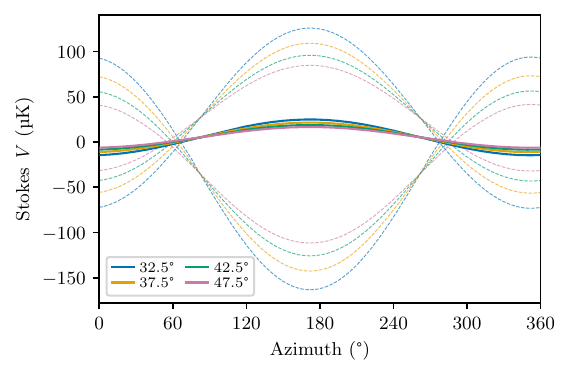}
    \caption{Predicted azimuthal dependence of atmospheric circular polarization. Solid lines show the nominal fit, while dashed lines indicate the 16th and 84th percentile bounds of the Monte Carlo predictions. It can be seen that the amplitude decreases with increasing elevation. Large uncertainties can be observed, reflecting the strong sensitivity of the predicted signal to variations in the detector band center. The model is as in Figure~\ref{fig:frequency_dependence}, additionally accounting for the measured bandwidth.}
    \label{fig:az_dependence}
\end{figure}

Due to its coupling with Earth's magnetic poles, the atmospheric circular polarization obtains a dipole-like structure which depends on the projection of the telescope's line-of-sight onto the local magnetic field vector. This produces a characteristic azimuthal dependence, with the signal largest when the line of sight is most aligned with the magnetic-field direction and smallest when it is perpendicular (see Figure~\ref{fig:az_dependence}). Further, the atmospheric signal amplitude decreases with increasing elevation due to the reduced atmospheric column density along the telescope's line of sight. At a fixed telescope pointing, this signal remains constant in the sky, but in the detector timestream it will get modulated to $4f_\text{HWP}$ such that the corresponding raw detector signal is not constant. Demodulation then removes the HWP-angle dependence, and recovers the constant atmospheric Stokes \textit{V} signal. 

We are able to mitigate any major contribution from the atmospheric circular polarization by a combination of polynomial and scan-synchronous signal filtering. We apply a polynomial filter both on the full timestream as well as individual subscans, as described in Section~\ref{sec:filter1}, and reapply it after demodulation (Section~\ref{sec:filter2}), where the atmospheric signal is no longer modulated. Over the limited azimuth range of an individual subscan, the atmospheric circular polarization signal is expected to vary slowly and can be approximated by a low-order expansion in azimuth. Therefore, the constant and first-order terms of the subscan polynomial filter remove much of this atmospheric contribution and any other slow varying offsets. The scan-synchronous signal filter provides an additional suppression of components that are stable in telescope azimuth. Since atmospheric circular polarization is tied to the telescope line of sight relative to Earth's magnetic field, rather than to fixed celestial coordinates, it is expected to appear primarily as an azimuth- and elevation-dependent signal and is therefore similarly able to be removed by the constant and low order terms of the scan-synchronous filter.

To describe the atmospheric circular polarization at the PB-2b site, we evaluate the empirical model of \citet{Petroff_2020} at \qty{90}{\giga\Hz}, taking both azimuthal and zenith angle dependence into account. This theoretical model has previously been confirmed at \qty{40}{\giga\Hz} by the CLASS experiment \citep{Petroff_2020} and at \qty{150}{\giga\Hz} by the \textsc{Polarbear} project \citep{Fujino_2025}. We use a band center of \qty{89.5 \pm 4.5}{\giga\Hz} and a fractional bandwidth of $0.324 \pm 0.032$, as defined in pre-deployment detector measurements \citep{Westbrook_2018}. To determine the uncertainty in the theoretical atmospheric emission, we draw Monte Carlo realizations of the band center and fractional bandwidth from independent Gaussian distributions. We then calculate the band-averaged emission for each realization and take the 16th--84th percentile range as the bandpass-induced uncertainty. The resulting uncertainty is much larger than the nominal signal, mainly due to the strong dependence of the atmospheric signal on observing frequency. Across the frequency ranges sampled by the Monte Carlo bandpasses, the atmospheric spectrum changes sign, therefore broadening the distribution of predicted band-averaged signals. Because the zenith angle dependence is an empirical approximation, we only evaluate the model within its calibrated range from 30 to 60 degrees. Further, as the Earth's magnetic field is not perfectly aligned with true north, we also must determine the magnetic declination. We calculate the declination using the International Geomagnetic Reference Field model \citep{Alken_IGRF13}, evaluated at a site elevation of \qty{5200}{\meter}. Since the main geomagnetic field varies slowly in time, we evaluate the field at a representative midpoint date of the observing period and approximate it as constant over the data set. The resulting atmospheric circular-polarization curve is therefore slightly offset from true north. As expected, the atmospheric circular polarization varies slowly in azimuth and as such is mitigated by the post-demodulation filtering of Section~\ref{sec:filter2}. Due to the limited azimuthal range of our scans, we do not currently detect any atmospheric emission.

\section{Gain Validation with J0440$-$4333}
\label{app:point_source_validation}

\begin{figure}[htbp]
    \centering
    \includegraphics[width=\linewidth]{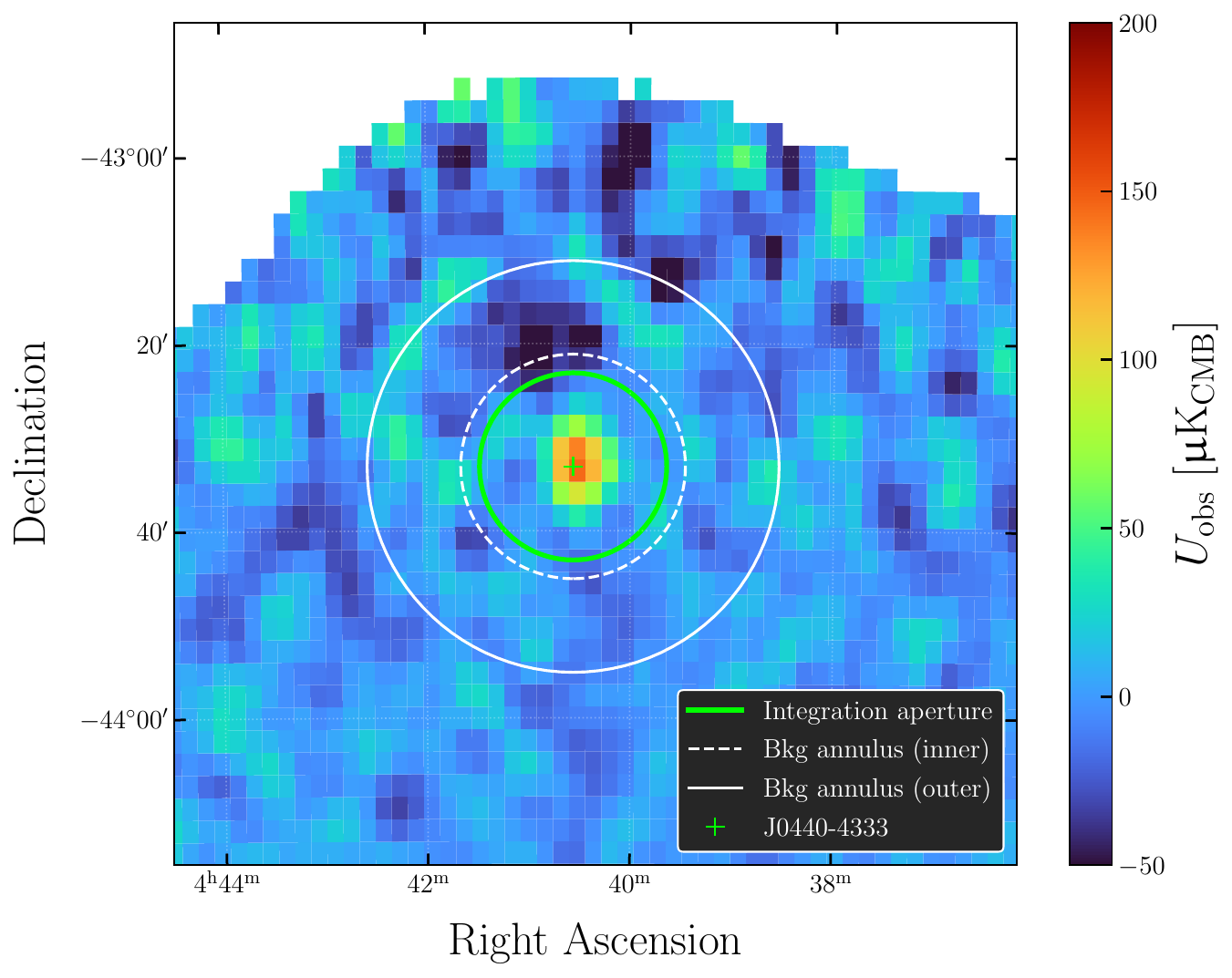}
    \caption{Zoomed-in section of the RA4.5 patch showing J0440$-$4333 with a \qty{-8}{\decibel} hit mask. The green circle is the $10'$ aperture and the region between the solid and dashed white circles is the background annulus; together they define the aperture estimator used as a cross-check on the matched filter of Equation~\eqref{eq:matched_filter}.}
    \label{fig:pointsource}
\end{figure}

The RA4.5 patch contains the bright, compact radio source J0440$-$4333 (redshift $z=2.852$, a Q-type blazar) \citep{shen_5-ghz_1998}, which provides a unique opportunity to validate our calibration and mapmaking procedures. We measure its flux in our map (Figure~\ref{fig:pointsource}) and compare it to a model prediction based on its cataloged intensity and expected polarization behavior.

The pipeline's response to a compact source is not simply the beam. The filtering of Section~\ref{sec:tod} suppresses the source core and surrounds it with a shallow negative trench from ${\sim}11'$ to ${\sim}32'$, so a $10'$ aperture with a $12'$--$22'$ background annulus measures the suppressed core against the source's own trench. Because that trench is coherent with the source rather than noise, it is properly treated as signal, and we estimate the flux with a matched filter built on the measured response,
\begin{equation}
    \hat{F}(\mathbf{x}) = \frac{\mathrm{IFFT}\!\left[\, t^{*}(\boldsymbol{\ell})\, d(\boldsymbol{\ell}) / P(\boldsymbol{\ell}) \,\right]\!(\mathbf{x})}
    {\sum_{\boldsymbol{\ell}} \left| t(\boldsymbol{\ell}) \right|^{2} / P(\boldsymbol{\ell})},
    \label{eq:matched_filter}
\end{equation}
where $\hat{F}(\mathbf{x})$ is the estimated source flux at map position $\mathbf{x}$, $\boldsymbol{\ell}$ is the two-dimensional Fourier wavevector, IFFT denotes the inverse fast Fourier transform, and the asterisk denotes complex conjugation. Here $d$ is the measured $U$ map (which contains the converted Stokes $V$ signal), $t$ is the point-source template, built from the same observations as the map, and $P$ is the map's own two-dimensional power spectrum, kept anisotropic because the residual noise is striped along the scan direction. Calibrating the response on a synthetic unit-flux source makes the estimator unbiased independently of Fourier conventions; it round-trips on simulations to $1.0098$. Evaluating $\hat{F}$ at the J2000 position of J0440$-$4333 gives the measured integrated flux $F_U^{\mathrm{meas}}$, and evaluating it at $N_{\mathrm{off}} = 5354$ off-source positions $\mathbf{x}_k$ in the same filtered map gives the reference set $F_k^{\mathrm{rand}} \equiv \hat{F}(\mathbf{x}_k)$, from which the uncertainty follows as $\sigma_F^{\mathrm{meas}} = 1.4826 \cdot \mathrm{MAD}\{F_k^{\mathrm{rand}}\}$, where MAD denotes the median absolute deviation.

Converting the filtered amplitude to a flux density requires the point-source flux transfer $\mathcal{F}_{\mathrm{PS}}$, the ratio of recovered to injected integrated flux. This differs from the transfer function $T_\ell$ in that it also carries the fraction of the source flux that the estimator collects, and the two effects cannot be separated within a single scalar. Analytically, a mask that is unity inside an aperture of radius $R_{\mathrm{ap}}$ and solid angle $\Omega_{\mathrm{ap}}$, and $-\Omega_{\mathrm{ap}}/\Omega_{\mathrm{ann}}$ within a background annulus $R_{\mathrm{in}} < r < R_{\mathrm{out}}$ of solid angle $\Omega_{\mathrm{ann}}$, has transform $\Omega_{\mathrm{ap}}K(\ell)$, giving
\begin{subequations}
\begin{align}
    \mathcal{F}_{\mathrm{PS}} &= \frac{\Omega_{\mathrm{ap}}}{2\pi} \int \mathrm{d}\ell \; \ell\, B_\ell\, T_\ell\, K(\ell), \\
    K(\ell) &= W(\ell R_{\mathrm{ap}}) - \frac{R_{\mathrm{out}}^{2} W(\ell R_{\mathrm{out}}) - R_{\mathrm{in}}^{2} W(\ell R_{\mathrm{in}})}{R_{\mathrm{out}}^{2} - R_{\mathrm{in}}^{2}},
\end{align}
\label{eq:aper_kernel}
\end{subequations}
where $J_1$ is the first-order Bessel function of the first kind and $W(u) = 2 J_1(u)/u$. Since $K(0)=0$, the annulus subtraction makes this a differential estimator, so the recovered fraction falls below unity even before filtering; the aperture efficiency matters here because the measured \qty{90}{\giga\Hz} beam has broad non-Gaussian wings that leave only \qty{89.5}{\percent} of the source flux within $10'$. We evaluate Equation~\eqref{eq:aper_kernel} over $\ell = 350$--$1950$ in $\Delta\ell = 100$ bins, holding $T_\ell$ constant outside that range, which gives $\mathcal{F}_{\mathrm{PS}} = 0.662 \pm 0.019$. The kernel places only ${\sim}\qty{1}{\percent}$ of its weight above $\ell = 2000$, since it first vanishes near $\ell \approx 1300$, so although $T_\ell$ is uncharacterized there, varying it over its full physical range shifts $\mathcal{F}_{\mathrm{PS}}$ by only $\pm 0.006$.

We nevertheless quote the value obtained end-to-end by injection, which assumes no model of the beam, filtering, or estimator. A source of exactly known integrated flux, carrying the measured beam profile, is added to the Stokes $V$ channel of an input sky map, propagated through the unmodified pipeline of Sections~\ref{sec:tod}--\ref{sec:mapmaking}, and photometered with the same matched filter, so $\mathcal{F}_{\mathrm{PS}}$ is the recovered-to-injected ratio by construction; differencing against a baseline over identical observations and CMB realization cancels the sky exactly. Over the RA4.5 observations, $\mathcal{F}_{\mathrm{PS}} = 0.6149 \pm 0.0016$, \qty{7}{\percent} below the analytic value, a shortfall expected in sign because Equation~\eqref{eq:aper_kernel} treats the map-making operator as diagonal in Fourier space. Because detector weights derive from each timestream's own noise and the simulations are far quieter per sample than the data, the injection must stay in the linear regime, which we verify through the ratio of total map weight to baseline, $0.9912$ here. The measured flux density is then $S_U^{\mathrm{meas}} = F_U^{\mathrm{meas}}/\mathcal{F}_{\mathrm{PS}}$, converted to Jansky.

Since circular polarization data for J0440$-$4333 is unavailable, we model its emission and propagate it through the same leakage and flux transfer chain,
\begin{equation}
    U_\text{obs}^\text{pred} = \mathcal{F}_{\mathrm{PS}} \left[ \eta_V \left(
              V_\text{b}
              + \epsilon^{UV} U_\text{b}
              + \epsilon^{QV} Q_\text{b} \right)
              + \epsilon_\text{tot}^{IU,\mathrm{eff}} I_\text{b}
          \right],
\label{eq:blazar_pred}
\end{equation}
where $b$ denotes the blazar. The total-intensity flux $I_\text{b}$ is estimated from ALMA catalog queries \citep{bonato_almacal_2018} and the circular polarization fraction assumed to be $p_V = 0.3\% \pm 0.2\%$ at \qty{15}{\giga\Hz} \citep{homan_mojave_2006}. Since the linear polarization angle is unconstrained, $Q_\text{b}$ and $U_\text{b}$ are marginalized uniformly on $[0,\pi)$, introducing a variance
\begin{equation}
    \sigma_{\mathrm{pol}}^{2}
    = \tfrac{1}{2}\,\eta_V^2 \left[ \left( \epsilon^{UV} \right)^{2} + \sigma^{2}_{\epsilon^{UV}} \right]
          \langle p^{2} \rangle\, I_\text{b}^{2},
    \quad
    \langle p^{2} \rangle = p_{\mathrm{lin}}^{2} + \sigma_{p}^{2},
\end{equation}
on the prediction, where $\sigma_{\epsilon^{UV}}$ and $\sigma_p$ are the uncertainties on the leakage $\epsilon^{UV}$ and on the linear polarization fraction, the latter taken as $p_{\mathrm{lin}} = 0.06 \pm 0.04$ scaled from the \qty{18}{\giga\Hz} measurements of \citet{massardi_polarization_2013}; $\epsilon^{QV}$ is neglected as small. The leakage spectrum $\epsilon_{\ell,\text{tot}}^{IU}$ of Equation~\eqref{eq:eps_ItoU} is collapsed with the same matched-filter weighting to $\epsilon_\text{tot}^{IU,\mathrm{eff}} = 0.01751$, a value insensitive to that weighting. It is measured from Jupiter, itself a point source, and is flat enough across the band ($0.0166$--$0.0188$) that any reasonable choice alters it by at most \qty{7}{\percent}.

\begin{figure}[htbp]
    \centering
    \includegraphics[width=\linewidth]{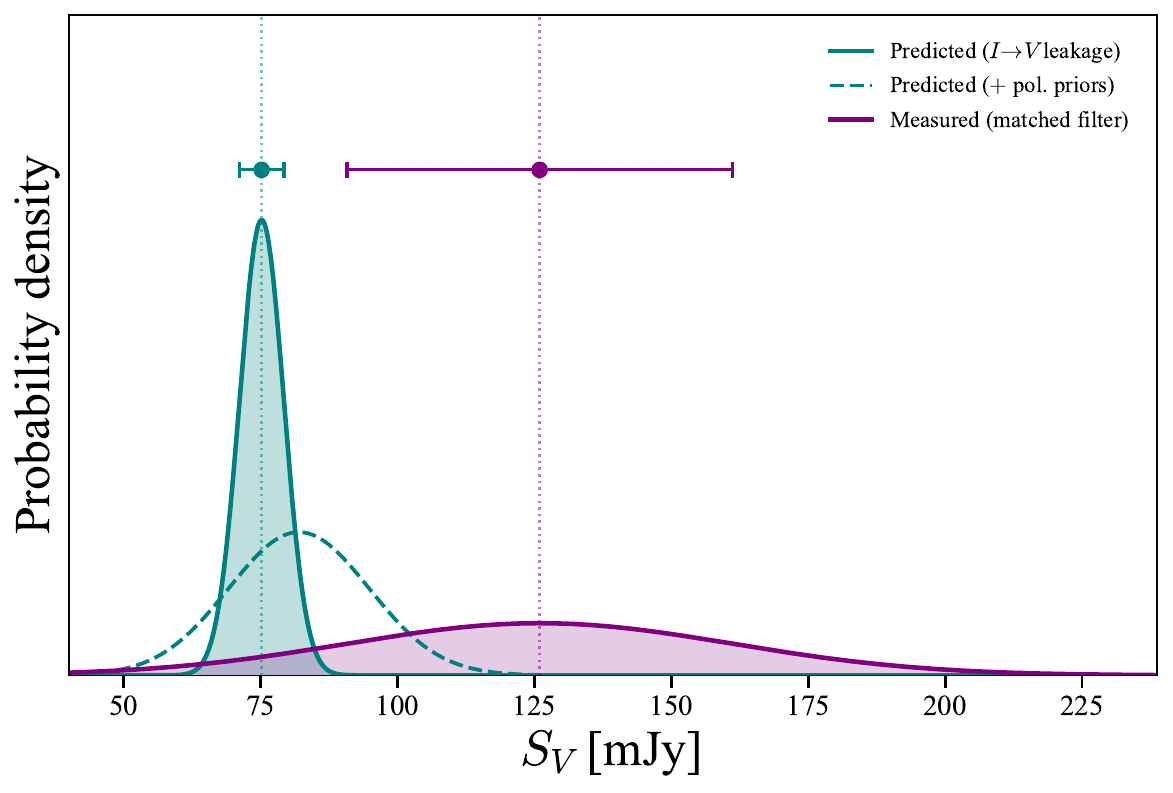}
    \caption{Circular polarization flux density of J0440$-$4333, measured against prediction. The measurement is the matched filter of Equation~\eqref{eq:matched_filter} applied to the coadded RA4.5 $U_\text{obs}$ map, divided by the injection-calibrated flux transfer $\mathcal{F}_{\mathrm{PS}}$. Two predictions are shown: one retaining only the instrumental leakage terms of Equation~\eqref{eq:blazar_pred}, and one additionally assuming the intrinsic polarization priors described in the text. Error bars are $1\sigma$, propagated at first order.}
    \label{fig:flux_dist}
\end{figure}

The comparison is shown in Figure~\ref{fig:flux_dist}. The matched filter gives $\qty{125.9 \pm 35.2}{\milli\jansky}$, a $3.6\sigma$ detection of the source in the circular polarization channel, while the $10'$ aperture on the same map returns $\qty{149.3 \pm 113.4}{\milli\jansky}$, consistent but $3.2$ times less precise. Retaining only the leakage terms of Equation~\eqref{eq:blazar_pred} predicts $\qty{75.2 \pm 4.0}{\milli\jansky}$, differing by $\qty{50.8 \pm 35.5}{\milli\jansky}$ ($1.4\sigma$), while adding the intrinsic polarization priors predicts $\qty{82.0 \pm 12.8}{\milli\jansky}$, differing by $\qty{43.9 \pm 37.5}{\milli\jansky}$ ($1.2\sigma$). Both are quoted because J0440$-$4333 is a variable FSRQ, so neither a single assumed pair of polarization fractions nor their omission is defensible alone; both move the comparison toward agreement, making the leakage-only prediction the conservative statement.

The statistical power of this check is limited and should be stated plainly. Taking the prediction with intrinsic polarization priors as truth, the measurement constrains the flux transfer to $\mathcal{F}_{\mathrm{PS}} = 0.94 \pm 0.30$, so it is sensitive only to errors above ${\sim}\qty{32}{\percent}$. The comparison therefore does not itself determine $\mathcal{F}_{\mathrm{PS}}$, which rests on Equation~\eqref{eq:aper_kernel} and the injection; what the agreement establishes is that no gross error is present in the calibration and mapmaking chain, complementing the characterization in \citet{farias_performance_2026} and \citet{sakaguri_development_2026}. This appendix does not bear on the science result, since Equation~\eqref{eq:forward_model} deconvolves $B_\ell^2$ and $T_\ell^2$ bandpower by bandpower rather than through any scalar collapse.

\bibliographystyle{aasjournalv7}
\bibliography{references_formatted}

\end{document}